\documentclass[twocolumn]{article}
\usepackage{preprint}
\usepackage{hyperref}
\usepackage[numbers,square]{natbib}

\usepackage{color}
\usepackage[utf8]{inputenc} 
\usepackage[T1]{fontenc}    
\usepackage{hyperref}       
\usepackage{url}            
\usepackage{booktabs}       
\usepackage{amsfonts}       
\usepackage{nicefrac}       
\usepackage{microtype}      
\usepackage{lipsum}		
\usepackage{graphicx}
\usepackage{natbib}
\usepackage{doi}
\usepackage{makecell}
\usepackage{wrapfig}
\usepackage{amsmath}
\usepackage{multirow}  
\usepackage{array}     
\usepackage[ruled]{algorithm2e} 
\SetAlFnt{\small}  

\SetAlCapHSkip{0pt}

\usepackage{geometry}
\usepackage{geometry}
\hypersetup{colorlinks=true, linkcolor=purple, urlcolor=blue, citecolor=cyan, anchorcolor=black}

\usepackage[utf8]{inputenc}	
\usepackage[T1]{fontenc}
\usepackage{xcolor}
\usepackage{lineno}					
\usepackage{tikz} 					

\usepackage{newfloat}
\DeclareFloatingEnvironment[name={Supplementary Figure}]{suppfigure}
\usepackage{sidecap}
\sidecaptionvpos{figure}{c}

\usepackage{float}  
\newfloat{teaserfigure}{t}{lop} 

\usepackage{titlesec}
\titlespacing\section{0pt}{12pt plus 3pt minus 3pt}{1pt plus 1pt minus 1pt}
\titlespacing\subsection{0pt}{10pt plus 3pt minus 3pt}{1pt plus 1pt minus 1pt}
\titlespacing\subsubsection{0pt}{8pt plus 3pt minus 3pt}{1pt plus 1pt minus 1pt}

\definecolor{lime}{HTML}{A6CE39}

\date{}

\title{Diffusing Winding Gradients (DWG): A Parallel and Scalable Method for 3D Reconstruction from Unoriented Point Clouds}

\author{Weizhou Liu$^*$\\School of Artificial Intelligence\\Beijing Normal University\\China\And
Jiaze Li\thanks{W. Liu and J. Li contribute equally to the project. }\\College of Computing and Data Science\\Nanyang Technological University\\Singapore 
\And Xuhui Chen\\Institute of Software\\Chinese Academy of Sciences\\China
\And Fei Hou\\Institute of Software\\Chinese Academy of Sciences\\China
\And Shiqing Xin\\School of Computer Science\\Shandong University\\China
\And Xingce Wang\\School of Artificial Intelligence\\Beijing Normal University\\China
\And Zhongke Wu\\School of Artificial Intelligence\\Beijing Normal University\\China
\And Chen Qian\\SenseTime Research\\China
\And Ying He\thanks{Corresponding author: Y. He (yhe@ntu.edu.sg) }\\S-Lab\\Nanyang Technological University\\Singapore
}
\renewcommand{\shorttitle}{\textit{arXiv} Template}

\begin{document}

\maketitle

\begin{abstract}
This paper presents a new method, Diffusing Winding Gradients (DWG), for reconstructing watertight 3D surfaces from unoriented point clouds. \textcolor{black}{Our method exploits the alignment between the gradients of the generalized winding number (GWN) field and globally consistent normals to orient points effectively. Starting with an unoriented point cloud, DWG initially assigns a random normal to each point. It computes the corresponding GWN field and extract a level set whose iso-value is the average GWN values across all input points. The gradients of this level set are then utilized to update the point normals. This cycle of recomputing the GWN field and updating point normals is repeated until the GWN level sets stabilize and their gradients cease to change. Unlike conventional methods, our method does not rely on solving linear systems or optimizing objective functions, which simplifies its implementation and enhances its suitability for efficient parallel execution. } Experimental results demonstrate that our method significantly outperforms existing methods in terms of runtime performance. For large-scale models with 10 to 20 million points, our CUDA implementation on an NVIDIA GTX 4090 GPU achieves speeds 30-120 times faster than iPSR, the leading sequential method, tested on a high-end PC with an Intel i9 CPU. \textcolor{black}{Additionally, by employing a screened variant of GWN, DWG demonstrates enhanced robustness against noise and outliers, and proves effective for models with thin structures and real-world inputs with overlapping and misaligned scans. For source code and more details, visit our project webpage: \url{https://dwgtech.github.io/}.}
\end{abstract}

\section{Introduction}
\label{sec:introduction}

\begin{figure*}
\centering
\newcommand{\teasereachsize}{0.0945}
\newcommand{\teasereachsizetwo}{0.105}
\setlength\tabcolsep{0pt}
\begin{scriptsize}
\begin{tabular}{cccccccccc}
\includegraphics[width=\teasereachsize\textwidth]{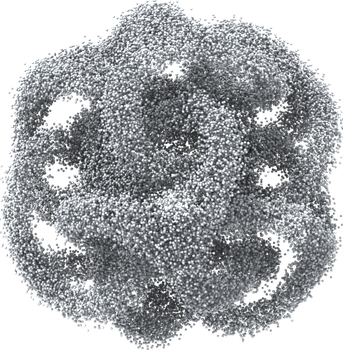}
&\includegraphics[width=\teasereachsize\textwidth]{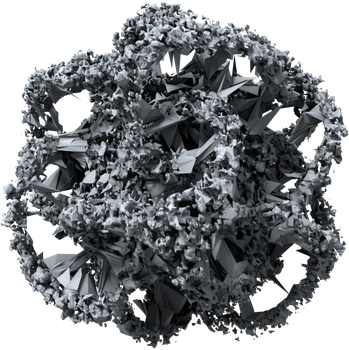} 
&\includegraphics[width=\teasereachsize\textwidth]{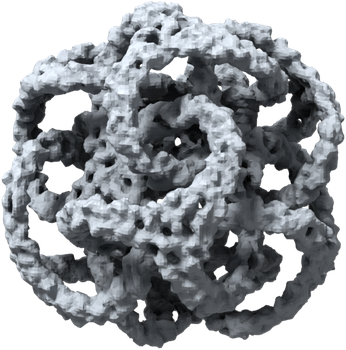}
&
\includegraphics[width=\teasereachsize\textwidth]{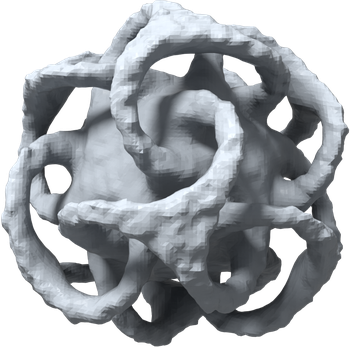}
&
\includegraphics[width=\teasereachsize\textwidth]{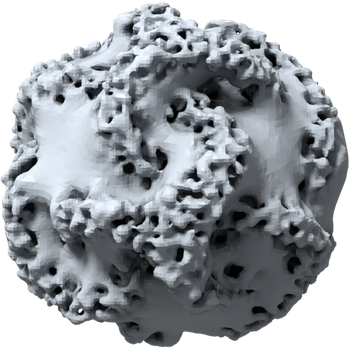}
&
\includegraphics[width=\teasereachsize\textwidth]{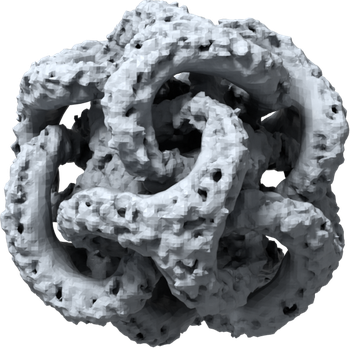}
&
\includegraphics[width=\teasereachsize\textwidth]{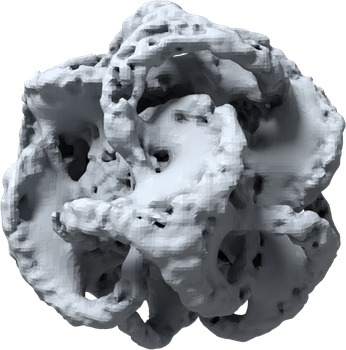}&
\includegraphics[width=\teasereachsize\textwidth]{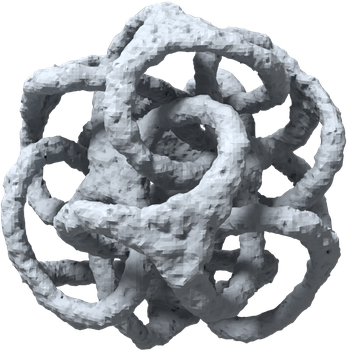}&
\includegraphics[width=\teasereachsize\textwidth]{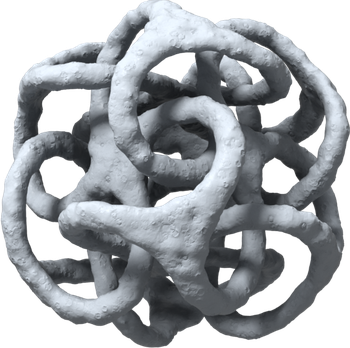}
&
\includegraphics[width=\teasereachsize\textwidth]{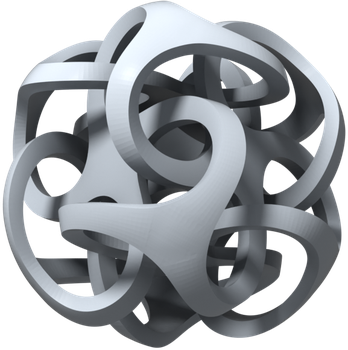}\\
     70K pts  & 00:00:39 & 00:03:09& 00:11:28& 23:09:49&  12:15:39 & 00:22:11 & 00:00:09 & 00:00:01 & \\ 
     0.75\% noise & 1451 MB  & 226 MB & 866 \& 22308 MB& 1132 MB& 180 MB & 5690 MB & 5021 \& 2239 MB & 729 MB & \\ 
      & 0.05592  & 0.01209 & 0.02887 & 0.02574 & 0.02838 &  0.02299 & 0.00879 & 0.00814 & \\      
        Input & Mullen et al. & iPSR& PGR& GCNO& BIM& \textcolor{black}{SNO} & \textcolor{black}{WNNC} & DWG& GT\\      
\end{tabular}
\begin{tabular}{ccccccccc}
\includegraphics[width=\teasereachsizetwo\textwidth]{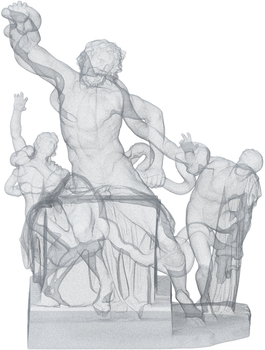}&
\includegraphics[width=\teasereachsizetwo\textwidth]{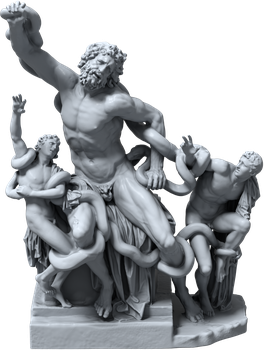}&
\includegraphics[width=\teasereachsizetwo\textwidth]{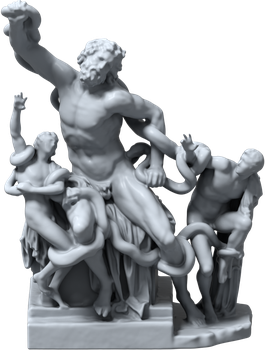}&
\includegraphics[width=\teasereachsizetwo\textwidth]{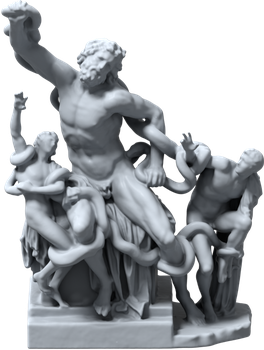}&
\includegraphics[width=\teasereachsizetwo\textwidth]{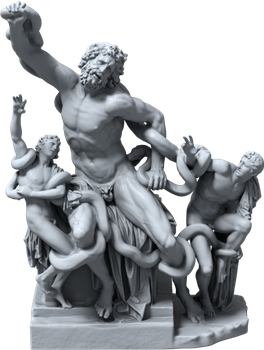}&
\includegraphics[width=\teasereachsizetwo\textwidth]{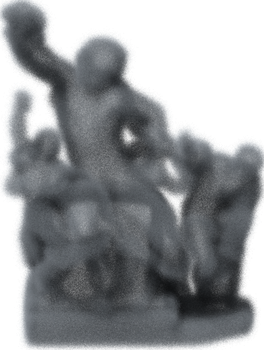}&
\includegraphics[width=\teasereachsizetwo\textwidth]{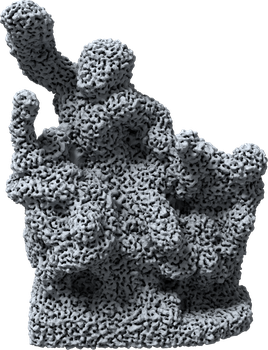}&
\includegraphics[width=\teasereachsizetwo\textwidth]{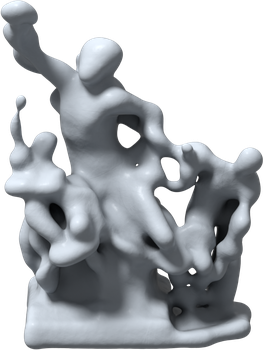}&
\includegraphics[width=\teasereachsizetwo\textwidth]{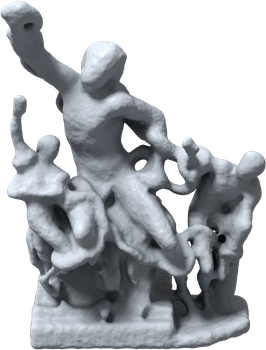}
\\
10M points & 08:59:54 & 00:18:02 & 00:04:00 & 00:12:47 & 10M points & 06:35:53 & 00:16:21 & 00:04:15 \\
 0\% noise & 34321 MB & 74952 \&  7212 MB &  24059 MB & 24059 \& 14001 MB & 0.75\% noise &  20972 MB & 31924 \& 6215 MB &  16553 MB \\
 & 0.00153 & 0.00181 & 0.00185 & 0.00156 & &0.01301   & 0.00823 & 0.00543 \\
  Input & iPSR & \textcolor{black}{WNNC} & DWG   & DWG+sPSR & Input  & iPSR& \textcolor{black}{WNNC} & DWG \\
\end{tabular}
\end{scriptsize}
  \caption{\textcolor{black}{DWG distinguishes itself from existing surface reconstruction methods for unoriented points, such as classic techniques \cite{kobbelt} and \cite{mullen2010signing}, and recent approaches including iPSR~\cite{hou2022iterative}, PGR~\cite{lin2022surface}, GCNO~\cite{xu2023globally},  BIM~\cite{BIM}, SNO~\cite{Huang2024Stochastic} and WNNC~\cite{Lin2024fast}. DWG operates without reliance on external numerical solvers for linear systems or optimization, and is fully parallel and scalable. By utilizing a screened variant of the generalized winding number, DWG achieves enhanced robustness against noise and outliers, and effectively handles models with thin structures.} We report the running time (h:m:s), peak memory consumption (MB) and Chamfer distance for each method. DWG consistently outperforms the others in terms of running time for both small-scale and large-scale models. Our tests were conducted on an NVIDIA RTX 4090 GPU with 24 GB of memory for the parallel methods, DWG and WNNC, and on an Intel i9 CPU with 120 GB of memory for the sequential methods. For noisy models, each point in the point cloud was subjected to Gaussian noise at a level of 0.75\% relative to the bounding box diagonal. The term ``DWG+sPSR'' refers to the application of normals predicted by DWG to the screened PSR solver to generate the final surface, effectively preserving fine geometric details in reconstructed surfaces under noise-free conditions.}  \label{fig:teaser}
\end{figure*}
 
3D reconstruction from point clouds is a fundamental problem in computer graphics and 3D vision. Early approaches based on computational geometry, such as Power Crust~\cite{Amenta2001} and Tight CoCone~\cite{Dey2003}, are efficient and capable of generating high-quality triangle meshes. However, they require the input point clouds to be clean and are sensitive to noise and outliers. In contrast, implicit function approaches~\cite{hoppe1992surface,Kazhdan2006PoissonSR,kazhdan2013screened} are known for their robustness. These methods compute an implicit function that characterizes the interior and exterior of the 3D surface and the target surface is then extracted using iso-surfacing methods, such as Marching Cubes~\cite{Lorensen1987}. While these approaches offer improved resilience compared to computational geometry methods, they often require the input point clouds to be oriented, meaning each point is associated with a globally consistent outward normal. However, point orientation is a challenging problem due to its combinatorial optimization nature.

\textcolor{black}{Early approaches for surface reconstruction from unoriented points~\cite{kobbelt,mullen2010signing} first compute an inner and outer envelopes, then estimate normal orientations between these envelopes using graph cuts or optimization techniques.} \textcolor{black}{These methods are elegant and perform well for clean models and those with moderate noise levels. However, they face significant challenges when applied to models with high noise levels, outliers or thin structures, primarily due to their reliance on voxelization. For noisy models, voxels that should represent the same geometric feature often end up being incorrectly separated; for models with thin structures, spatially-adjacent-but-topologically-distinct voxels may be incorrectly merged. Adjusting voxel resolution alone does not resolve the issues for inputs with complex geometries and non-uniform sampling rates.}

\textcolor{black}{
Recent trends indicate a shift from classic methods, which often involve techniques like graph cuts and Voronoi diagrams, toward more robust approaches that utilize emerging computational tools like generalized winding number (GWN)~\cite{jacobson2013winding}, \cite{Barill2018FastWN} and screened Poisson equation~\cite{kazhdan2013screened}.} 
Starting with randomly initialized normals, these methods iteratively refine the implicit function until convergence is reached. The key distinction between these approaches lies in their formulation of the implicit functions.
Iterative Poisson surface reconstruction (iPSR)~\cite{hou2022iterative} utilizes PSR's indicator function~\cite{Kazhdan2006PoissonSR}, which is computed by solving Poisson's equation. This approach aligns the gradient of the indicator function with the iteratively updated normal fields. Parametric Gaussian reconstruction (PGR)~\cite{lin2022surface}, the globally consistent normal orientation (GCNO) method~\cite{xu2023globally} and the boundary integration method (BIM)~\cite{BIM} all use the GWN~\cite{jacobson2013winding,Barill2018FastWN} as their foundation. In this computational framework, GWN, which measures the number of times a point is enclosed by a closed and orientable 3D surface, serves as a robust computational tool for inside-outside segmentation. 
Specifically, PGR~\cite{lin2022surface} and GCNO~\cite{xu2023globally} directly regularize the GWN values, while BIM~\cite{BIM} optimizes the boundary energy derived from the Dirichlet energy of GWN. As PGR requires solving a dense linear system and GCNO and BIM necessitate the incorporation of additional geometric structures, such as Voronoi cells, to facilitate the discretization of space, these methods are computationally intensive and exhibit limited scalability, constraining their applications to small- and medium-scale models.

This paper aims to address the challenges of poor scalability and runtime performance in the current GWN-based 3D reconstruction methods. Towards this goal, we propose a new method, called diffusing winding gradients (DWG).  Starting with randomly initialized normals, our method iteratively updates the normal for each point by diffusing the normalized gradient of the GWN field associated with the current normals. Upon convergence, the target surface can be directly extracted from the GWN field via Marching Cubes. We also adopt a screened variant of the conventional GWN for making DWG robust against noise, outliers and thin structures. Our method is conceptually simple and easy to implement, as it does not rely on any numerical solvers, which are indispensable to all other approaches. Additionally, it is parallel and scalable, making it suitable for handling large-scale models on both CPUs and GPUs. Our CUDA implementation achieves high-quality reconstruction results for models with 10 to 20 million points on an NVIDIA RTX 4090 GPU in just 4 to 8 minutes. This performance is 30 to 120 times faster than iPSR~\cite{hou2022iterative} and \textcolor{black}{4 to 10 times faster than WNNC~\cite{Lin2024fast}}, the state-of-the-art techniques for 3D reconstruction from unoriented point clouds. See Figure~\ref{fig:teaser} for a comparative illustration of the results generated by DWG and other methods.

Our contributions are threefold: \textcolor{black}{1) We introduce a novel perspective to the well-explored domain of 3D reconstruction by presenting a parallel and scalable computational framework that constructs watertight 3D surfaces from unoriented point clouds. Unlike existing methods, our approach features a straightforward, solver-free implementation that avoids complex numerical solvers. It is characterized by low computational cost and a highly parallelized architecture. 2) We develop a CUDA program that excels in efficiency and scalability, significantly outperforming existing methods in terms of speed across on both small- and large-scale models. 3) By utilizing a screened variant of GWN, DWG enhances robustness against  challenges such as noise, outliers, thin structures, and registration misalignments. We demonstrate its effectiveness on real-world scans and point clouds generated from 3D Gaussian Splatting. DWG's high performance and reliability push the boundaries of what is achievable in 3D surface reconstruction. }

\section{Related Work}
\label{sec:relatedwork}

Reconstructing 3D surfaces from point clouds has been studied extensively in the last three decades. A popular technique for surface reconstruction from oriented point clouds 
is to establish a non-degenerate scalar field, where a specified level-set corresponds to the desired surface~\cite{kolluri2008imls, oztireli2009RIMLS, shen2004interpolating, schroers2014HessianIMLS, ohtake2003multi,SSD}. These scalar fields can take various forms, including signed distance fields (SDF)~\cite{hoppe1992surface,carr2001reconstruction,ohtake2003multi}, inside-outside indicator fields ~\cite{Kazhdan2006PoissonSR,kazhdan2013screened,Kazhdan2020PoissonSR,SPSR2022}, occupancy fields~\cite{Occupancy_Networks}, and winding number fields~\cite{Barill2018FastWN}; here, accurate normal orientations play a crucial role in defining the implicit field.

Unoriented point clouds present significant challenges to surface reconstruction. Achieving normal recovery is relatively straightforward if the point cloud conforms to the strict local feature size condition~\cite{alliez2007voronoi,dey2005adaptive}, but this condition is not universally applicable. A commonly employed technique for orienting point clouds involves estimating initial normals via local methods, such as local fitting or principal component analysis, and then adjusting their orientations using propagation rules to ensure coherent orientation across the cloud ~\cite{hoppe1992surface,Huang2009,Huang2013,metzer2021orienting}. However, these propagation-based methods often struggle with models that have complex topologies and geometries, particularly those featuring protruding parts and thin structures. 

Given the global and combinatorial optimization challenges involved in establishing globally consistent orientations, there is a growing research trend focused on methods that bypass the direct computation of normals. \textcolor{black}{Classic methods such as ~\cite{mullen2010signing}~\cite{kobbelt} utilize voxelization to discretize the space and employ inner and outer envelopes to guide the surface reconstruction process.} The VIPSS method \cite{VIPSS} employs Duchon's energy to define the desired signed function and is effective for sparse point clouds. The PGR method \cite{lin2022surface} treats both the surface normals and the areas of surface elements as variables, applying the Gauss formula to conceptualize the indicator function as part of a defined parametric function space. Both VIPSS and PGR involve solving dense linear systems, which limits their applicability to small-scale models. The iterative PSR (iPSR) method \cite{hou2022iterative} begins with randomly oriented normals and progressively refines the surface by incorporating the newly computed normals with each iteration. This method inherits the efficiency of traditional PSR solver~\cite{kazhdan2005reconstruction,kazhdan2013screened}, making it suitable for large-scale models. Our method, which leverages a parallel diffusion process, achieves even greater efficiency and scalability than iPSR. 

The generalized winding number, effective in distinguishing interior and exterior spaces, has gained increasing attention for its utility in orienting point clouds. \citet{xu2023globally} observed that an optimal arrangement of normals yields a winding-number field that is binary-valued—either 0 or 1. This property facilitates the identification of unknown normals through the regularization of the winding number distribution. Building on this, \citet{BIM} introduced the boundary integration method, which regularizes the GWN field by aligning its gradient with the point normals. 

\textcolor{black}{Using Stokes' theorem, \citet{Kai_linear} reduced the point orientation problem to solving a sparse linear system with binary solutions. To avoid trivial solutions due to the under-constrained nature of this system, they introduced additional constraints. However, implementing these regularizations requires computing several cross sections on the 3D model, a task equivalent to solving a traveling salesman problem. Consequently, they opted for an approximate solution. Finally, they determined the orientation of the input point cloud by rounding the least square solution of the over-constrained linear system. Given that the computation of the cross-section constraints is computationally expensive, particularly for models with complex geometry and topologies, their method is unsuitable for large-scale or complex models. Their reported results limited to small-scale models with up to 10K points, featuring simple geometry and topology.}

\textcolor{black}{There are a few concurrent works related to normal orientation and surface reconstruction from unoriented point clouds. Anisotropic Gauss Reconstruction (AGR)~\cite{Ma2024APGR}, an extension of PGR~\cite{lin2022surface},  incorporates a convection term into the traditional Laplace equation to introduce anisotropy, which enhances handling of models with thin structures. However, AGR remains computationally intensive for large-scale models. WNNC~\cite{Lin2024fast} accelerates PGR~\cite{lin2022surface} by employing winding number normal consistency, significantly improving its runtime performance. Despite these advancements, both AGR and WNNC assume that the GWN at the point cloud is precisely $\frac{1}{2}$--a condition often compromised by noise or registration misalignments, limiting their robustness against defective inputs.} \textcolor{black}{\citet{Huang2024Stochastic} employ a stochastic probability model to introduce a signed uncertainty function that differentiates between the inside and outside of a point cloud model. Their method also incorporates statistical uncertainty information to supplement global estimates with local directional estimations. It is simple and elegant, however, its reliance on intensive optimization makes it impractical for handling large-scale models. }

\textcolor{black}{
Our approach also utilizes GWN regularization but stands out from existing GWN-based methods, including ~\cite{lin2022surface},~\cite{xu2023globally}, \cite{BIM},~\cite{Kai_linear},~\cite{Ma2024APGR} and~\cite{Lin2024fast}. Unlike these methods, which rely on computationally intensive optimization processes suitable only for small- to middle-scale models, our approach leverages full parallelization and scalability to efficiently handle large-scale models on GPUs. Additionally, our method employs gradients of the GWN rather than its values, enhancing its resistance to noise and registration misalignments. By incorporating a screened variant of GWN that decays faster than the standard form, our method demonstrates superior performance on models with thin structures. Our results consistently show improvements over all existing GWN-based methods in speed, accuracy and robustness. }

Deep learning-based techniques have recently gained popularity in surface reconstruction. Supervised methods, in particular, excel at processing noisy and incomplete inputs without requiring explicit point normals and orientations~\cite{Occupancy_Networks, DeepLS, jiang2020local, SA-ConvONet, park2019deepsdf, Points2surf, tang2021octfield, wang2022dual, lin2023patchgrid, peng2021shape, huang2022neural, Peng20ConvONet, POCO}. However, these methods typically require expensive hardware setups and extensive training. Their performance heavily relies on the quality and diversity of the training datasets, and they often struggle with unseen models, necessitating frequent adjustments of hyperparameters for new categories. Unsupervised and self-supervised methods~\cite{CAP-UDF, PCP, SAL, SALD, ben2022digs, wang2023neural} alleviate reliance on extensive datasets but remain computationally demanding. Despite these advancements, no existing learning-based method can handle large-scale models as efficiently as our approach.

\section{Preliminaries}
\label{sec:preliminaries}

This section provides a brief review of the background knowledge on generalized winding numbers. For more details, we refer readers to the seminal papers~\cite{jacobson2013winding,Barill2018FastWN} and  many applications that involve winding numbers, including 
reconstruction~\cite{lin2022surface, BIM}, orientation \cite{xu2023globally}, boolean operation~\cite{WindingNumberOnDiscreteSurface}, inside/outside segmentation~\cite{jacobson2013winding}, point location~\cite{PointinPolygonStrategies}, just name a few.

The winding number $w$ in 2D is a signed, integer-valued property of a query point $\mathbf{q}\in\mathbb{R}^3$ with respect to a closed curve $\mathcal{C}$, quantifying the number of times $\mathbf{q}$ is enclosed by $\mathcal{C}$. When parameterized using polar coordinates, $w(\mathbf{q})$ is defined as $w(\mathbf{q})=\frac{1}{2\pi}\oint_\mathcal{C}\mathrm{d}\theta$, where $\theta$ is the polar angle. By replacing the polar angle $\theta$ with the solid angle $\Omega$, the winding number can be generalized to $\mathbb{R}^3$~\cite{jacobson2013winding} as 
\begin{equation}
    w(\mathbf{q})=\frac{1}{4\pi}\oint_{\mathcal{S}}\mathrm{d}\Omega=\oint_{\mathcal{S}}\frac{\langle\mathbf{x}-\mathbf{q},\mathbf{n}_{\mathbf{x}}\rangle}{4\pi\|\mathbf{x}-\mathbf{q}\|^3}\mathrm{d}\mathbf{x},
    \label{eqn:continuous_gwn}
\end{equation} where $\mathcal{S}$ is a closed orientable 3D surface, $\mathbf{n}_{\mathbf{x}}$ is the outward facing unit normal vector at point $\mathbf{x}\in\mathcal{S}$, and $\langle,\rangle$ is the vector dot product.

For discrete objects such as point clouds, the generalized winding numbers can be computed by discretizing the above surface integral. Noting that the integrand is the normal derivative of Green's function of Laplace's equation in $\mathbb{R}^3$, \citet{Barill2018FastWN} approximated the GWN for an oriented point cloud $\mathcal{P} = \{\mathbf{p}_i\}_{i=1}^n$ as follows:
\begin{equation}
w(\mathbf{q}) \approx \sum_{i=1}^n a_i P_{\mathbf{p}_i}(\mathbf{q}),
\label{eqn:discete_gwn}
\end{equation}
where $a_i$ is the geodesic Voronoi area of the point $\mathbf{p}_i$ on the surface and $\mathbf{n}_i$ is the outward unit normal of $\mathbf{p}_i$, and 
\begin{equation}
\label{eqn:poissonkernel}
    P_\mathbf{x}(\mathbf{y}) = \frac{1}{4\pi} \frac{\langle\mathbf{x}-\mathbf{y},\mathbf{n}_\mathbf{x}\rangle}{\|\mathbf{x}-\mathbf{y}\|^3} 
\end{equation} is the Poisson kernel of point $\bf x$. To avoid division by zero, the evaluation of the generalized winding number for an input point $\mathbf{p}_i$ involves adding a small constant to the denominator.



\section{Diffusing Winding Gradients}

\subsection{Motivation}
\label{subsec:motivation}

Given a smooth, closed, orientable manifold surface $\mathcal{S}$ and an arbitrary normal field $\mathbf{n}=\{\mathbf{n}_\mathbf{p}|\mathbf{p}\in\mathcal{S}\}$, 
we extend the definition of the generalized winding number $w(\mathbf{q})$ to include $\bf n$ as a variable: 
$$w(\mathbf{q}, \mathbf{n})=\oint_{\mathcal{S}}\frac{\langle\mathbf{x}-\mathbf{q},\mathbf{n}_{\mathbf{x}}\rangle}{4\pi\|\mathbf{x}-\mathbf{q}\|^3}\mathrm{d}\mathbf{x},~~~~~\forall \mathbf{q}\in\mathbb{R}^3.$$ Additionally, we treat point normals as time-varying variables. For convenience, we denote  $\mathbf{n}(0)=\left\{\mathbf{n}_{\mathbf{p}}{(0)}\left|\right.\mathbf{p}\in\mathcal{S}\right\}$ as the randomly assigned normals and $\mathbf{n}{(\infty)}=\left\{\mathbf{n}_{\mathbf{p}}{(\infty)}\left|\right.\mathbf{p}\in\mathcal{S}\right\}$ as the globally consistent outward normals for the input points $\{\mathbf{p}_i\}_{i=1}^n$. We compute the gradient of the GWN field $\nabla w(\mathbf{q},\mathbf{n})$ with $\bf n$ held fixed, indicating that the derivatives are taken with respect to the coordinates of the query point $\mathbf{q}\in\mathbb{R}^3$.

\begin{figure}[htbp]
\centering
\includegraphics[width=3.10in]{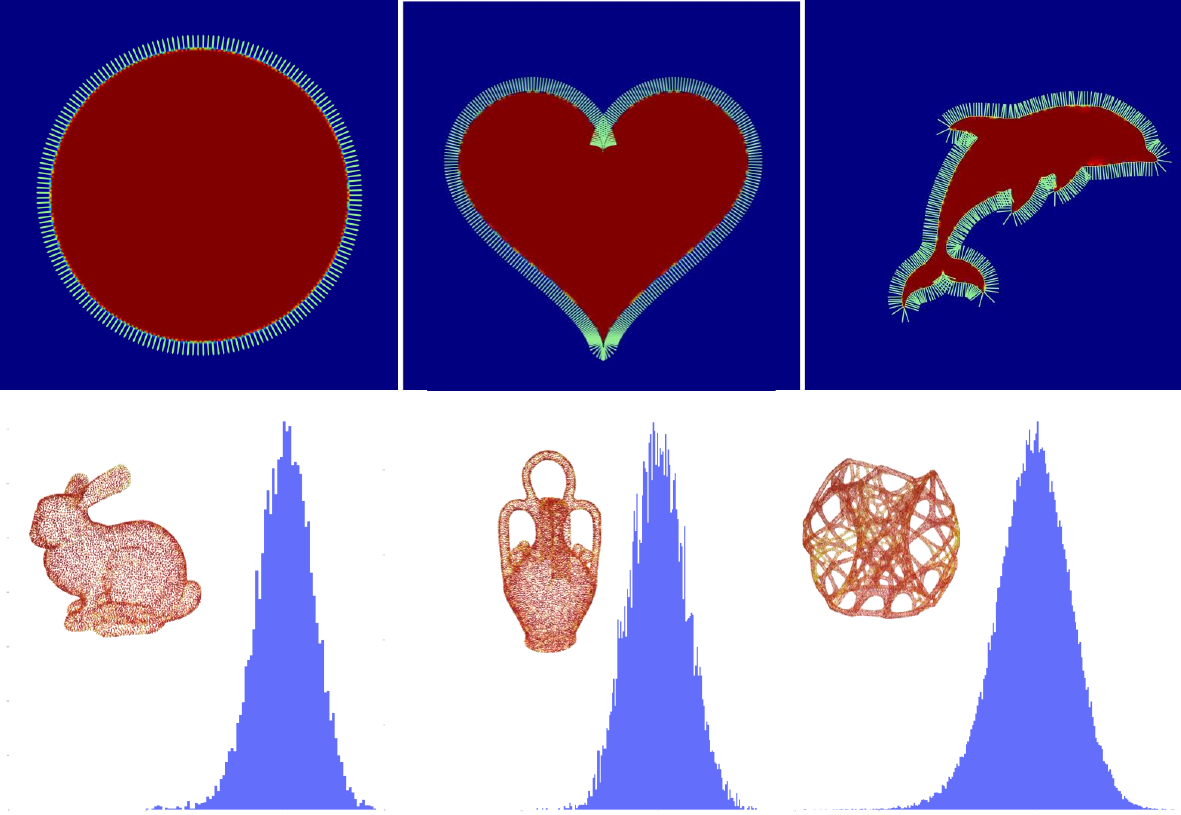}  
\caption{Illustration of alignment between GWN gradient and globally consistent normals. With globally consistent outward normals $\mathbf{n}{(\infty)}$, the gradient $\nabla w(\mathbf{p}_i, \mathbf{n}{(\infty)})$ at each point $\mathbf{p}_i$ closely aligns with its normal $\mathbf{n}_{i}{(\infty)}$, i.e., $\nabla w(\mathbf{p}_i) \approx \mathbf{n}_{i}{(\infty)}$.  The top row visualizes the GWN fields for three 2D curves, with gradients numerically calculated and depicted as short green line segments. Red and blue indicate GWN values of 1 and 0, respectively. The bottom row shows the distribution of dot products $\nabla w(\mathbf{p}_i) \cdot \mathbf{n}_{i}{(\infty)}$ for three point cloud models, with the mean values and the standard deviations for each: 1.007 and 0.104 (Bunny), 0.999 and 0.104 (Bojito), and 0.987 and 0.120 (Bottable). The distributions with mean values close to one and small standard deviations confirm the alignment between the normals and the GWN gradients.}
\label{fig:alignment}
\end{figure}

A key observation is that with globally consistent outward normals $\mathbf{n}{(\infty)}$, the gradient of the GWN at any surface point aligns with its outward normal\footnote{Given that the gradient of the GWN is measured in units of 1/length, by applying appropriate scaling, these gradients can be transformed into unit vectors.}, i.e., 
\begin{equation}
\mathbf{n}_{\mathbf{p}}{(\infty)}\approx\nabla w\left(\mathbf{p}, \mathbf{n}{(\infty)}\right),~\forall \mathbf{p}\in\mathcal{S}.
\label{eqn:observation}
\end{equation}
Figure~\ref{fig:alignment} provides 2D and 3D examples to illustrate this observation.

\textcolor{black}{This alignment of point normals with GWN gradients motivates us to use these gradients for orienting points. Starting with an unoriented point cloud, we initially assign a random normal to each point. We  compute the corresponding GWN field and extract a level set whose iso-value aligns with the average GWN values across all input points. The gradients of this level set are then utilized to update the point normals. This process is repeated by recomputing the GWN field with the updated normals and then updating the point normals using the gradients from the level set of the corresponding GWN field. The cycle continues until the GWN level set stabilizes and its gradients no longer change. Throughout this iterative process, the point normals gradually align with the GWN gradients. }

\begin{figure}[!htbp]
   \centering
   \begin{minipage}{0.98\linewidth}%
    \includegraphics[width=0.125\linewidth]{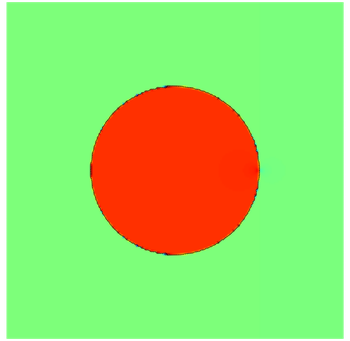}
    \includegraphics[width=0.125\linewidth]{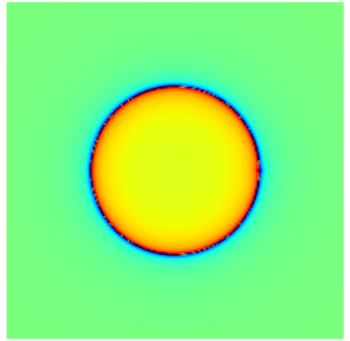}
    \includegraphics[width=0.125\linewidth]{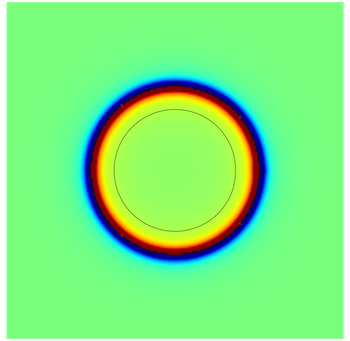}
    \includegraphics[width=0.125\linewidth]{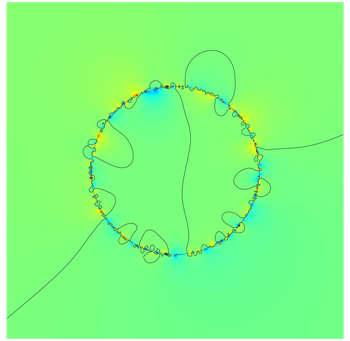}
    \includegraphics[width=0.125\linewidth]{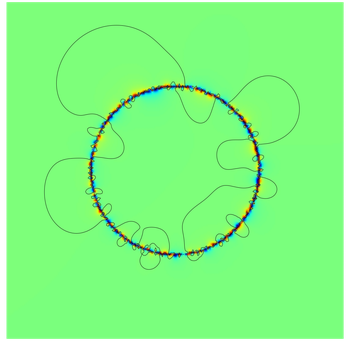}
    \includegraphics[width=0.125\linewidth]{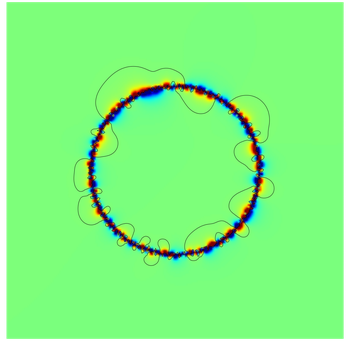}\\
    \includegraphics[width=0.125\linewidth]{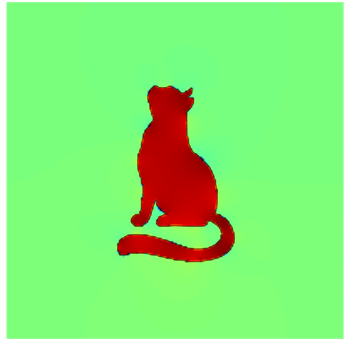}
    \includegraphics[width=0.125\linewidth]{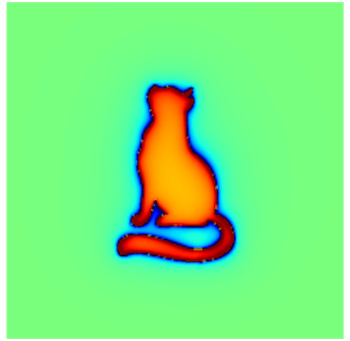}
    \includegraphics[width=0.125\linewidth]{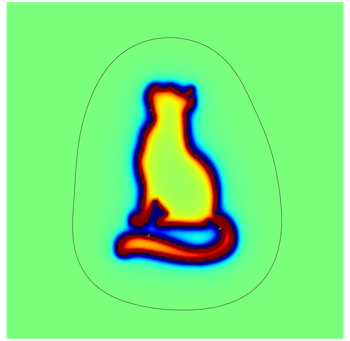}
    \includegraphics[width=0.125\linewidth]{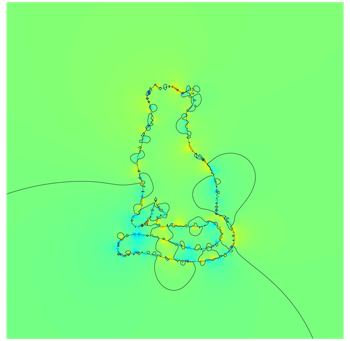}
    \includegraphics[width=0.125\linewidth]{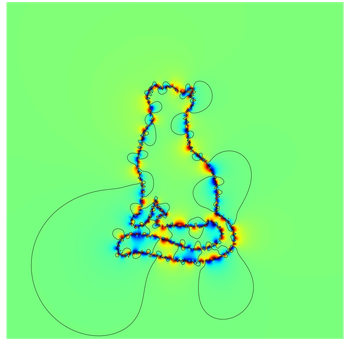}
    \includegraphics[width=0.125\linewidth]{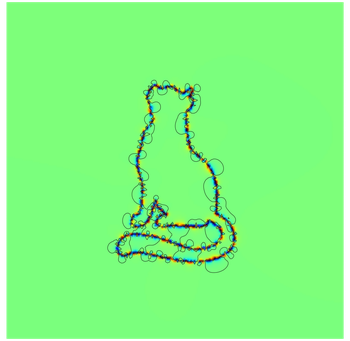}\\
   \makebox[0.125\linewidth]{$\lambda = 0$}
   \makebox[0.125\linewidth]{$\lambda = 10$}
   \makebox[0.125\linewidth]{$\lambda = 400$}
    \makebox[0.125\linewidth]{$\lambda = 0$}
   \makebox[0.125\linewidth]{$\lambda = 10$}
   \makebox[0.125\linewidth]{$\lambda = 400$}\\
   \makebox[0.39\linewidth]{With consistent normals}
   \makebox[0.39
   \textwidth]{With random normals}\\
   \end{minipage}
   \hspace{-0.5in}
   \begin{minipage}{0.01\linewidth} 
        \makebox[0.16\linewidth]{ $>0.2$} \\ \vspace{0.0em}
        \includegraphics[width=\linewidth,height=20\linewidth]{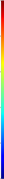} \\ \vspace{-0.6em}
        \makebox[0.16\linewidth]{ $<-0.2$} \\
    \end{minipage}

    \includegraphics[width=0.108\textwidth]{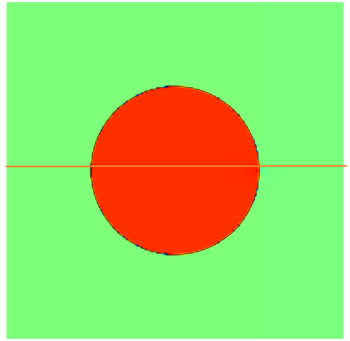}
    \includegraphics[width=0.108\textwidth]{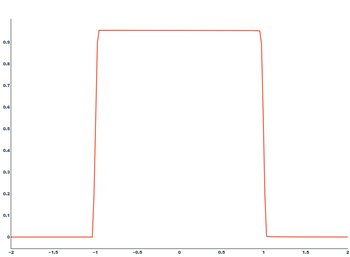}
    \includegraphics[width=0.108\textwidth]{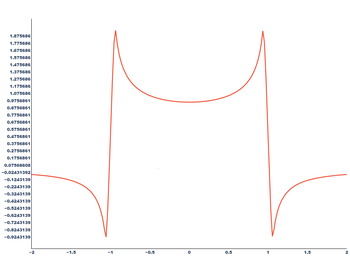}
    \includegraphics[width=0.108\textwidth]{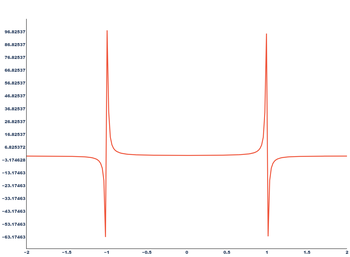}
    \\
    \includegraphics[width=0.108\textwidth]{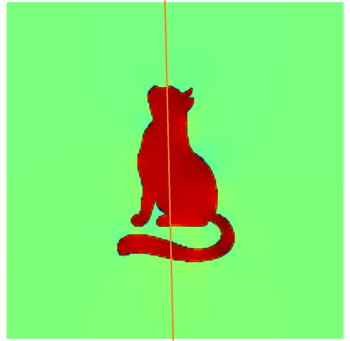}
    \includegraphics[width=0.108\textwidth]{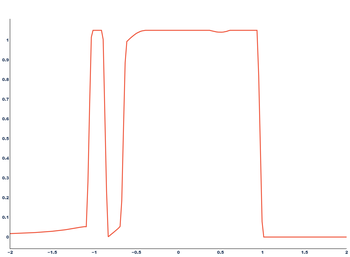}
    \includegraphics[width=0.108\textwidth]{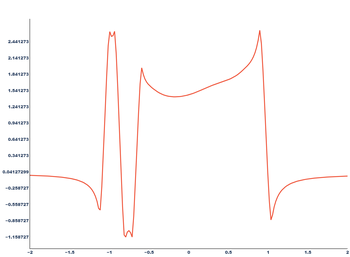}
    \includegraphics[width=0.108\textwidth]{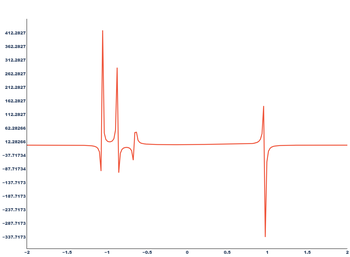}
    \\
    \makebox[0.108\textwidth]{2D models}
    \makebox[0.108\textwidth]{$\lambda = 0$}
    \makebox[0.108\textwidth]{$\lambda = 10$}
    \makebox[0.108\textwidth]{$\lambda = 400$}
    
\caption{Screened GWN fields. Top: Visualization of 2D screened GWN fields applied to two toy models--a circle and a cat. The solid curves show the $\overline{w}_{\lambda}$-level sets of the sGWN fields. This screening process results in iso-curves that are smoother and closer to the input points compared to those generated without screening. \textcolor{black}{Bottom: Illustration of the fast decaying property of screened GWNs. The values of $w$ are visualized along a horizontal line across the circle and a vertical line through the cat. As $\lambda$ increases, $w$ decays more rapidly, pulling the isocurve closer to the input points. However, an excessively large $\lambda$ can cause oscillation around the target shape, resulting in multiple connected components.} }
\label{fig:sgwn}
\end{figure}

\definecolor{RED}{rgb}{1,0,0}
\subsection{Screened GWNs}
\label{subsec:screenedgwn}

\textcolor{black}{A core component of our framework is the alternate computation of the GWN field and the utilization of its level set gradients to update normals for \textbf{nearby} points. For this iterative process to be effective, it is highly desired that the GWN field's level set remains in close proximity to the input points. }

To achieve this, we introduce a variant of the generalized winding number, termed screened GWN (sGWN), which incorporates a screening function to ensure that the influence of distant points decays more rapidly. Consider the screened Poisson equation~\cite{craneMiller2023BVC}: $$\Delta w - \lambda w = f,$$
\textcolor{black}{with jump boundary conditions defined as:
$w^+ - w^- = k$ and 
$\partial w^+ / \partial \mathbf{n}^+ = \partial w^- / \partial \mathbf{n}^-$,
where $f$ represents a source function, $\lambda$ is a screening constant that moderates the equation's smoothing effect, and $k$ serves as a constant boundary condition. The boundary surface $S$ is treated as a two-sided surface with $^+$ and $^-$ denoting the interior and exterior sides, respectively. }

For the homogeneous case $f=0$ \textcolor{black}{with the jump constraint $k=1$}, the solution, denoted as $w_\lambda$, to screened Poisson equation can be formulated using the boundary element method as: 
$$w_\lambda(\mathbf{q}) = \int_{\partial\Omega} \frac{\partial G_\lambda(\mathbf{x}, \mathbf{q})}{\partial \mathbf{n}_\mathbf{x}}\mathrm{d}\mathbf{x} = \int_{\partial\Omega}e^{-r\sqrt{\lambda}}\left(r\sqrt{\lambda}+1\right)\frac{\langle\mathbf{n}_\mathbf{x},\mathbf{x} - \mathbf{q}\rangle}{4\pi r^3},$$ 
where $G_\lambda$ represents Green's function of the screened Poisson equation and
$r = \|\mathbf{x} - \mathbf{q}\|$. Discretizing the surface integral yields
\begin{equation}
w_{\lambda}(\mathbf{q}) \approx \sum_{\mathbf{p}_i \in \mathcal{P}}a_ie^{-\|\mathbf{p}_i-\mathbf{q}\|\sqrt{\lambda}}\left(\|\mathbf{p}_i-\mathbf{q}\|\sqrt{\lambda}+1\right)\frac{\langle\mathbf{n}_\mathbf{i},\mathbf{p}_i - \mathbf{q}\rangle}{4\pi \|\mathbf{p}_i-\mathbf{q}\|^3}.
\label{eqn:sGWN}
\end{equation}
Unlike standard GWN, screened GWN incorporates a decay factor, $e^{-r\sqrt{\lambda}}\left(r\sqrt{\lambda}+1\right)$, which diminishes rapidly as the distance $r$ increases. This property is highly beneficial for dealing with models that exhibit noise and/or feature thin structures.  

\textcolor{black}{The concept of screened generalized winding numbers is inspired by the techniques used in screened Poisson surface reconstruction (sPSR)~\cite{kazhdan2013screened} that enhance the accuracy and robustness of PSR~\cite{kazhdan2005reconstruction}. In sGWN, the screening coefficient $\lambda$ controls the decaying speed of the GWN values for query positions away from the input points. Figure~\ref{fig:sgwn} visualizes the screened GWN field on two 2D examples, under both globally consistent normals and random normals. We observe that increasing the screening coefficient $\lambda$ effectively narrows the regions with non-zero GWN, resulting in a more stable $\overline{w}_\lambda(t)$-level set that is closer to the input points. This characteristic is particularly advantageous for our method when managing noisy inputs and handling models with thin structures. }



\subsection{Algorithm}
\label{subsec:algorithm}

Our algorithm takes an unoriented point cloud $\mathcal{P}=\{\mathbf{p}_i\}_{i=1}^n$ as input. Similar to \cite{kazhdan2013screened}, 
we employ an octree to adaptively discretize the space around $\mathcal{P}$. We define the set of \textbf{non-empty} leaf nodes as $\mathcal{L}=\{\mathbf{s}_j\}_{j=1}^{l}$, where each node $\mathbf{s}_j$ is closely associated with the input point cloud $\mathcal{P}$. Empirically, the number of such nodes, $l$, scales linearly with $n$, i.e., $l=O(n)$.

We associate each input point $\mathbf{p}_i$ a dynamic normal $\mathbf{n}_i^{(t)}$. These normals evolve over time, starting from a random initialization $\mathbf{n}_i^{(0)}$ and converging towards as a globally consistent outward normal $\mathbf{n}_i^{(\infty)}$. 
With the randomly initialized normals $\mathbf{n}^{(0)}$, our method proceeds iteratively. At iteration $t$, it starts by computing the screened GWN field $w^{(t)}_\lambda$ using the current normal field $\mathbf{n}^{(t-1)}$ and the user-specified screening parameter $\lambda$. This involves evaluating the screened winding number $w_{\lambda}$ for the octree nodes $\{\mathbf{s}_j\}_{j=1}^l$ using Equation~(\ref{eqn:sGWN}).  
Subsequently, the algorithm extracts the iso-surface, denoted as $\mathcal{M}^{(t)}$, from the computed sGWN field. The iso-value is set to the average winding number $\overline{w}_\lambda^{(t)}$ for all $\mathbf{s}_j$, defined as  
\begin{equation}
\overline{w}_\lambda^{(t)}=\frac{1}{l}\sum_{j=1}^{l}w^{(t)}_\lambda(\mathbf{s}_i). 
\label{eqn:averagew}
\end{equation} For iso-surface extraction, we simply employ the standard Marching Cubes algorithm. 

\textcolor{black}{
After extracting the $\overline{w}_\lambda^{(t)}$-level set $\mathcal{M}^{(t)}$, we proceed to update the point normals $\left\{\mathbf{n}_i^{(t)}\right\}_{i=1}^{n}$ by utilizing the GWN gradients. Specifically, 
for each triangular face $f$ within $\mathcal{M}^{(t)}$, we approximate the GWN gradient $\nabla w_\lambda^{(t)}(f)$ using the vector cross product and then identify the $k$-nearest points in the input point cloud $\mathcal{P}$, denoted as 
$P(f)$. Next, we use the gradient $\nabla w_\lambda^{(t)}(f)$ to update each of these neighboring points. }

\textcolor{black}{In this updating process, it is common for an input point $\mathbf{p}_j$ to be associated with multiple faces of the level set $\mathcal{M}^{(t)}$, resulting in multiple updates. The gradients from all nearby faces within the $\overline{w}^{(t)}$-level set are used to update the normal of $\mathbf{p}_j$. This process is mathematically expressed as follows: 
\begin{equation}
\mathbf{n}_j^{(t+1)}
= \mathrm{normalize}\left(\sum\limits_{f\in F(\mathbf{p}_j)}\nabla w^{(t)}(f)\right),
\label{eqn:update1}
\end{equation}
where $F(\mathbf{p}_j)$ represents the set of triangular faces from $\mathcal{M}^{(t)}$ influencing point $\mathbf{p}_j$. } 
\textcolor{black}{Unlike the set $P(f)$, which maintains a fixed size $k$ due to $k$-nearest neighbor searching, the set $F(\mathbf{p}_j)$ can vary in size and may sometimes be empty. While it is uncommon, if $|F(\mathbf{p}_j)|=0$, $\mathbf{p}_j$ does not undergo normal updates during the current iteration. Nevertheless, as the overall level set progressively aligns more closely with the input point cloud $\mathcal{P}$, these points will be updated in future iterations. }

We continue the iterative process until the average difference in point normals between two consecutive iterations falls below a user-specified threshold $\epsilon$. 

Since the iterative process gradually aligns point normals and GWN gradients through a diffusion-like smoothing of the gradients, we call our method Diffusing Winding Gradients (DWG). Upon completion, DWG outputs both the $\overline{w}_\lambda^{(\infty)}$-level set of the steady GWN field $w_\lambda^{(\infty)}$ and the final point orientations $\mathbf{n}^{(\infty)}$. 

\begin{figure*}[htbp] \centering
\newcommand{\ninesize}{0.0475}
\newcommand{\ninesizecontour}{0.0505}
\newcommand{\ninesizenormals}{0.0325}
\setlength\tabcolsep{1pt}
   \begin{small}
\begin{tabular}{cccccc|cccccc|cccccc}

    \multicolumn{6}{c|}{0\% noise, $\lambda=0$} &
    \multicolumn{6}{c|}{0\% noise, $\lambda=10$} &
    \multicolumn{6}{c}{0.75\% noise, $\lambda=100$}\\

    \includegraphics[width=\ninesize\textwidth]{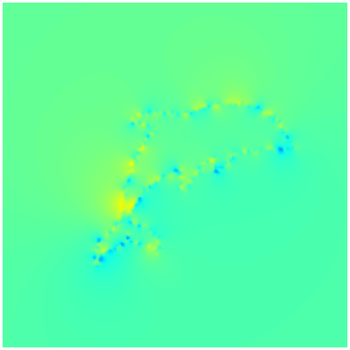} &
    \includegraphics[width=\ninesize\textwidth]{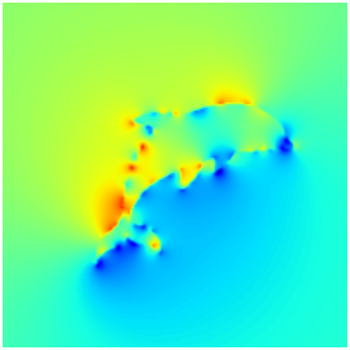} &
    
    \includegraphics[width=\ninesize\textwidth]{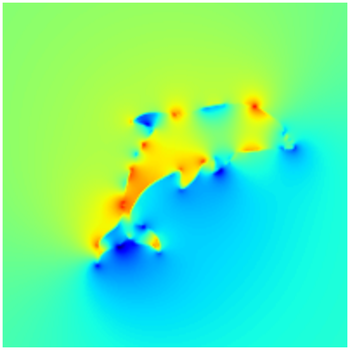} &
    
    \includegraphics[width=\ninesize\textwidth]{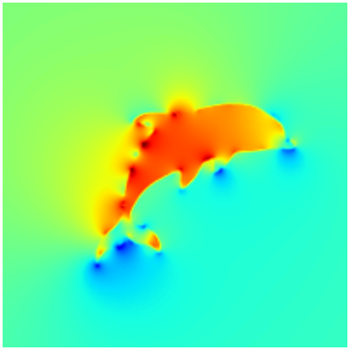} &
    
    \includegraphics[width=\ninesize\textwidth]{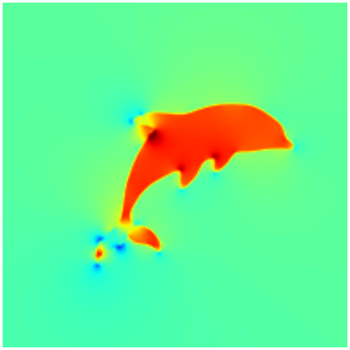} &
    
    \includegraphics[width=\ninesize\textwidth]{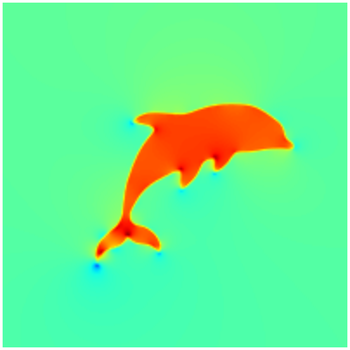} &

    \includegraphics[width=\ninesize\textwidth]{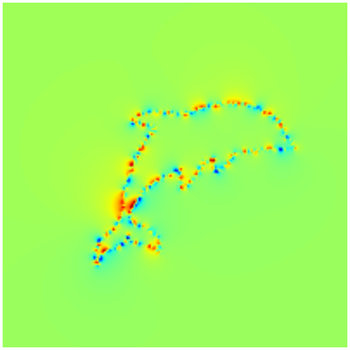} & 
    \includegraphics[width=\ninesize\textwidth]{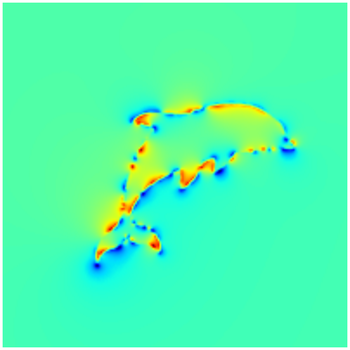} &
    \includegraphics[width=\ninesize\textwidth]{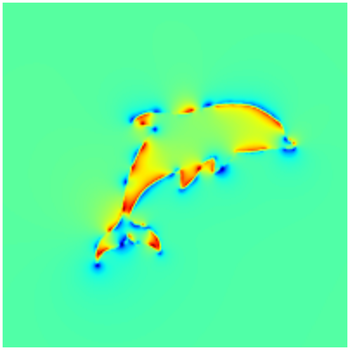} &
    \includegraphics[width=\ninesize\textwidth]{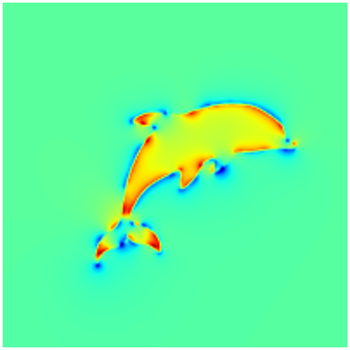} &
    \includegraphics[width=\ninesize\textwidth]{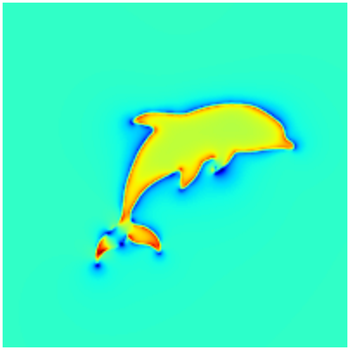} &
    
    \includegraphics[width=\ninesize\textwidth]{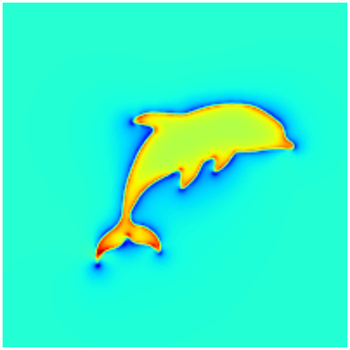} &

    \includegraphics[width=\ninesize\textwidth]{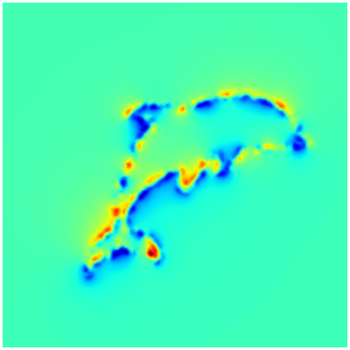} &
    \includegraphics[width=\ninesize\textwidth]{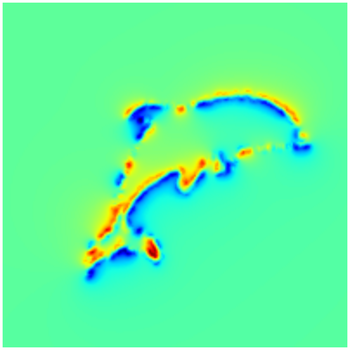} &
    \includegraphics[width=\ninesize\textwidth]{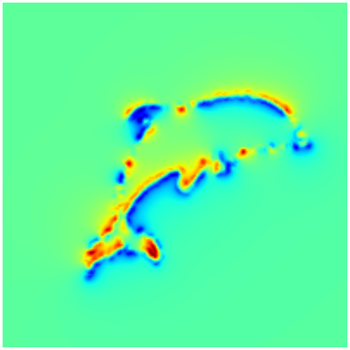} &
    
    \includegraphics[width=\ninesize\textwidth]{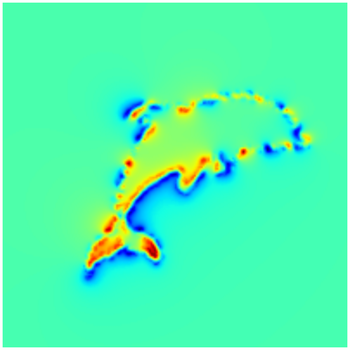} &
    
    \includegraphics[width=\ninesize\textwidth]{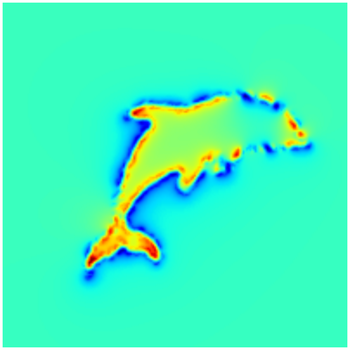} &
    
    \includegraphics[width=\ninesize\textwidth]{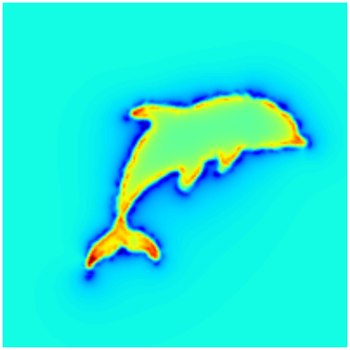}
    \\

    \includegraphics[width=\ninesizecontour\textwidth]{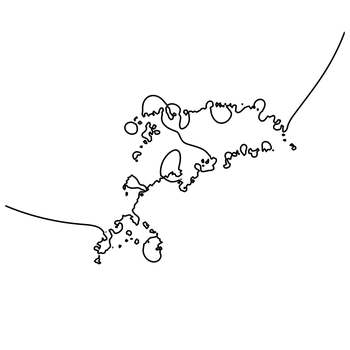} &
    \includegraphics[width=\ninesizecontour\textwidth]{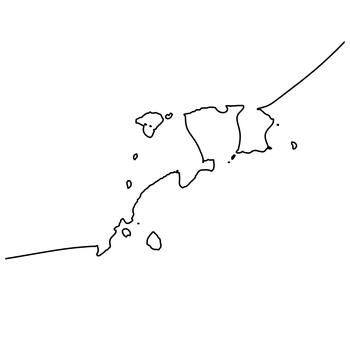} &
    \includegraphics[width=\ninesizecontour\textwidth]{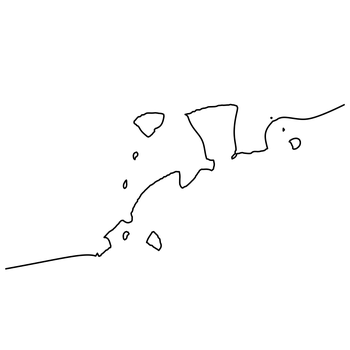} &
    
    \includegraphics[width=\ninesizecontour\textwidth]{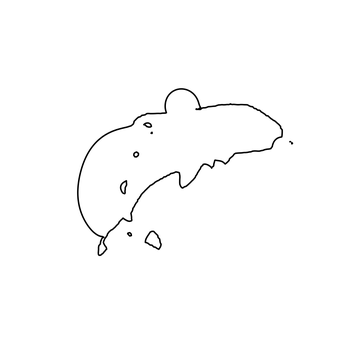} &
    \includegraphics[width=\ninesizecontour\textwidth]{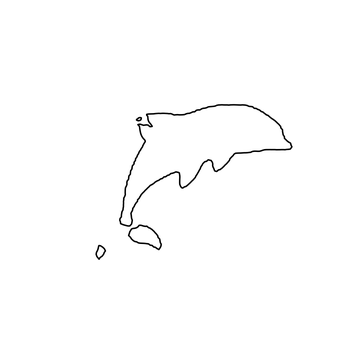} &
   
    \includegraphics[width=\ninesizecontour\textwidth]{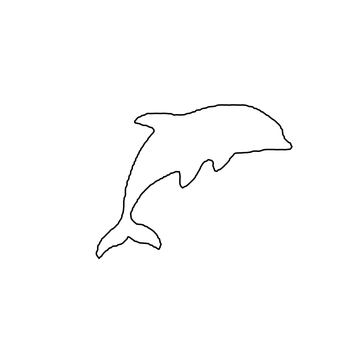} &

    \includegraphics[width=\ninesizecontour\textwidth]{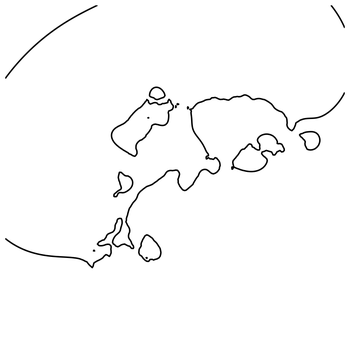} &
    \includegraphics[width=\ninesizecontour\textwidth]{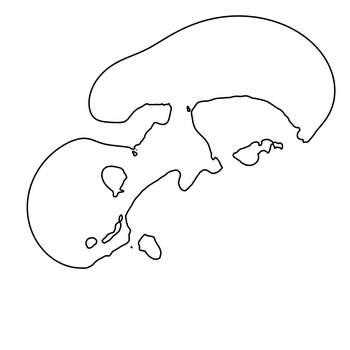} & 
    \includegraphics[width=\ninesizecontour\textwidth]{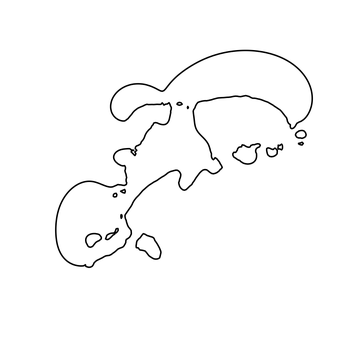} &
    
    \includegraphics[width=\ninesizecontour\textwidth]{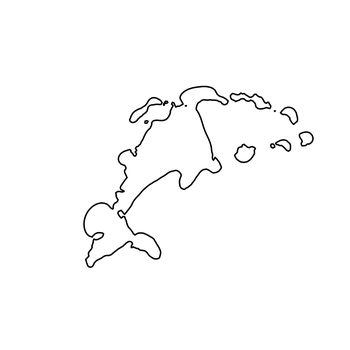} &
    
    \includegraphics[width=\ninesizecontour\textwidth]{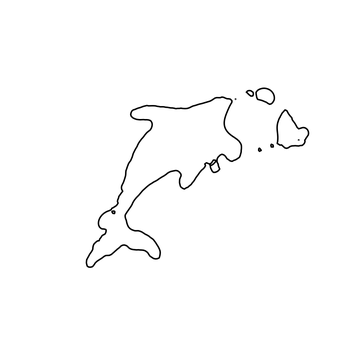} &
    
    \includegraphics[width=\ninesizecontour\textwidth]{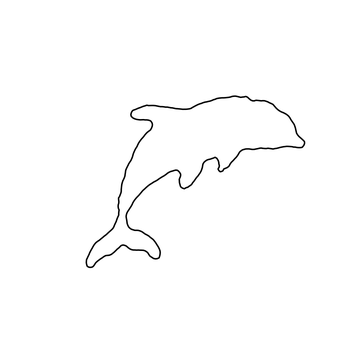} &
    \includegraphics[width=\ninesizecontour\textwidth]{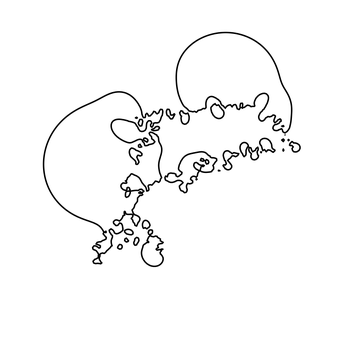} &
    \includegraphics[width=\ninesizecontour\textwidth]{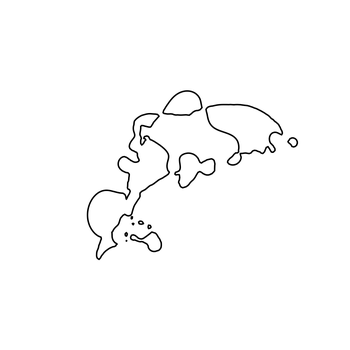} &
    \includegraphics[width=\ninesizecontour\textwidth]{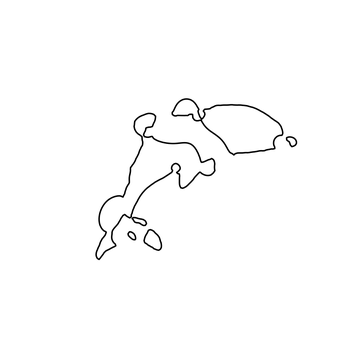} &
    \includegraphics[width=\ninesizecontour\textwidth]{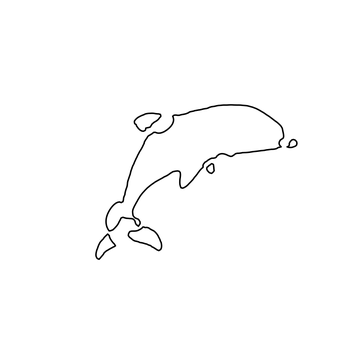} &
    \includegraphics[width=\ninesizecontour\textwidth]{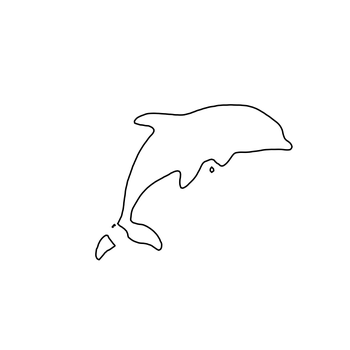} & 
    \includegraphics[width=\ninesizecontour\textwidth]{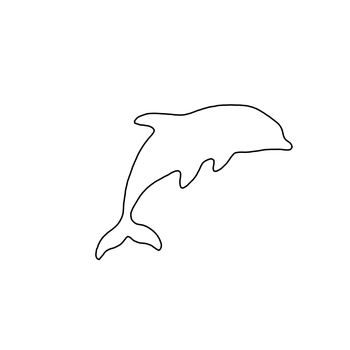}    \\
    \includegraphics[width=\ninesizenormals\textwidth]{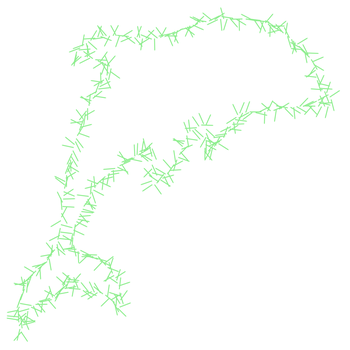} &
\includegraphics[width=\ninesizenormals\textwidth]{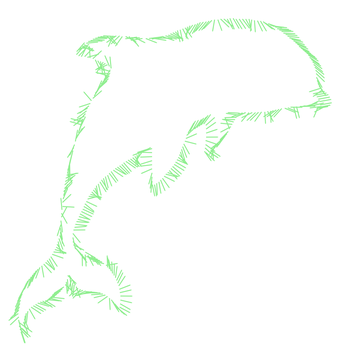} & \includegraphics[width=\ninesizenormals\textwidth]{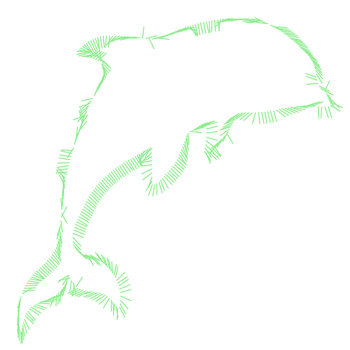}& \includegraphics[width=\ninesizenormals\textwidth]{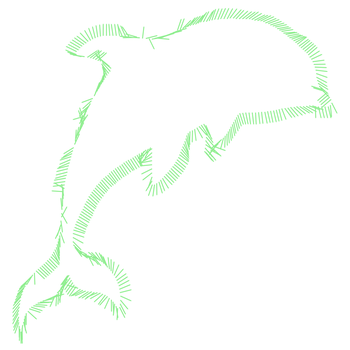} & \includegraphics[width=\ninesizenormals\textwidth]{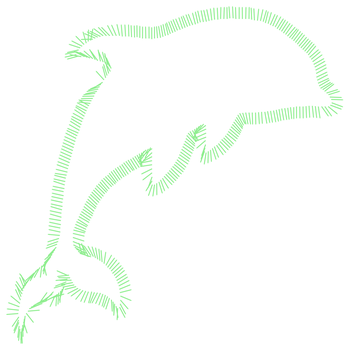} & \includegraphics[width=\ninesizenormals\textwidth]{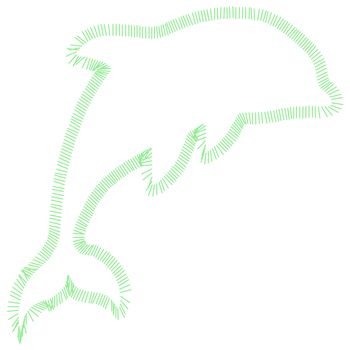} &
\includegraphics[width=\ninesizenormals\textwidth]{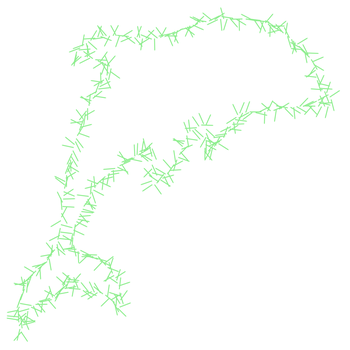} &
\includegraphics[width=\ninesizenormals\textwidth]{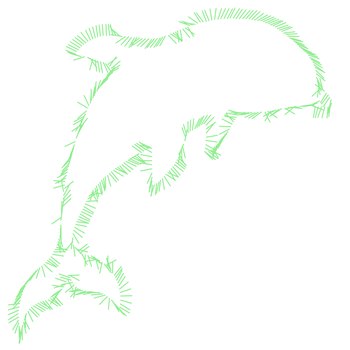} &
\includegraphics[width=\ninesizenormals\textwidth]{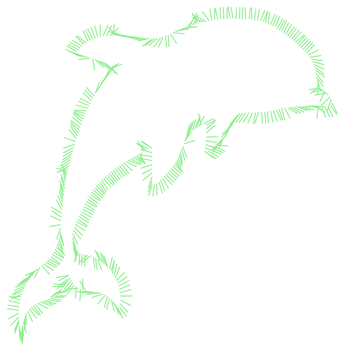} &
\includegraphics[width=\ninesizenormals\textwidth]{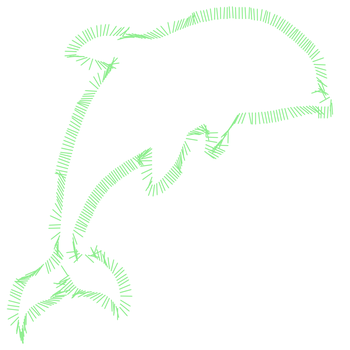} &
\includegraphics[width=\ninesizenormals\textwidth]{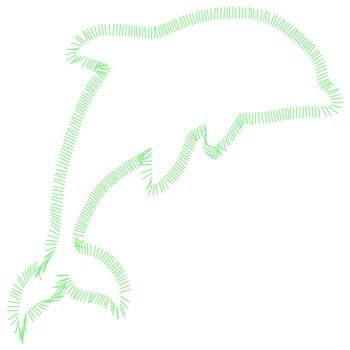} &
\includegraphics[width=\ninesizenormals\textwidth]{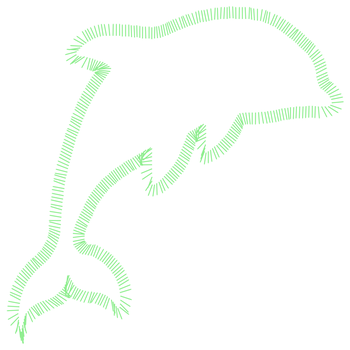} &
\includegraphics[width=\ninesizenormals\textwidth]{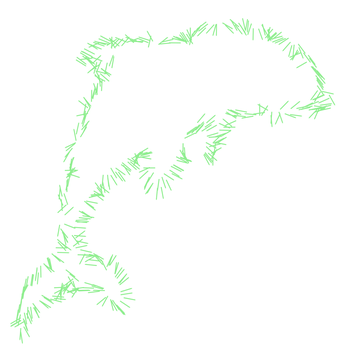} &
\includegraphics[width=\ninesizenormals\textwidth]{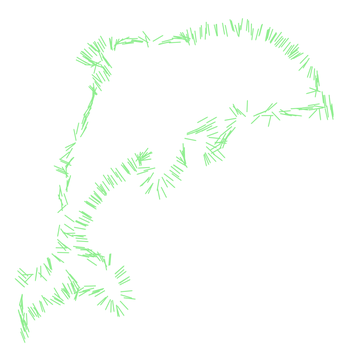} &
\includegraphics[width=\ninesizenormals\textwidth]{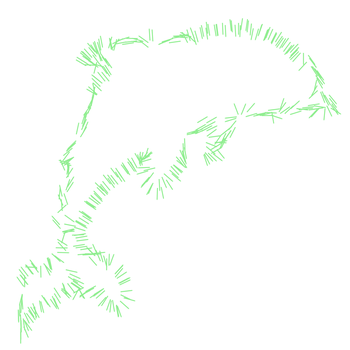} &
\includegraphics[width=\ninesizenormals\textwidth]{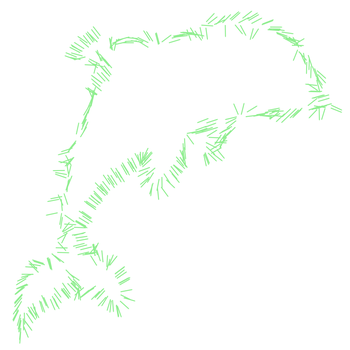} &
\includegraphics[width=\ninesizenormals\textwidth]{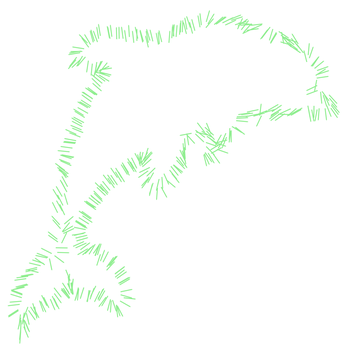} &
\includegraphics[width=\ninesizenormals\textwidth]{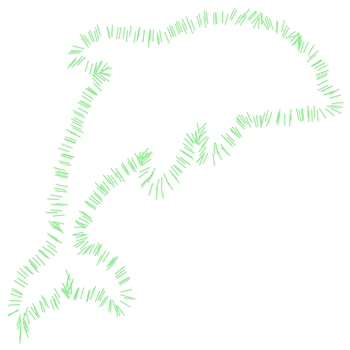}
    \\
 
    $t=0$ &  $t=1$ & $t=4$ & $t=6$ & $t=11$ & $t=21$ & $t=0$ & $t=1$ & $t=2$ &  $t=4$ &  $t=6$ &  $t=7$ & $t=0$ &  $t=1$ & $t=2$ & $t=4$ & $t=6$ & $t=8$  
\\
    \end{tabular}
        \end{small}

    \caption{\textcolor{black}{Illustration of the algorithmic pipeline using a 2D example under three different scenarios: 0\% noise with $\lambda=0$, 0\% noise with $\lambda=10$ and $0.75\%$ noise with $\lambda=100$. Each sequence displays the GWN field $w^{(t)}$ (top), the $\overline{w}^{(t)}$-level set (middle), and the normals $\mathbf{n}_i$ (bottom) across various iterations $t$, from random initialization to convergence. A higher value of the screening coefficient $\lambda$ accelerates the decay rate of the screened GWN at input points, resulting in faster convergence and enhanced resilience against noise. The screened GWN fields are visualized using a heat color map, with GWN values ranging from -2 to 2, under the three different settings.} }
    \label{fig:2dpipeline}
\end{figure*}

\begin{figure*}[!htbp]
    \centering
    \includegraphics[width=0.1575\textwidth]{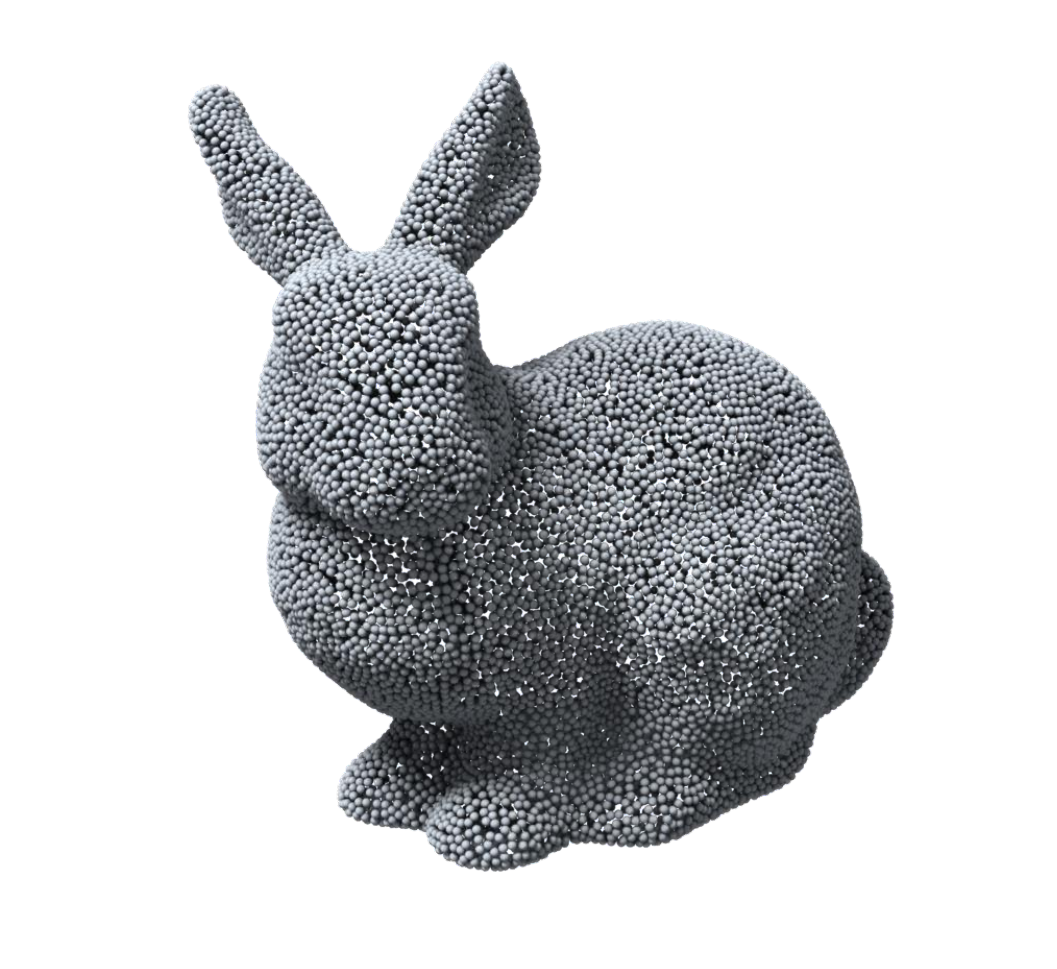}
    \includegraphics[width=0.1575\textwidth]{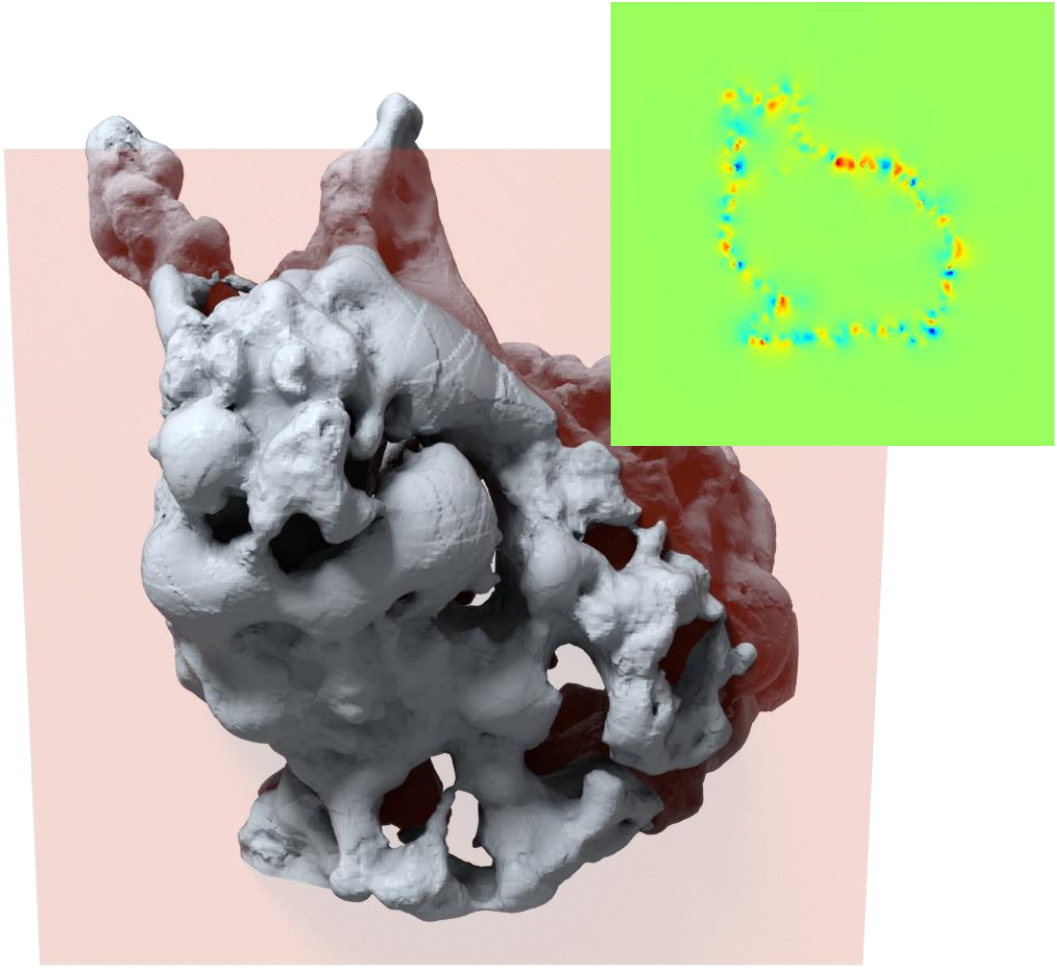}
    \includegraphics[width=0.1575\textwidth]{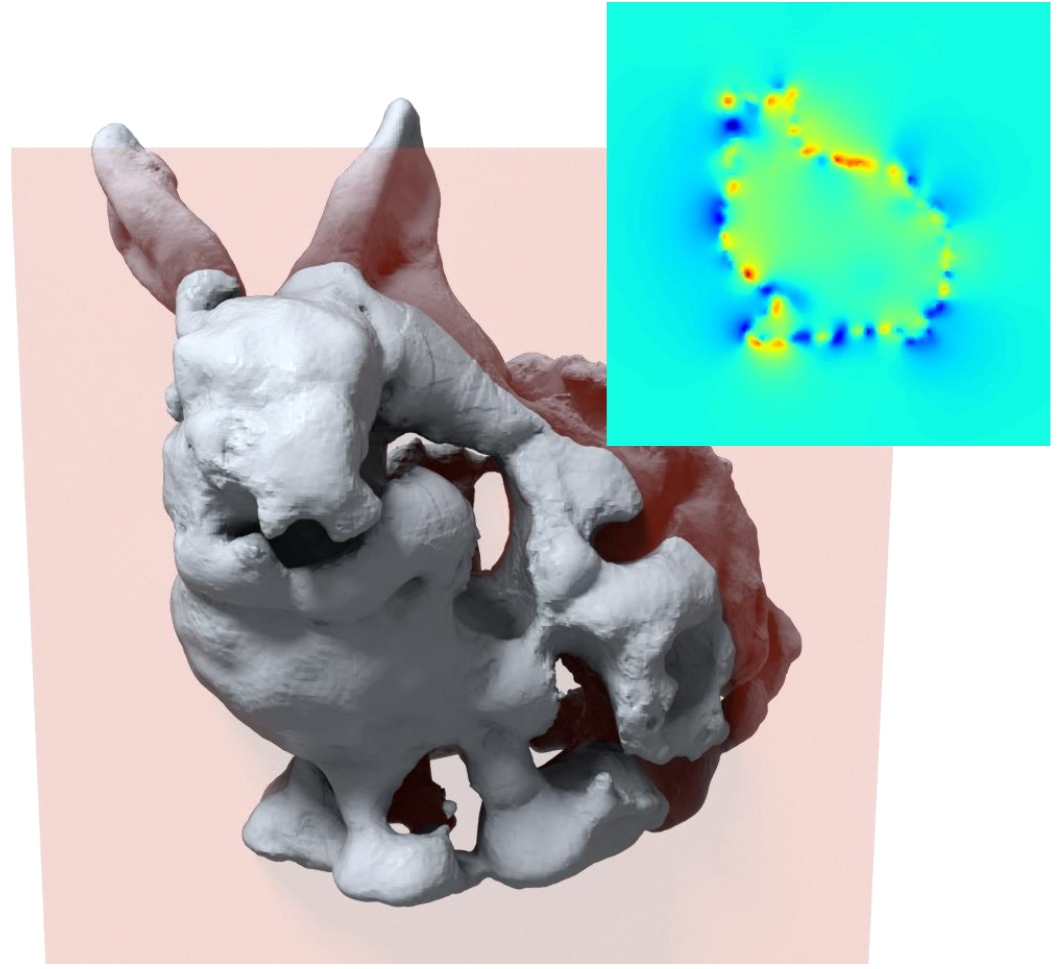}
    \includegraphics[width=0.1575\textwidth]{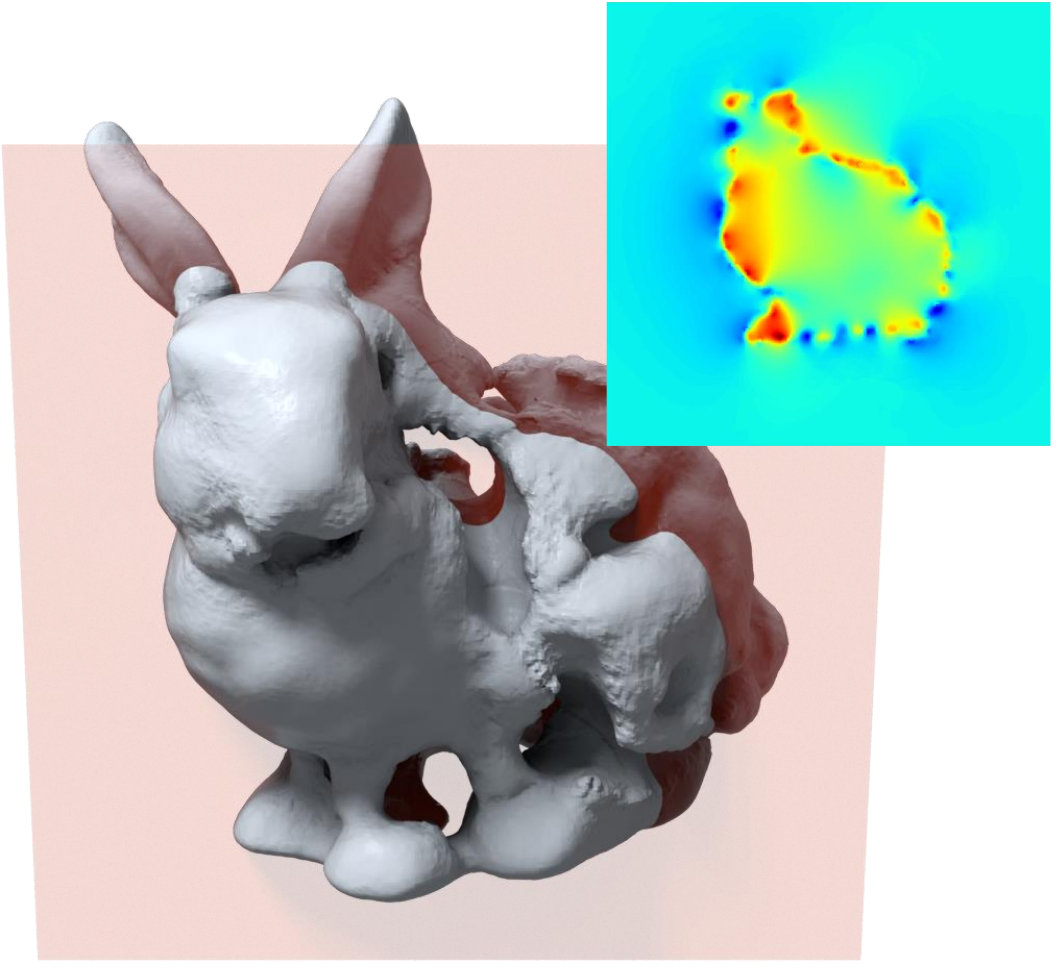}
    \includegraphics[width=0.1575\textwidth]{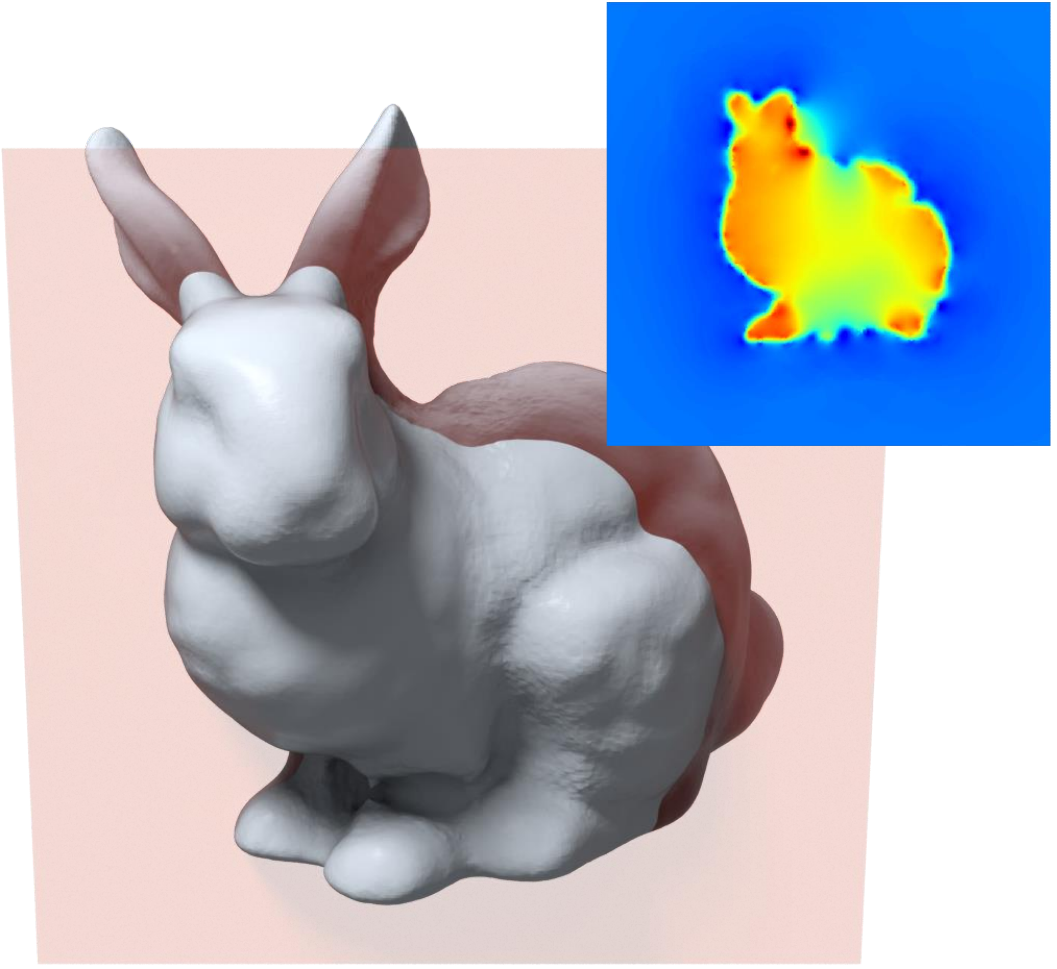}
    \includegraphics[width=0.1575\textwidth]{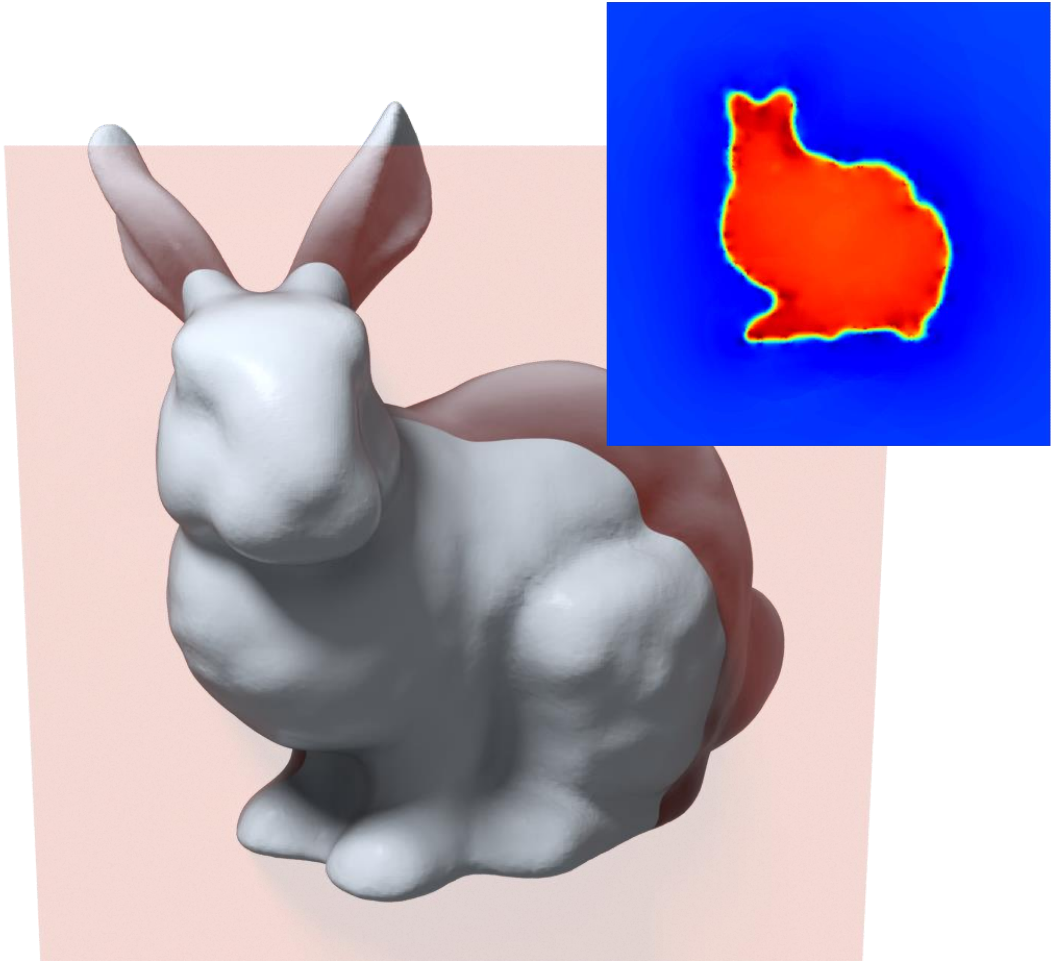}
    \\
    \makebox[0.1575\textwidth]{Input $n=10$K}
    \makebox[0.1575\textwidth]{$t=0$}
    \makebox[0.1575\textwidth]{$t=1$}
    \makebox[0.1575\textwidth]{$t=2$}
    \makebox[0.1575\textwidth]{$t=4$}
    \makebox[0.1575\textwidth]{$t=7$}\\
    \caption{3D Illustration of the algorithmic pipeline. Starting with an unoriented point cloud as input, our method initializes each point with a randomly assigned normal. The point normals are then iteratively updated by diffusing the GWN gradients in a fully parallel manner. We display the $\overline{w}^{(t)}$-level set of the computed GWN field $w^{(t)}$ and visualize the GWN field along the cross-sections using a heat color map. For better visualization of the GWN field, we set the screening coefficient $\lambda=0$. See also the accompanying video for the entire diffusion process. }
    \label{fig:3dpipeline}
\end{figure*}

To ease understanding, we provide the pseudo-code of DWG in Algorithm~\ref{alg:pipeline} and refer readers to Figures~\ref{fig:2dpipeline} and~\ref{fig:3dpipeline} for illustrations of the iterative diffusion process applied to a 2D toy model and the 3D Bunny model, respectively. 

\subsection{Discussions}
\label{subsec:features}

\paragraph{Numerical Solver-Free Feature}
\textcolor{black}{
Our algorithm alternates between computing the screened GWN field using point normals and updating those normals using the gradients of the sGWN level set. The normal updating process is done through mutually linking input points with triangular faces within the level set $\mathcal{M}^{(t)}$. Both the computation of the sGWN field and the updating of point normals in our method are executed in a fully parallel manner. Unlike existing methods, DWG does not rely on numerical solvers such as linear systems or optimization algorithms. This independence from complex solvers significantly enhances the method's computational efficiency and scalability, making it well-suited for deployment on both CPUs and GPUs.  } 

\textcolor{black}{\paragraph{Space Complexity} DWG exhibits a space complexity of  $O(n\log n)$ due to the octree nodes created during space discretization. Of these nodes, empirically, only $O(n)$ non-empty leaf nodes are actively used as query points in the computation of the screened GWN field. }

\paragraph{Time Complexity}
In the initialization stage, we construct a kd-tree for efficient local searching and an octree for adaptively partitioning the space around the input point cloud $\mathcal{P}$; both structures are built in $O(n\log n)$ time \cite{Barill2018FastWN}. Empirically, the number of non-empty leaf nodes in the octree is $O(n)$, scaling linearly with the number of input points $n$. Consequently, the computation of the GWN for each octree node takes $O(n\log n)$ time, following the fast computation strategy proposed by \citet{Barill2018FastWN}. The subsequent extraction of the $\overline{w}^{(t)}_\lambda$-level set using the Marching Cubes algorithm also takes $O(n)$ time. For each triangular face $f\in\mathcal{M}^{(t)}$, searching the $k$-nearest neighboring points in $\mathcal{P}$ incurs $O(\log n+k)$ time with the kd-tree. Given that the extracted level set $\mathcal{M}^{(t)}$ has $O(n)$ faces, the cost per iteration 
amounts to $O\left(n\log n+n+n(\log n +k)\right)=O(n\log n)$ given that $k$ is a small constant. 
Thus, the overall time complexity for DWG is empirically estimated as $O(In\log n)$, where $I$ is the iteration number.

\begin{algorithm}
    \caption{\textcolor{black}{Diffusing Winding Gradients}}
    \SetKwInput{Input}{Input}
\SetKwInput{Output}{Output}

         \Input{An unoriented point cloud $\mathcal{P} = \{\mathbf{p}_i\}_{i=1}^n$
        sampled from a closed, orientable, manifold surface  $\mathcal{S}$; the octree depth $d$; the screening coefficient $\lambda$; and the convergence criteria $\epsilon$ }
         \Output{Globally consistent orientations and a watertight surface approximating $\mathcal{S}$}

Construct an octree of depth up to $d$ for adaptively sampling $\mathcal{P}$\\
Collect the set $\mathcal{L}$ of non-empty leaf nodes $\{\mathbf{s}_j\}_{j=1}^l$ of the octree\\
\For{each input point $\mathbf{p}_i\in\mathcal{P}$}{
Compute the coefficient $a_i$\\ 
Assign an initial unit normal vector $\mathbf{n}_i^{(0)}$ \\
}
$t\leftarrow 1$\\

    
\Repeat{the average normal difference $\leq \epsilon$}{

\For{every octree node $\mathbf{s}_j$ }
{
Compute the sGWN $w^{(t)}_\lambda(\mathbf{s}_j)$ 
using $\mathbf{n}^{(t-1)}$ as in Eqn. (\ref{eqn:sGWN})\\
}

        Compute the average winding number $\overline{w}^{(t)}_\lambda$ as in Eqn. (\ref{eqn:averagew}) \\

        Extract the $\overline{w}^{(t)}_\lambda$-level set  $\mathcal{M}^{(t)}$ using Marching Cubes\\

        \For{each face $f\in \mathcal{M}^{(t)}$}   {
        Compute $\nabla w^{(t)}_\lambda(f)$ using the vector cross product\\
        }
        
        \For{\textcolor{black}{each point $p_i \in \mathcal{P}$}}{
            \textcolor{black}{$\mathbf{n}_i^{(t)} \leftarrow \mathbf{0}$}\\
        }
        
       \For{\textcolor{black}{each face $f\in\mathcal{M}^{(t)}$}}
        {
            \textcolor{black}{ Find the set of $k$-nearest points $P(f)\subset \mathcal{P}$}\\

            \textcolor{black}{\For{each point $\mathbf{p}_j\in P(f)$}   {
                $\mathbf{n}^{(t)}_{j} +=  \nabla w_\lambda^{(t)}(f)$ 
            }
            }
        }
   
        \For{\textcolor{black}{each point $\mathbf{p}_i \in \mathcal{P}$}}{
               \textcolor{black}{ $\mathbf{n}_i \leftarrow \mathrm{normalize}(\mathbf{n}_i)$}
        
        }
        
        $t\leftarrow t+1$\\
}

\textbf{return} $\mathbf{n}^{(\infty)}$ and $\mathcal{M}^{(\infty)}$ \\

    \label{alg:pipeline}
\end{algorithm}

\section{Implementation Details}
\label{sec:implementation}

\subsection{Initialization}
\label{subsec:initialization}

\paragraph{Space Discretization} In the initialization stage, we construct an octree of depth up to a user-specified maximum depth $d$ to adaptively partition the space around the input point cloud $\mathcal{P}$. For this, we employ\footnote{\url{https://libigl.github.io/}} and store the octree as a linear data structure to facilitate efficient traversal on GPUs. In addition, we construct a kd-tree to facilitate efficient nearest neighbor searching, which is crucial in DWG. Finding the nearest neighbors in the point cloud $\mathcal{P}$ for a given query point $\mathbf{q}$ is critical both for computing the coefficient $a_i$ for the discrete GWN formula during the initialization stage and for calculating the Laplacian of GWN gradients throughout the diffusion process. Our implementation utilizes a GPU-based kd-tree\footnote{\url{https://github.com/ingowald/cudaKDTree}}, which enables fast retrieval of the $k$-nearest points to any query point.

\paragraph{Computing Area Weights $a_i$} In the discrete GWN formula, each input point $\mathbf{p}_i$ is assigned a coefficient $a_i$, representing its contribution to the discretized surface integral. Ideally, $a_i$ should reflect the geodesic Voronoi area associated with $\mathbf{p}_i$. However, computing a high-quality geodesic Voronoi diagram on point clouds is time consuming. For large-scale, densely sampled point clouds, we observe that simply adopting uniform coefficients $a_i\equiv 1$ leads to high-quality GWN, sufficient for DWG to successfully reconstruct 3D surfaces. However, for low-resolution models or models with uneven sampling rates, uniform $a_i$ typically fails. This is because the geodesic Voronoi areas $a_i$ become position-dependent. To address this issue, we employ a simplified method for approximating geodesic Voronoi areas as in~\cite{Barill2018FastWN,BIM}. For each input point $\mathbf{p}_i$, we identify $m$ nearest neighbors and locally fit a tangent plane at $\mathbf{p}_i$. We then project these neighbors onto the plane and compute a 2D Voronoi diagram using the projected points as seeds. The area of the 2D Voronoi cell corresponding to $\mathbf{p}_i$ is then used as its approximated geodesic Voronoi area $a_i$. In our implementation, we empirically set $m=15$. 

\textcolor{black}{\paragraph{Normal Initialization} 
Our method provides three options for initializing point normals: random, principal component analysis (PCA) and the Gauss map. For PCA initialization, we compute the covariance matrix from the $k$ nearest neighbors of each input point $\mathbf{p}_i$ and derive the normal vector from the eigenvector corresponding to the smallest eigenvalue. This eigenvector best approximates the local surface normal. For Gauss map initialization, we approximate the normal at each point $\mathbf{p}_i$ by normalizing its coordinate vector, projecting the point onto a unit sphere centered at the origin. Figure~\ref{fig:Ablationstudy-lambda-init} displays the convergence plots for the three initialization strategies applied to three representative models. Each model underwent testing in both noise-free conditions and with a noise level of 0.75\%. The tests evaluated various values of the screening coefficient $\lambda$ to assess the impact on the convergence behavior of each initialization method. Based on the test results, we recommend specific normal initialization strategies depending on the model's characteristics. For models without thin structures, initializing normals using either the Gauss map or PCA significantly reduces the number of iterations required for convergence. For example, while DWG with random initialization takes 30 iterations to converge on the Lucy model, using PCA can accelerate convergence dramatically, requiring only one iteration. Similarly, initializing with the Gauss map also leads to convergence in 21 iterations, which is quicker compared to random initialization. 
For models featuring thin structures, we recommend using random initialization. This preference arises because PCA and Gauss map initializations can lead to locally accurate but globally inconsistent normals. In cases like thin plates, these initializations might yield similar normals for both the top and bottom sides, leading to global inconsistencies. Consequently, DWG requires more iterations to correct these inaccuracies compared to starting from a random initialization.} 

\paragraph{Model-dependent Parameters}
Our algorithm involves two model-dependent parameters: the octree depth, $d$,  and the screening coefficient, $\lambda$. These parameters are essential for 
adaptively discretizing the space around the point cloud $\mathcal{P}$ and managing noisy inputs, respectively, affecting both the convergence rate and the quality of the reconstructed surfaces. Based on our convergence tests illustrated in Figure~\ref{fig:Ablationstudy-lambda-init}, we recommend the following parameter settings: 1) \textbf{Octree Depth ($d$)}: For models with up to {1 million} points, we recommend an octree depth of $d \in [8, 10]$, and for models exceeding 1 million points, $d \in [11, 13]$ is advisable. In the presence of low-level noise, decrease the recommended octree depth by one or two levels. For  higher noise levels, a reduction of two to three levels is advisable. 2) \textbf{Screening Coefficient ($\lambda$)}: We recommend setting $\lambda$ between 10 and 100, adjusting within this range based on the specific noise level of the model.

\paragraph{Model-independent Parameters}
Other parameters, such as the convergence criterion $\epsilon$ (the average normal difference between consecutive iterations), the number of nearest neighbors, $k$, used for updating the point normals, are set as fixed, model-independent constants for simplicity and consistency. For the convergence threshold, we note that while the normals for the majority of input points converge quickly, certain regions require additional iterations to align correctly. To avoid premature termination, we do not calculate the average difference across all normals between two iterations. Instead, we focus on the top 1\% of the largest normal differences and compute their average. This strategy helps in identifying significant discrepancies that could impact the overall quality of the reconstruction. In our implementation, we empirically set $\epsilon=0.1^\circ$ and $k=10$, respectively.

\subsection{Parallel Diffusion}
\label{subsec:paralleldiffusion}

\paragraph{Octree Traversal} During the diffusion stage, we compute the GWN for each octree node in parallel by adopting the fast GWN method from ~\citep{Barill2018FastWN}. To enhance the runtime performance, we modify the original recursive implementation with a non-recursive one. Starting from the root node, we perform a depth-first search on the octree. For each query point $\bf q$, if an octree node $c$ is in close proximity, satisfying $\|\mathbf{q} - c.\mathbf{m}\| \leq 2.3r$ (where $c.\mathbf{m}$ is the mass center of the node and $r$ is the maximum distance from the points in the node to the mass center), we delve further into this child node. Otherwise, we represent all points inside node $c$ using the area-weighted normal and position of the node, thus avoiding point-by-point enumeration. If all children of a node have been traversed or if the node is a leaf, we return to the parent node and proceed to the next sibling node, continuing until we return to the root node.

\begin{table}[!htbp]
    \centering
    \setlength\tabcolsep{1.4pt}
        \caption{Qualitative comparison of \textcolor{black}{implicit function-based} methods for 3D reconstruction from unoriented points. Here, $n$ represents the number of unoriented points in the input point cloud, $m~(>>n)$ is the number of Voronoi vertices used in GCNO and BIM, $I$ refers to the iteration count for the iterative methods. ``SLS'' and ``DLS'' represent sparse  and dense linear systems, respectively. ``UDF'' indicates unsigned distance fields. }
    \begin{scriptsize}
     \begin{tabular}{c|c|c|c|c|c}
    \hline
         Method   & Implicit & Discreti- & Time& Space & Numerical \\
            & function & zation & complexity & complexity & solver \\

         \hline
         \hline
         \textcolor{black}{\citet{kobbelt}}  & UDF & Voxels
         & $O(n^2)$ & $O(n)$ & SLS \\ \hline
         \textcolor{black}{\citet{mullen2010signing} }& UDF & Delaunay & $O(n^2)$ & $O(n\log n)$ & SLS \\ \hline
         
         VIPSS~\cite{VIPSS} & Duchon & -- & $O(n^3)$ & $O(n^2)$ & DLS \\ \hline
         iPSR~\cite{hou2022iterative} & Poisson & Octree& $O(In^2)$ & $O(n\log n)$ & SLS\\ \hline
         PGR~\cite{lin2022surface} & GWN & Octree & $O(n^3)$ & $O(n^2)$ & DLS \\ \hline 
         GCNO~\cite{xu2023globally}  & GWN  & Voronoi & $O(Imn)$ & $O(m+n)$ & L-BFGS  \\ \hline 
         BIM~\cite{BIM} & GWN  & Voronoi &  $O(I(n^2+m))$ & $O(m+n)$ & L-BFGS \\    \hline 
         \textcolor{black}{\citet{Kai_linear}} &  Stokes & -- & $O(n^3)$ & $O(n^2)$ & SLS \\   \hline
         \textcolor{black}{AGR~\cite{Ma2024APGR}} & GWN & Octree & $O(n^3)$& $O(n^2)$ & DLS \\ \hline
         \textcolor{black}{WNNC~\cite{Lin2024fast}} & GWN & Octree &  $O(In^2)$ & $O(n\log n)$ & SLS \\ \hline 
         \textcolor{black}{SNO~\cite{Huang2024Stochastic}} &  Poisson & Octree & $O(n^3)$ & $O(n\log n)$ & DLS \\ \hline
         Ours & sGWN  &  Octree & $O(In\log n)$  & $O(n\log n)$ & --  \\ \hline
    \end{tabular}
    \end{scriptsize}

    \label{tab:comparison}
\end{table}

\paragraph{Poisson Kernels}
When the query point $\mathbf{q}$ is in close proximity to the surface, direct evaluation of the Poisson kernel results in singularities due to the zero denominator $\|\mathbf{p}_i-\mathbf{q}\|^3$. This often leads to the creation of zigzagged iso-surfaces. To reduce
these artifacts, we employ a modified Poisson kernel as introduced in \cite{lin2022surface}, which adjusts the kernel's behavior near the surface to avoid these singularities. Specifically, if the distance between $\mathbf{p}_i$ and $\mathbf{q}$ is less than a threshold $\delta_i$, i.e.,  $\|\mathbf{p}_i - \mathbf{q}\| < \delta_i$, we adjust the denominator to $\delta_i^3$. Following \cite{lin2022surface}, we determine the threshold $\delta_i$ as follows: First, identify the $k$-nearest neighbors of $\mathbf{p}_i$. Then, filter out neighbors whose distances to $\mathbf{p}_i$ are either less than a minimum value $v_{\min}$ or greater than a maximum value $d_{\max}$. After that, compute $\delta_i$ as the average distance from $\mathbf{p}_i$ to the filtered set of neighbors. 
We set $k = 10$, $v_{\min} = 0.0015$, and $v_{\max} = 0.015$ in our implementation. Figure~\ref{fig:combiningdwgandpsr} illustrates the effectiveness of using the modified Poisson kernel in producing  smoother iso-surfaces compared to those generated using the original Poisson kernel.

\newcommand{\sevensize}{0.08050}
\newcommand{\sevensizeone}{0.06075}
\begin{figure*}[t] \centering
   
    \includegraphics[width=\sevensizeone\textwidth]{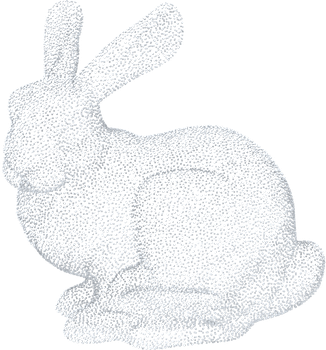}
    \includegraphics[width=\sevensize\textwidth]{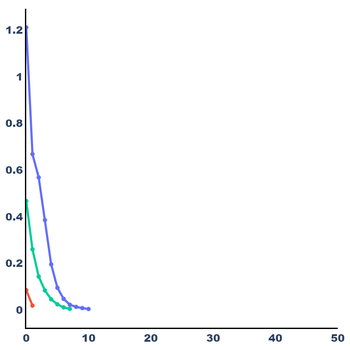}
    \includegraphics[width=\sevensize\textwidth]{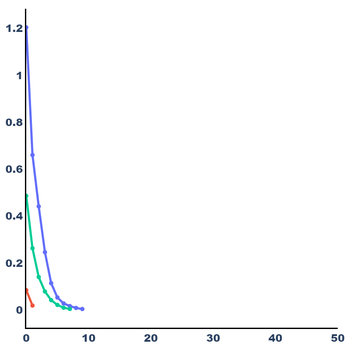}
    \includegraphics[width=\sevensize\textwidth]{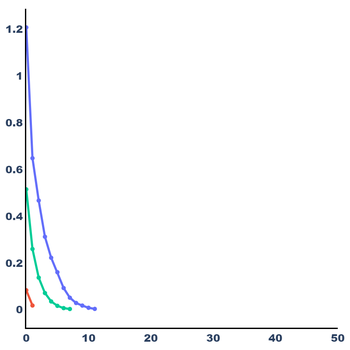}
    \includegraphics[width=\sevensize\textwidth]{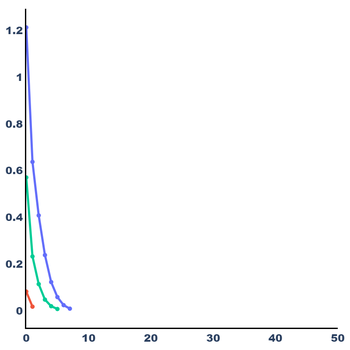}
    \includegraphics[width=\sevensize\textwidth]{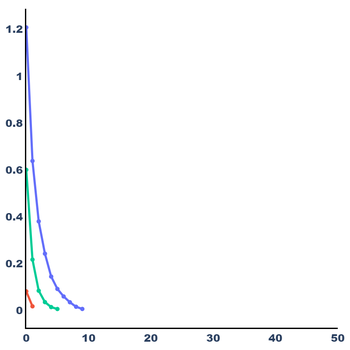} \includegraphics[width=\sevensizeone\textwidth]{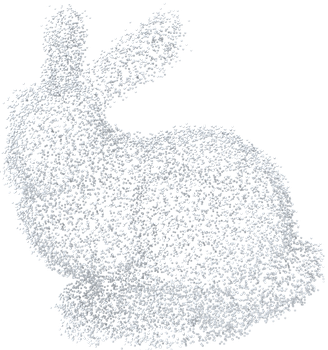}
    \includegraphics[width=\sevensize\textwidth]{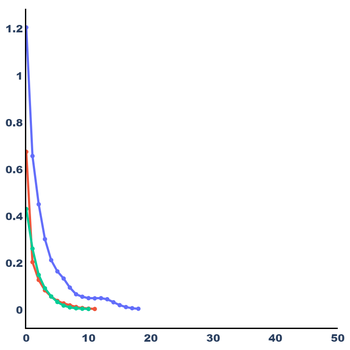}
    \includegraphics[width=\sevensize\textwidth]{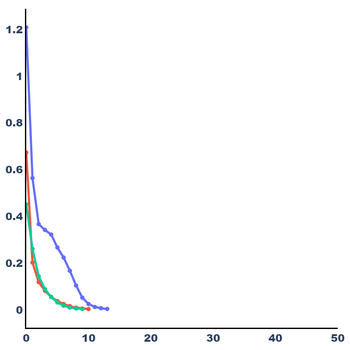}
    \includegraphics[width=\sevensize\textwidth]{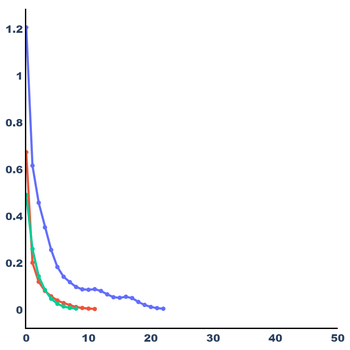}
    \includegraphics[width=\sevensize\textwidth]{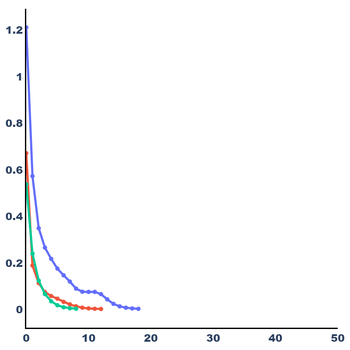}
    \includegraphics[width=\sevensize\textwidth]{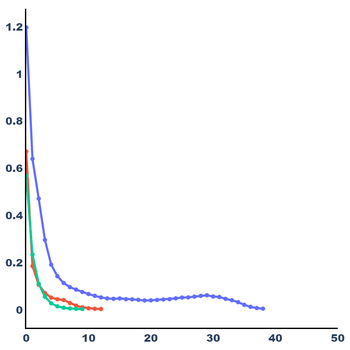}
    \\    
    \makebox[\sevensizeone\textwidth]{\tiny }
    \makebox[\sevensize\textwidth]{\tiny (6.53, 6.46, 6.54)}
    \makebox[\sevensize\textwidth]{\tiny (6.51, 6.42, 6.50)}
    \makebox[\sevensize\textwidth]{\tiny (6.47, 6.38, 6.46)}
    \makebox[\sevensize\textwidth]{\tiny (6.78, 6.71, 6.79)}
    \makebox[\sevensize\textwidth]{\tiny (7.10, 7.01, 7.09)}
    \makebox[\sevensizeone\textwidth]{\tiny }
    \makebox[\sevensize\textwidth]{\tiny (9.32, 9.33, 9.31)}
    \makebox[\sevensize\textwidth]{\tiny (9.25, 9.26, 9.23)}
    \makebox[\sevensize\textwidth]{\tiny (9.15, 9.17, 9.18)}
    \makebox[\sevensize\textwidth]{\tiny (9.74, 9.72, 9.74)}
    \makebox[\sevensize\textwidth]{\tiny (10.43, 10.32, 10.41)}\\
    \includegraphics[width=\sevensizeone\textwidth]{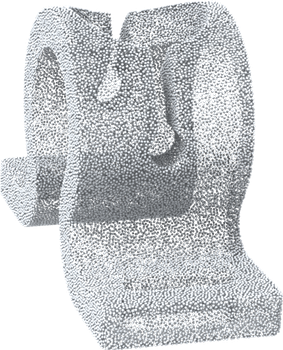}
    \includegraphics[width=\sevensize\textwidth]{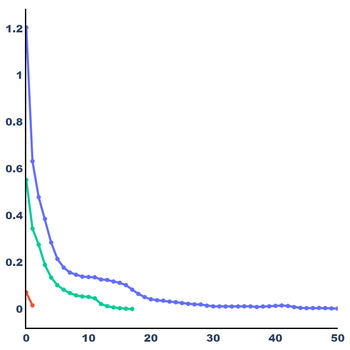}
    \includegraphics[width=\sevensize\textwidth]{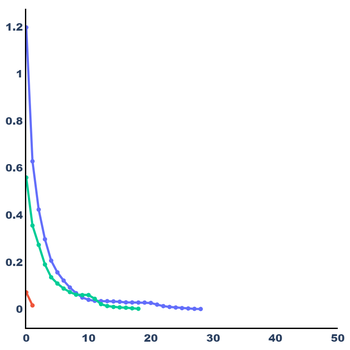}
    \includegraphics[width=\sevensize\textwidth]{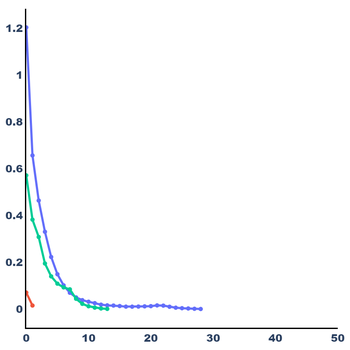}
    \includegraphics[width=\sevensize\textwidth]{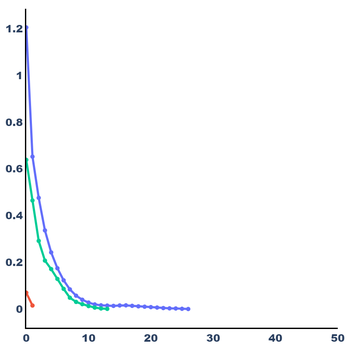}
    \includegraphics[width=\sevensize\textwidth]{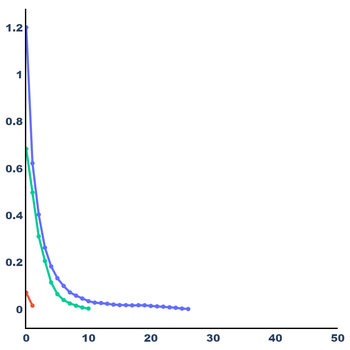}
    \includegraphics[width=\sevensizeone\textwidth]{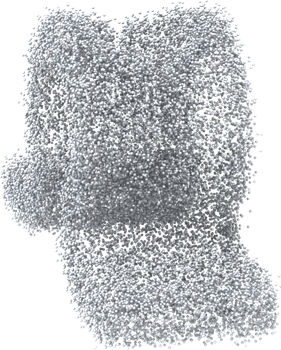}
    \includegraphics[width=\sevensize\textwidth]{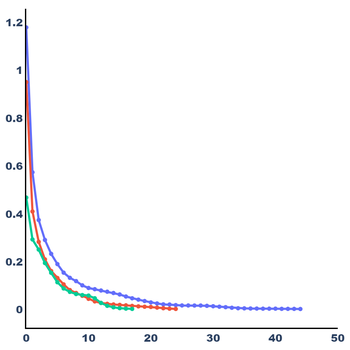}
    \includegraphics[width=\sevensize\textwidth]{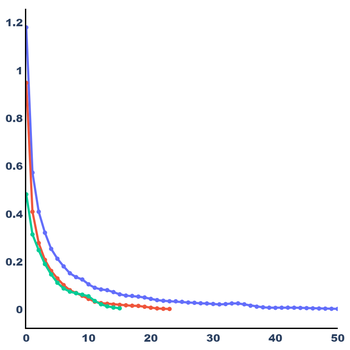}
    \includegraphics[width=\sevensize\textwidth]{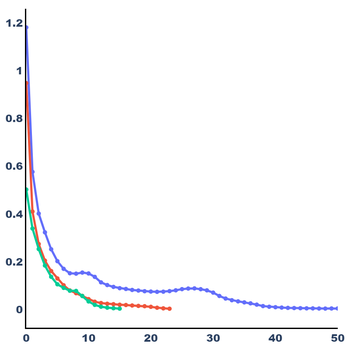}
    \includegraphics[width=\sevensize\textwidth]{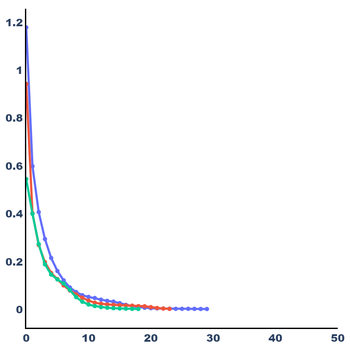}
    \includegraphics[width=\sevensize\textwidth]{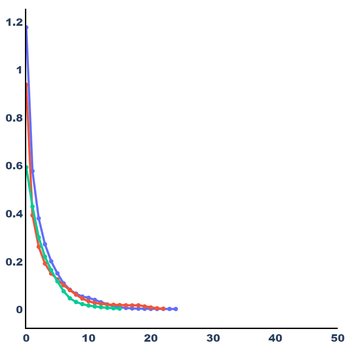}
    \\
    \makebox[\sevensizeone\textwidth]{\tiny }
    \makebox[\sevensize\textwidth]{\tiny (5.62, 5.58, 5.61)}
    \makebox[\sevensize\textwidth]{\tiny (5.61, 5.57, 5.60)}
    \makebox[\sevensize\textwidth]{\tiny (5.57, 5.53, 5.58)}
    \makebox[\sevensize\textwidth]{\tiny (5.56, 5.52, 5.57)}
    \makebox[\sevensize\textwidth]{\tiny (5.59, 5.58, 5.61)}
    \makebox[\sevensizeone\textwidth]{\tiny }
    \makebox[\sevensize\textwidth]{\tiny (9.63, 9.34, 9.21)}
    \makebox[\sevensize\textwidth]{\tiny (9.29, 9.28, 9.16)}
    \makebox[\sevensize\textwidth]{\tiny (9.08, 9.10, 8.99)}
    \makebox[\sevensize\textwidth]{\tiny (9.05, 8.94, 8.89)}
    \makebox[\sevensize\textwidth]{\tiny (9.03, 9.02, 8.97)}\\
 \includegraphics[width=\sevensizeone\textwidth]{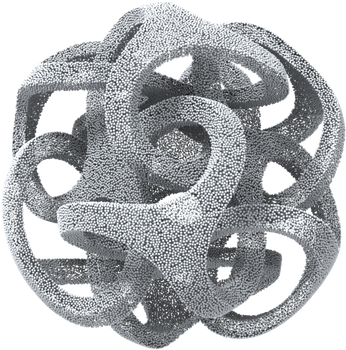}
    \includegraphics[width=\sevensize\textwidth]{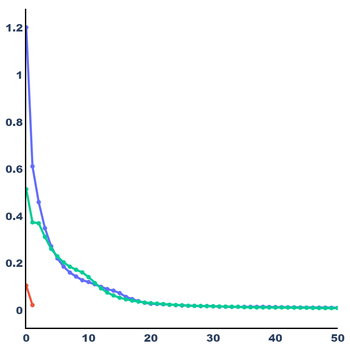}
    \includegraphics[width=\sevensize\textwidth]{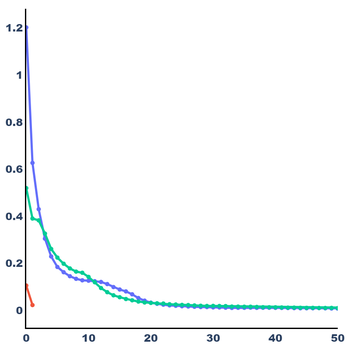}
    \includegraphics[width=\sevensize\textwidth]{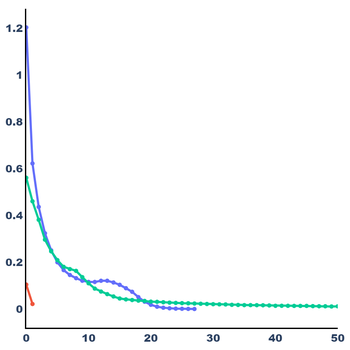}
    \includegraphics[width=\sevensize\textwidth]{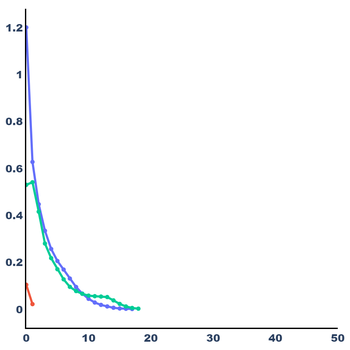}
    \includegraphics[width=\sevensize\textwidth]{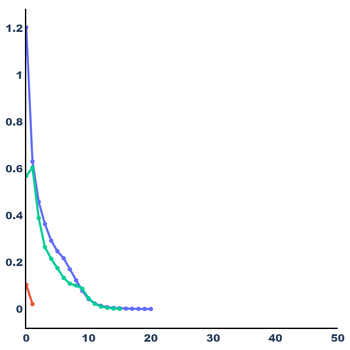}
\includegraphics[width=\sevensizeone\textwidth]{figures/0.75_noise/0.75_noise_pc/54725_7w_0.75.png}
    \includegraphics[width=\sevensize\textwidth]{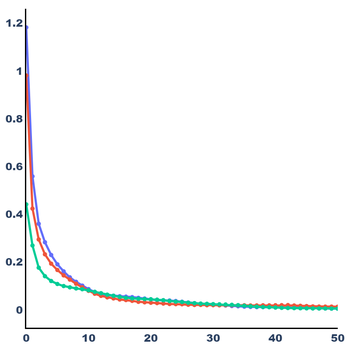}
    \includegraphics[width=\sevensize\textwidth]{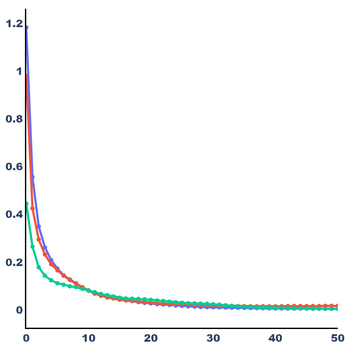}
    \includegraphics[width=\sevensize\textwidth]{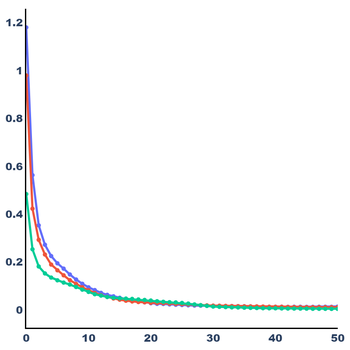}
    \includegraphics[width=\sevensize\textwidth]{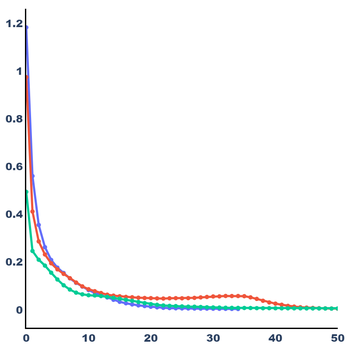}
    \includegraphics[width=\sevensize\textwidth]{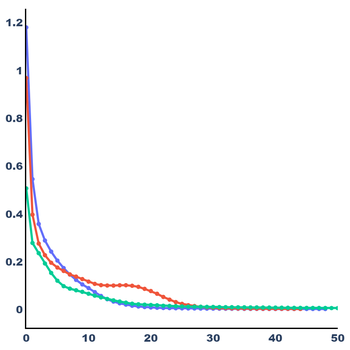}
    \\
    \makebox[\sevensizeone\textwidth]{\tiny }
    \makebox[\sevensize\textwidth]{\tiny (11.85, 7.10, 11.66)}
    \makebox[\sevensize\textwidth]{\tiny (11.86, 7.08, 11.52)}
    \makebox[\sevensize\textwidth]{\tiny (7.09, 7.07, 11.34)}
    \makebox[\sevensize\textwidth]{\tiny (7.07, 7.04, 7.07)}
    \makebox[\sevensize\textwidth]{\tiny (7.05, 7.04, 7.07)}
    \makebox[\sevensizeone\textwidth]{\tiny }
    \makebox[\sevensize\textwidth]{\tiny (17.49, 36.42, 17.64)}
    \makebox[\sevensize\textwidth]{\tiny (17.28, 36.27, 17.49)}
    \makebox[\sevensize\textwidth]{\tiny (19.20, 31.21, 17.13)}
    \makebox[\sevensize\textwidth]{\tiny (11.26, 16.48, 11.29)}
    \makebox[\sevensize\textwidth]{\tiny (11.15, 16.16, 11.07)}
 \\
    \makebox[\sevensizeone\textwidth]{\footnotesize Input}
    \makebox[\sevensize\textwidth]{\scriptsize $\lambda$=0}
    \makebox[\sevensize\textwidth]{\scriptsize $\lambda$=4}
    \makebox[\sevensize\textwidth]{\scriptsize $\lambda$=10}
    \makebox[\sevensize\textwidth]{\scriptsize $\lambda$=50}
    \makebox[\sevensize\textwidth]{\scriptsize $\lambda$=100}
      \makebox[\sevensizeone\textwidth]{\footnotesize Input}
    \makebox[\sevensize\textwidth]{\scriptsize $\lambda$=0}
    \makebox[\sevensize\textwidth]{\scriptsize $\lambda$=4}
    \makebox[\sevensize\textwidth]{\scriptsize $\lambda$=10}
    \makebox[\sevensize\textwidth]{\scriptsize $\lambda$=50}
    \makebox[\sevensize\textwidth]{\scriptsize $\lambda$=100}
    \\
    \includegraphics[width=0.35\textwidth]{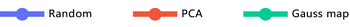}
    \\
    \caption{\textcolor{black}{Convergence tests on three representative models--Bunny (simple topology), CAD (simple topology with thin structure), and Metatron (complex topology with thin structure)-- varying the screening coefficient $\lambda$ and normal initialization. Each model was tested at both 0\% and 0.75\% noise levels. In each plot, the $x$-axis represents the number of iterations, and the $y$-axis shows the average of normal changes for all points, i.e., $\frac{1}{n}\sum_{i=1}^n\left\|\mathbf{n}_i^{(t+1)}-\mathbf{n}_i^{(t)}\right\|$. We explored $\lambda$ values of 0, 4, 10, 50 and 100, using three initialization methods: random, PCA, and Gauss map. DWG converges under all testing configurations. Below each plot, a triplet indicates the Chamfer distance (scaled by $10^3$) for the reconstructed surfaces upon convergence for each initialization method. \textbf{Noise-free scenarios:} PCA initialization significantly reduces the number of iterations, allowing DWG to converge within just three iterations across all models. Gauss map initialization also accelerates convergence for the Bunny and CAD models compared to random initialization, though it shows less effectiveness for the Metatron model. \textbf{Noisy scenarios:} The distinctions among the three initialization methods become less significant. Generally, higher values of $\lambda$ lead to more robust reconstructions.}}
    \label{fig:Ablationstudy-lambda-init}
\end{figure*}

\subsection{Outputs}

\begin{figure*}[!htbp]
\newcommand{\dragonwidth}{0.1595}
\includegraphics[width=\dragonwidth\textwidth]{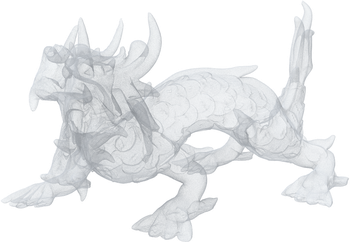}
\includegraphics[width=\dragonwidth\textwidth]{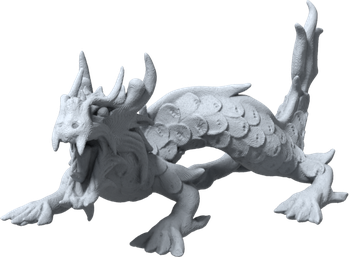}
\includegraphics[width=\dragonwidth\textwidth]{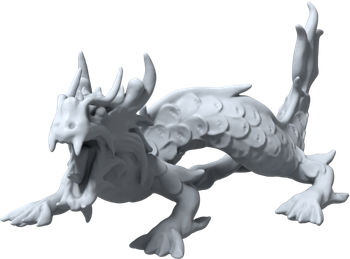}
\includegraphics[width=\dragonwidth\textwidth]{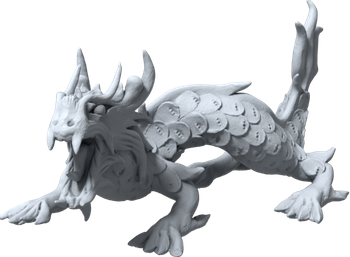}
\includegraphics[width=\dragonwidth\textwidth]{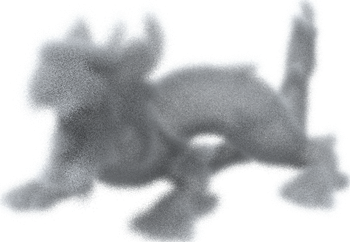}
\includegraphics[width=\dragonwidth\textwidth]{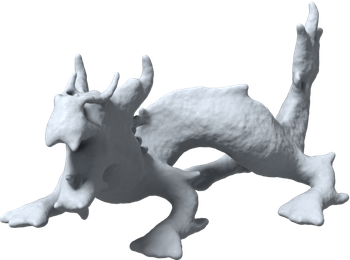} 
\\
\includegraphics[width=\dragonwidth\textwidth]{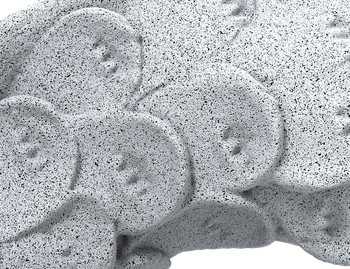}
\includegraphics[width=\dragonwidth\textwidth]{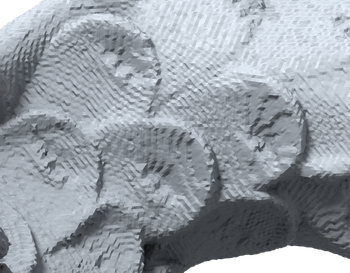}
\includegraphics[width=\dragonwidth\textwidth]{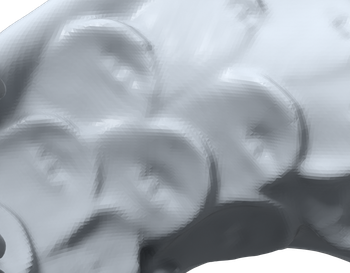}
\includegraphics[width=\dragonwidth\textwidth]{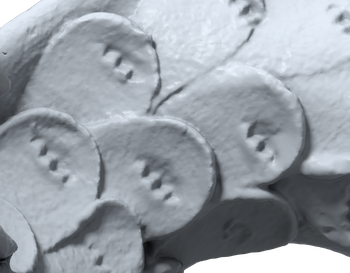}
\includegraphics[width=\dragonwidth\textwidth]{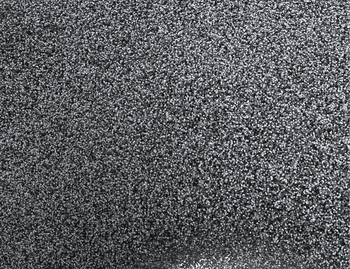}
\includegraphics[width=\dragonwidth\textwidth]{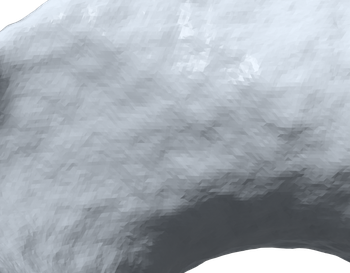} 
\\
\begin{scriptsize}
\makebox[\dragonwidth\textwidth]{(a) Input (noise free)}
\makebox[\dragonwidth\textwidth]{(b) Standard Poisson kernel}
\makebox[\dragonwidth\textwidth]{(c) Modified Poisson kernel}
\makebox[\dragonwidth\textwidth]{(d) DWG+sPSR}
\makebox[\dragonwidth\textwidth]{(e) Input (0.75\% noise)}
\makebox[\dragonwidth\textwidth]{(f) Modified Poisson kernel }\\
\end{scriptsize}
\caption{Poisson kernels and surface smoothness illustrated on the Dragon model with 5 million points. (a) Utilizing standard Poisson kernels often results in zigzagged non-smoothness artifacts in the GWN level sets. (b) Modifying the Poisson kernel greatly reduces these artifacts, producing a smoother GWN field, though it may induce excessive smoothness. (c) Leveraging the normals predicted by DWG in conjunction with screened Poisson surface reconstruction effectively preserves the fine details of the input models. (d)-(e) For point clouds with high levels of noise, utilizing the GWN level set directly is preferred, yielding more stable and reliable results compared to the combination of DWG and sPSR.}
    \label{fig:combiningdwgandpsr}
\end{figure*}

\textcolor{black}{Upon convergence, DWG produces both the level set $\overline{w}^{(\infty)}$ and the point normals $\mathbf{n}^{(\infty)}$. While the GWN level sets $\overline{w}^{(\infty)}$ capture the overall geometry effectively, they do not represent fine geometric details. Since methods in the PSR family excel at reconstructing fine details when provided with accurate point orientations, we recommend combining DWG with sPSR for models requiring detailed geometry. By using the oriented normals $\mathbf{n}^{(\infty)}$ from DWG as input to sPSR, we can achieve a final surface with enhanced detail. For models with high noise levels and real scans which often exhibit overlapping and misaligned scans, sPSR often yields double-layered outputs. Consequently, we 
recommend directly utilizing the GWN level $\overline{w}^{(\infty)}$ as the output for these scenarios. See Figures~\ref{fig:combiningdwgandpsr} and~\ref{fig:scan} for visual examples. }

\section{Experimental Results \& Comparisons}
\label{sec:results}

\paragraph{Qualitative Comparison} Table~\ref{tab:comparison} presents a qualitative comparison of DWG and \textcolor{black}{other implicit function-based methods} for 3D reconstruction from unoriented points. \textcolor{black}{Classic approaches~\cite{kobbelt}and~\cite{mullen2010signing} rely on unsigned distance fields, while most contemporary approaches, including ours, utilize GWN, demonstrating the robustness and global nature of this emerging computational tool. } Our method sets itself apart in terms of low time complexity,  parallelization,  scalability, and enhanced robustness, which will be further evidenced in the subsequent quantitative evaluations.

\renewcommand{\tiny}{\fontsize{5.1}{8}\selectfont}

\begin{table*}[htbp]
\setlength\tabcolsep{1pt}
    \centering
\caption{Statistics for small-scale models with up to 100K points (top) and large-scale models with the number of points exceeding 10 million (bottom). Each model is tested under noise-free and noisy conditions, with noise levels of 0.75\% for small- and large-scale models, respectively. The evaluation metrics include running time (h:m:s), peak memory consumption (in MB), and Chamfer distance (CD) scaled by $10^{-3}$, computed against high-resolution ground-truth meshes to assess reconstruction quality. \textcolor{black}{For consistency in runtime performance comparison, iPSR, WNNC, and DWG were all tested using identical settings for octree depth.} An NVIDIA RTX 4090 GPU with 24GB of memory was used for models up to 15M points, while an NVIDIA L20 GPU with 48GB of memory was used for larger models. Sequential methods were tested on an Intel i9 CPU equipped with 120GB of memory. PGR and WNNC, which have stages executed separately on CPUs and GPUs, report memory consumption for each stage. WNNC's reported times include both point orientation and subsequent surface reconstruction. DWG, which is operated on GPUs and can produce both point orientations and reconstructed surfaces simultaneously, reports the peak GPU memory consumption along with running times for initialization and parallel diffusion. The best results are highlighted in bold. 'DNF' indicates that the method did not complete within 24 hours.} 
\begin{tiny}
\begin{tabular}{cc|ccc|ccc|ccc|ccc|ccc|ccc|ccc|ccc|ccc}
\hline
\multicolumn{2}{c|}{\multirow{2}{*}{Model ($n$, noise level)}} & \multicolumn{3}{c|}{Hornung \& Kobbelt} & \multicolumn{3}{c|}{Mullen et al.} & \multicolumn{3}{c|}{iPSR} & \multicolumn{3}{c|}{PGR} & \multicolumn{3}{c|}{GCNO} & \multicolumn{3}{c|}{BIM} & 
\multicolumn{3}{c|}{\textcolor{black}{WNNC}} &
\multicolumn{3}{c|}{\textcolor{black}{SNO}} &
\multicolumn{3}{c}{DWG}                            \\ \cline{3-29} 
\multicolumn{2}{c|}{}                                          
& Time    & Mem.       & CD       
& Time    & Mem.       & CD          
& Time    & Mem.       & CD               
& Time    & Mem.   & CD               
& Time    & Mem.       & CD            
& Time    & Mem.       & CD            
& Time    & Mem.       & CD
& Time    & Mem.       & CD 
& Time   & Mem. & CD             \\ \hline

\multicolumn{1}{c|}{Bunny}                   & 0\%             
& 00:10:35       & 3542       & 7.15      
& 00:00:19           & 526          & 7.14      
& 00:00:16     & 153             & 1.84         
& 00:00:21     & 347 \& 1518     & \textbf{1.69}
& 00:30:11         & 215          & 2.10        
& 00:12:27      & \textbf{114} & 1.97          
& 00:00:06 & 4965 \& 2228 & 1.91
& 00:03:13 & 1304 & 2.02
& \textbf{00:00:01}  & 584  & 2.23            \\ \cline{2-29} 
\multicolumn{1}{c|}{10K}                     & 0.75\%         
& 00:10:34       & 4259       & 71.87   
& 00:00:17           & 504          & 12.97      
& 00:00:35     & 123             & 8.48           
& 00:00:23     & 342 \& 1520     & 12.18            
& 01:06:19         & 212          & 13.35        
& 00:23:01      & \textbf{113} & 8.27        
& 00:00:06 &  4951 \& 2228 & 9.75
& 00:07:10 & 1422 & 18.06
& \textbf{00:00:01}   & 582  & \textbf{7.70}  \\ \hline
\multicolumn{1}{c|}{CAD}          & 0\%             
& & Failed &   
& 00:00:35           & 603          & 6.90       
& 00:00:15     & 180             & 1.72          
& 00:01:07     & 349 \& 5036     & \textbf{1.66}   
& 02:10:05         & 495          & 24.09     
& 01:05:21      & \textbf{134} & 1.81          
& 00:00:07&  4996 \& 2232 & 1.81
& 00:05:41 & 2135 & 1.79
& \textbf{00:00:01}    & 645  & 2.00           \\ \cline{2-29} 
\multicolumn{1}{c|}{30K}                     & 0.75\%          
& & Failed &   
& 00:00:22           & 727          & 13.26       
& 00:01:58     & 171             & 12.10           
& 00:01:16     & 357 \& 5036     & 15.40         
& 03:57:52         & 486          & 31.33        
& 01:33:31      & \textbf{135} & 12.24        
& 00:00:07 & 4991 \& 2232 & 8.37 
& 00:09:18 & 2344 & 16.77
& \textbf{00:00:01}   & 634  & \textbf{7.23}  \\ \hline
\multicolumn{1}{c|}{LinkCup}              & 0\%             
& & Failed &  
& 00:00:44          & 970          & 48.55       
& 00:00:28     & 236             & 2.11           
& 00:06:42     & 518 \& 14542    & \textbf{1.91}           
& 05:22:38         & 963          & 17.91        
& 03:12:08      & \textbf{152} & 2.25 
& 00:00:11 & 5148 \& 2302 & 1.96
& 00:15:33 & 4620 & 1.97
& \textbf{00:00:01}   & 722  & 2.66           \\ \cline{2-29} 
\multicolumn{1}{c|}{50K}                     & 0.75\%          
& & Failed &  
& 00:00:36           & 1111          & 49.56       
& 00:02:37     & 171             & 11.38        
& 00:06:50     & 503 \& 14918    & 18.41           
& 20:50:33         & 904          & 28.17       
& 05:03:41      & \textbf{160} & 16.14     
&00:00:09 & 4993 \& 2239 & 11.03
& 00:16:17 & 4433 & 13.45
&  \textbf{00:00:01}   & 683  & \textbf{10.80} \\ \hline
\multicolumn{1}{c|}{Metratron}               & 0\%             
& 00:13:52       & 5764       & 17.34    
& 00:00:49           & 1273          & 38.76       
& 00:00:49     & 292             & 2.36           
& 00:11:10     & 710 \& 22387    & \textbf{2.15}    
& 22:20:07         & 1146         & 15.97  
& 08:31:25      & \textbf{181} & 2.73      
& 00:00:11 & 5042 \& 2239 & 2.23
& 00:11:10 & 4926 & 2.49
& \textbf{00:00:01}   & 827  & 2.29           \\ \cline{2-29} 
\multicolumn{1}{c|}{70K}                     & 0.75\%          
& & Failed &  
& 00:00:39           & 1451          & 55.92       
& 00:03:09     & 226             & 12.09          
& 00:11:28     & 866 \& 22308    & 28.87          
& 23:09:49         & 1132         & 25.74       
& 12:15:39      & \textbf{180} & 28.38  
& 00:00:09 & 5021 \& 2239 & 8.79
& 00:22:11 & 5690 & 22.99
& \textbf{00:00:01}   & 729  & \textbf{8.14} \\ \hline
\multicolumn{1}{c|}{Dino}                    & 0\%             
& 00:49:23       & 25913      & 47.90    
& 00:01:01           & 960          & 20.85     
& 00:01:58     & 424             & \textbf{1.74}    
& \multicolumn{3}{c|}{Out of memory}             
& & DNF &  
& 18:20:05      & \textbf{208} & 2.07
& 00:00:12 & 5029 \& 2251 & 1.81
& 00:19:10 & 7061 & 3.14
& \textbf{00:00:01}   & 890  & 2.27          \\ \cline{2-29} 
\multicolumn{1}{c|}{100K}                    & 0.75\%          
& & Failed &  
& 00:00:51           & 1340          & 23.62       
& 00:04:09     & \textbf{184}    & 13.69            
& \multicolumn{3}{c|}{Out of memory}             
& & DNF &  
& 10:58:24 & 202          & 17.48        
& 00:00:12 & 5017 \& 2251 & 22.12
& 00:33:53 & 8242 & 16.04
& \textbf{00:00:02}   & 739  & \textbf{10.23} \\ \hline
\end{tabular}
\begin{tabular}{c|ccc|ccc|ccc|ccc|ccc|ccc|ccc|ccc|ccc}
\hline
\multirow{3}{*}{Model ($n$)} & \multicolumn{9}{c|}{iPSR~\cite{hou2022iterative}} 
& \multicolumn{9}{c|}{\textcolor{black}{WNNC~\cite{Lin2024fast}}}
& \multicolumn{9}{c}{DWG} \\ \cline{2-28} 
& \multicolumn{3}{c|}{0.75\% noise, $d=10$} & \multicolumn{3}{c|}{0\% noise, $d=12$} & \multicolumn{3}{c|}{0\% noise, $d=13$} 
& 
\multicolumn{3}{c|}{0.75\% noise, $d=10$} & \multicolumn{3}{c|}{0\% noise, $d=12$} & \multicolumn{3}{c|}{0\% noise, $d=13$} & \multicolumn{3}{c|}{0.75\% noise, $d=10$}& \multicolumn{3}{c|}{0\% noise, $d=12$} & \multicolumn{3}{c}{0\% noise, $d=13$} 
\\ \cline{2-28} 
& Time   & Mem.   & \multicolumn{1}{c|}{CD}  
& Time & Mem. & \multicolumn{1}{c|}{CD} 
& Time &  Mem.    & CD  
& Time   & Mem.   & \multicolumn{1}{c|}{CD}  
& Time & Mem. & \multicolumn{1}{c|}{CD} 
& Time &  Mem.    & CD  & Time & Mem. & \multicolumn{1}{c|}{CD} 
& Time & Mem. & \multicolumn{1}{c|}{CD} 
& Time  & Mem.    & CD  \\ \hline
Laocoon (10M)&  06:35:53 & 20972   &  13.01
& 03:40:30& 20155  &  \textbf{1.53}    
& 08:59:54& 34321 &  \textbf{1.53}    

& 00:16:21 &  31924 \& 6215 & 8.23
& 00:14:53 &  47992 \& 6448 & 1.81
& 00:18:02 &  74952 \& 7217 & 1.81

& \textbf{00:04:15} & 16553  &  \textbf{5.43}
& \textbf{00:03:53} & 23981  &  1.91
& \textbf{00:04:00}  & 24059 &  1.85
\\ \hline
Ariadne (10M) & 12:10:48        & 24012         &  12.61                    
& 02:51:41      & 21274       &    \textbf{1.63}                           
& 05:28:04     & 27324  &   \textbf{1.63}     

& 00:16:02 & 37161 \& 6202 & 6.03
& 00:11:56 & 35078 \& 6439 & 1.74
& 00:16:24 & 63163 \& 6621 & 1.74

& \textbf{00:04:29}    & 17405        & \textbf{4.72}       
& \textbf{00:03:38}  & 22377       & 1.75        
& \textbf{00:03:44}  & 22134       &  1.74    
\\ \hline
Lucy (15M) & 01:59:32       & 11179         & 10.60    
& 03:17:19      & 29882       &   \textbf{1.01}                    
& 04:24:47     & 33105        &   \textbf{1.00}     

& 00:20:36 &  22188 \& 8116 &  8.28
& 00:19:07 & 53734 \& 8522 &  1.38
& 00:22:04 & 56263 \& 8522 & 1.38

& \textbf{00:02:47} & 15057       & \textbf{5.04}                  
& \textbf{00:03:37} & 23503       & 1.73                  
& \textbf{00:04:13} & 30977        & 1.72   
\\ \hline
\textcolor{black}{Alien (15M)} &   03:00:11     & 16722         &  12.47    
&05:47:39    & 39523       &  \textbf{1.16}                  
& 06:41:41     & 43047      &  \textbf{1.15}    

& 00:22:16 &  27815 \& 8087 & 9.20
& 00:21:25 &  67926 \& 8496 & 1.56
& 00:24:05 &  68345 \& 8496 & 1.55

& \textbf{00:04:23} &    19408     & \textbf{5.66}             
& \textbf{00:04:15} &    27559    &    1.45
& \textbf{00:04:53} &    34325     &  1.41       
\\ \hline
Thai Statue (20M) & 04:01:01       & 15933         &  11.21
& 11:09:16     & 44391       &   \textbf{1.22}            
& 11:48:38     & 49332       &  \textbf{1.20}    

& 00:28:35 & 30463 \& 10119 & 8.62
& 00:27:41 & 75282 \& 10644 & 1.71
& 00:28:10 & 75431 \& 10644 & 1.71  

& \textbf{00:03:51}  & 21048        &  \textbf{5.79}           
& \textbf{00:04:52}  & 33161       & 1.67  
& \textbf{00:05:53}  & 42151       & 1.64   
\\ \hline
\end{tabular}
\end{tiny}   
\label{tab:statistics}
\end{table*}

\begin{figure*}[!htbp] \centering
    \newcommand{\sixsize}{0.105}
    \includegraphics[width=\sixsize\textwidth]{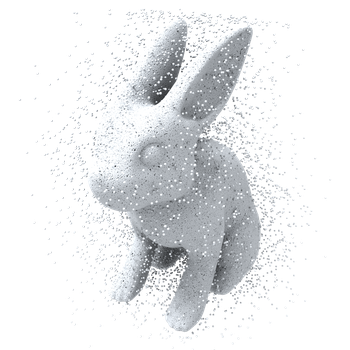}
    \includegraphics[width=\sixsize\textwidth]{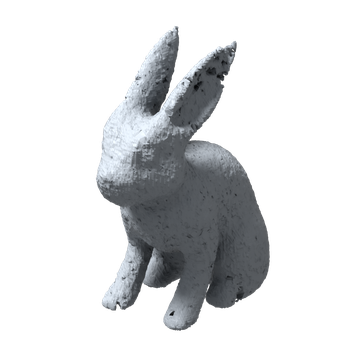}
    \includegraphics[width=\sixsize\textwidth]{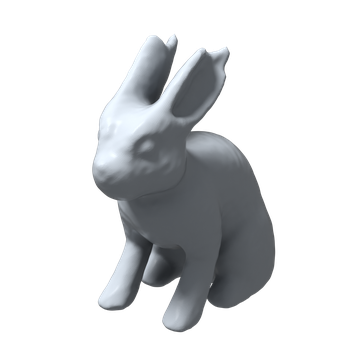}
    \includegraphics[width=\sixsize\textwidth]{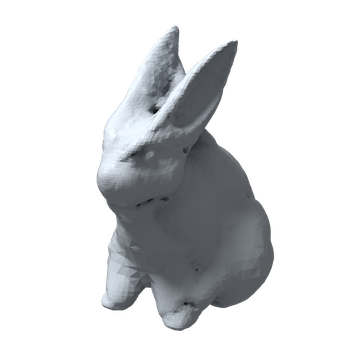}
    \includegraphics[width=\sixsize\textwidth]{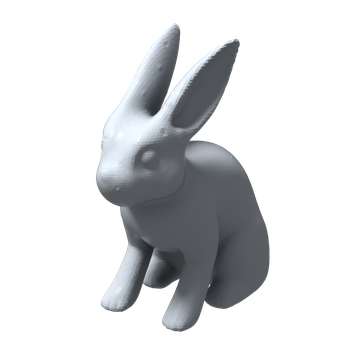}
    \includegraphics[width=\sixsize\textwidth]{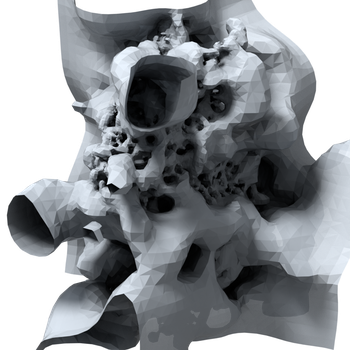}
    \includegraphics[width=\sixsize\textwidth]{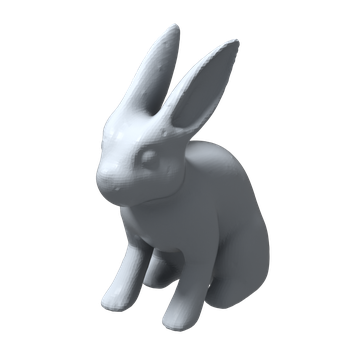}
    \includegraphics[width=\sixsize\textwidth]{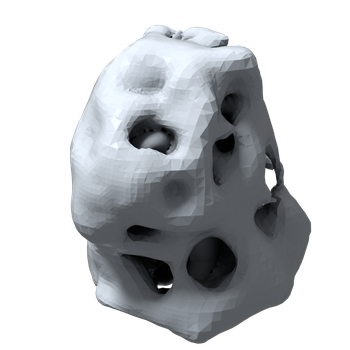}
    \includegraphics[width=\sixsize\textwidth]{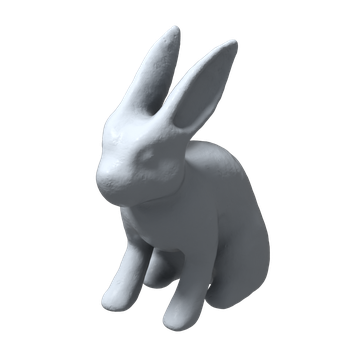}
    \\
    \includegraphics[width=\sixsize\textwidth]{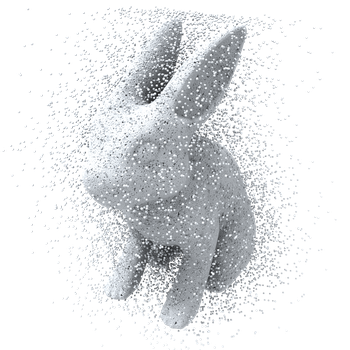}
    \includegraphics[width=\sixsize\textwidth]{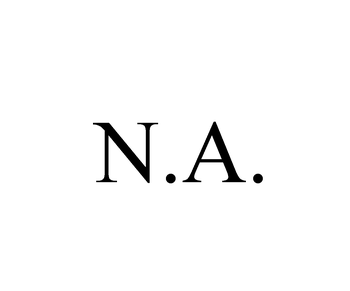}
    \includegraphics[width=\sixsize\textwidth]{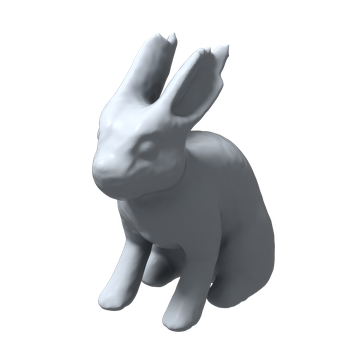}
    \includegraphics[width=\sixsize\textwidth]{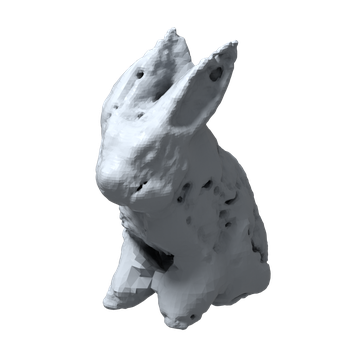}
    \includegraphics[width=\sixsize\textwidth]{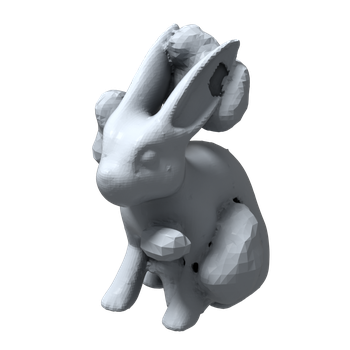}
    \includegraphics[width=\sixsize\textwidth]{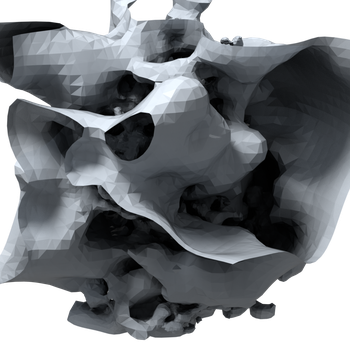}
    \includegraphics[width=\sixsize\textwidth]{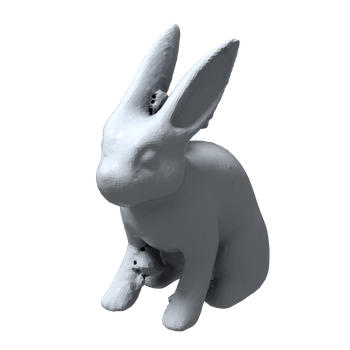}
    \includegraphics[width=\sixsize\textwidth]{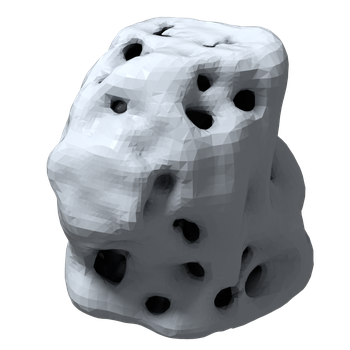}
    \includegraphics[width=\sixsize\textwidth]{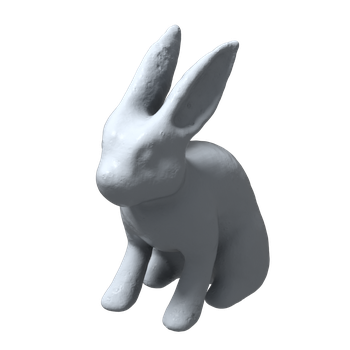}
    \\
    \includegraphics[width=\sixsize\textwidth]{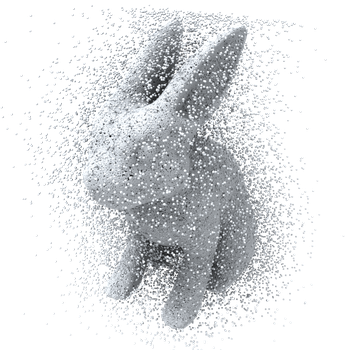}
    \includegraphics[width=\sixsize\textwidth]{figures/NA.png}
    \includegraphics[width=\sixsize\textwidth]{figures/NA.png}
    \includegraphics[width=\sixsize\textwidth]{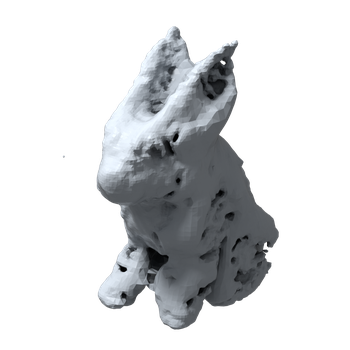}
    \includegraphics[width=\sixsize\textwidth]{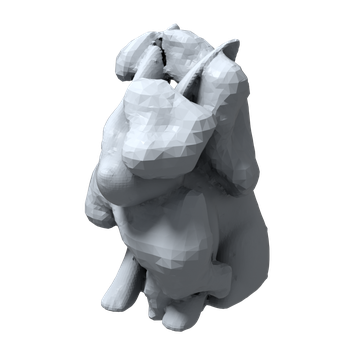}
    \includegraphics[width=\sixsize\textwidth]{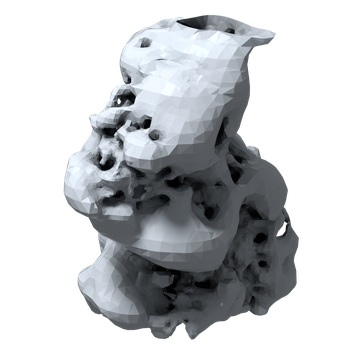}
    \includegraphics[width=\sixsize\textwidth]{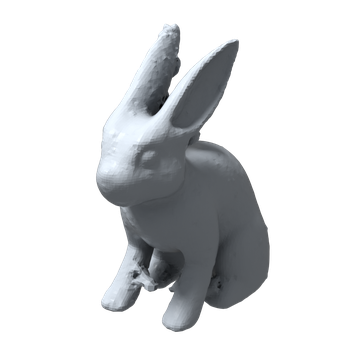}
    \includegraphics[width=\sixsize\textwidth]{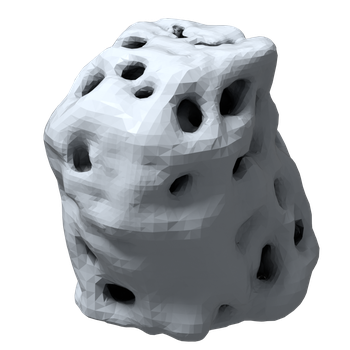}
    \includegraphics[width=\sixsize\textwidth]{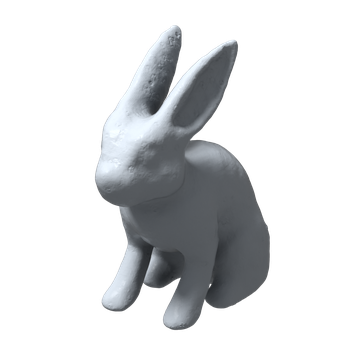}
    \\
    \includegraphics[width=\sixsize\textwidth]{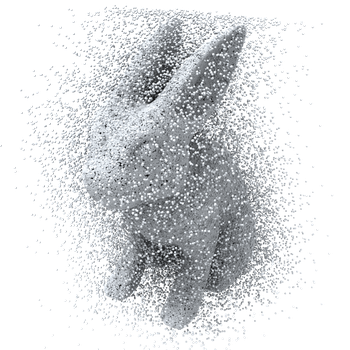}
    \includegraphics[width=\sixsize\textwidth]{figures/NA.png}
    \includegraphics[width=\sixsize\textwidth]{figures/NA.png}
    \includegraphics[width=\sixsize\textwidth]{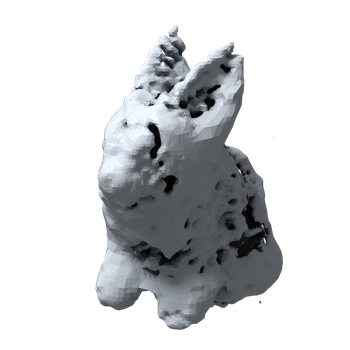}
    \includegraphics[width=\sixsize\textwidth]{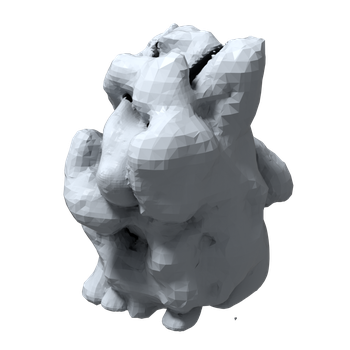}
    \includegraphics[width=\sixsize\textwidth]{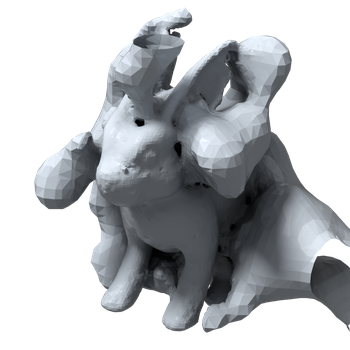}
    \includegraphics[width=\sixsize\textwidth]{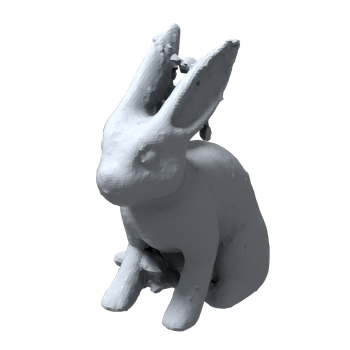}
    \includegraphics[width=\sixsize\textwidth]{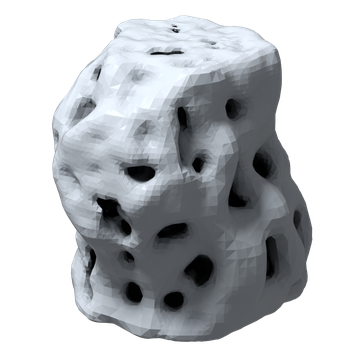}
    \includegraphics[width=\sixsize\textwidth]{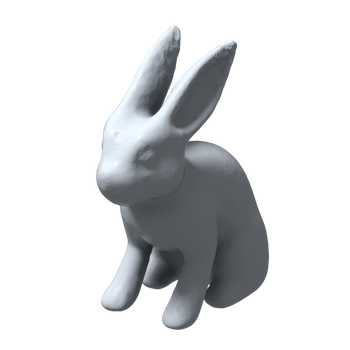}
    \\
    \includegraphics[width=\sixsize\textwidth]{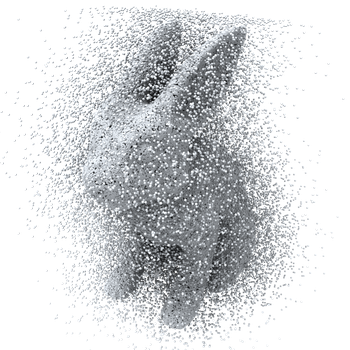}
    \includegraphics[width=\sixsize\textwidth]{figures/NA.png}
    \includegraphics[width=\sixsize\textwidth]{figures/NA.png}
    \includegraphics[width=\sixsize\textwidth]{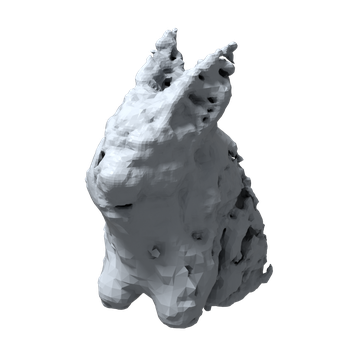}
    \includegraphics[width=\sixsize\textwidth]{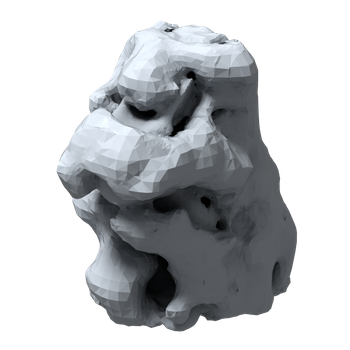}
    \includegraphics[width=\sixsize\textwidth]{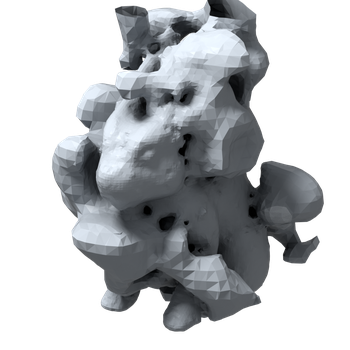}
    \includegraphics[width=\sixsize\textwidth]{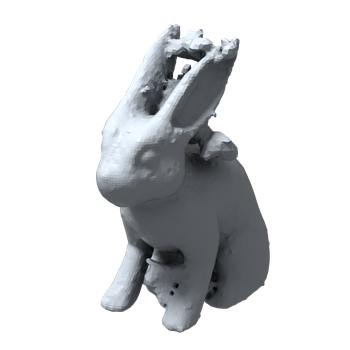}
    \includegraphics[width=\sixsize\textwidth]{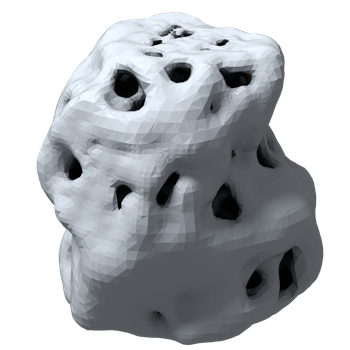}
    \includegraphics[width=\sixsize\textwidth]{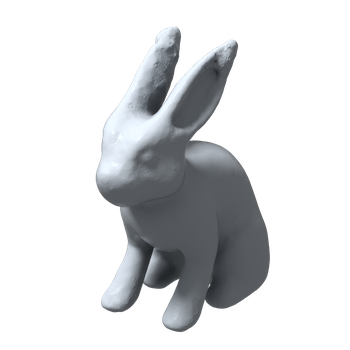}
    \\
    \makebox[\sixsize\textwidth]{\scriptsize Input}
    \makebox[\sixsize\textwidth]{\scriptsize  \textcolor{black}{Mullen et al.}}
    \makebox[\sixsize\textwidth]{\scriptsize \textcolor{black}{Hornung and Kobbelt}}
    \makebox[\sixsize\textwidth]{\scriptsize PGR}
    \makebox[\sixsize\textwidth]{\scriptsize iPSR}
    \makebox[\sixsize\textwidth]{\scriptsize GCNO}
    \makebox[\sixsize\textwidth]{\scriptsize BIM}
    \makebox[\sixsize\textwidth]{\scriptsize \textcolor{black}{SNO}}
    \makebox[\sixsize\textwidth]{\scriptsize DWG}
    \\
    \caption{\textcolor{black}{Robustness tests against outliers. From top to bottom, outlier ratios of 5\%, 10\%, 15\%, 20\%, and 25\% were added to the Rabbit model. Our method demonstrates greater robustness against outliers compared to baseline approaches. ``N.A.'' indicates cases where the method fails to produce any results.}} 
    \label{fig:outliers}
\end{figure*}

\begin{figure*}[!ht] \centering
    \newcommand{\eightsize}{0.102750}
    \newcommand{\eightsizedoghead}{0.094500}    
    \newcommand{\eightsizephone}{0.102750}
    \newcommand{\eightsizetrebol}{0.10275}
    \newcommand{\eightsizenormals}{0.102750}
    \includegraphics[width=\eightsizedoghead\textwidth]{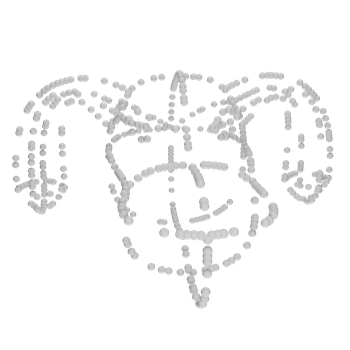}  \includegraphics[width=\eightsize\textwidth]{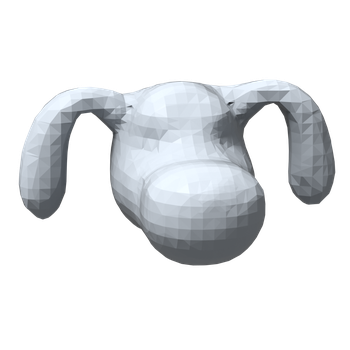}   \includegraphics[width=\eightsize\textwidth]{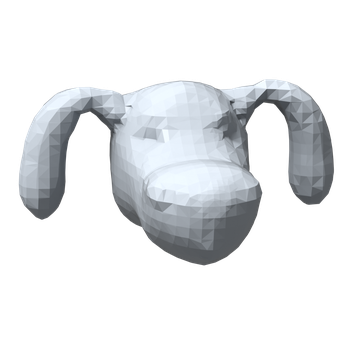}    \includegraphics[width=\eightsize\textwidth]{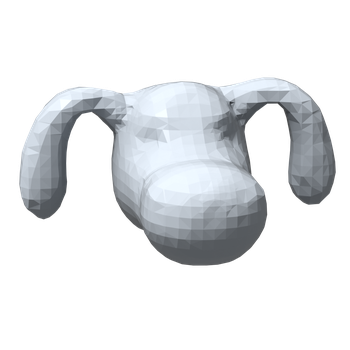}    \includegraphics[width=\eightsize\textwidth]{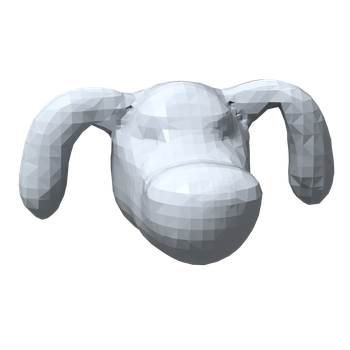}
    \includegraphics[width=\eightsize\textwidth]{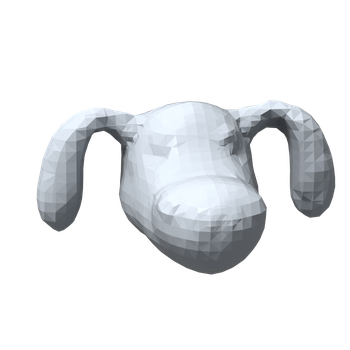}
    \includegraphics[width=\eightsize\textwidth]{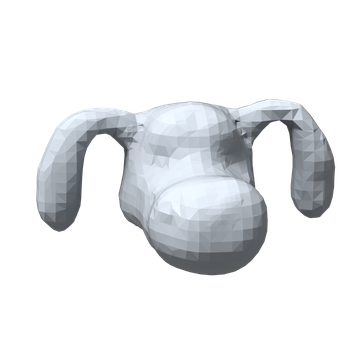}
    \includegraphics[width=\eightsize\textwidth]{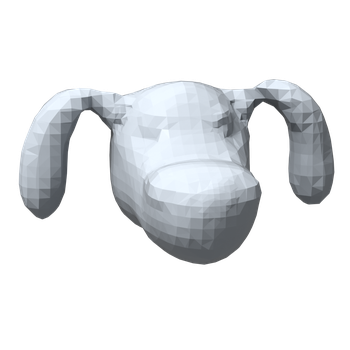}
    \includegraphics[width=\eightsize\textwidth]{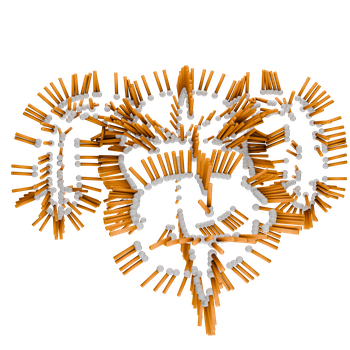}\\
    \makebox[\eightsize\textwidth]{\footnotesize 500 points }
    \makebox[\eightsize\textwidth]{\footnotesize 00:03:41}
    \makebox[\eightsize\textwidth]{\footnotesize 00:00:11}
    \makebox[\eightsize\textwidth]{\footnotesize 00:00:28}
    \makebox[\eightsize\textwidth]{\footnotesize 00:00:21}
    \makebox[\eightsize\textwidth]{\footnotesize 00:01:06}
    \makebox[\eightsize\textwidth]{\footnotesize 00:00:01}  
    \makebox[\eightsize\textwidth]{\footnotesize 00:00:01}
    \makebox[\eightsize\textwidth]{\footnotesize }\\
    \includegraphics[width=\eightsizephone\textwidth]{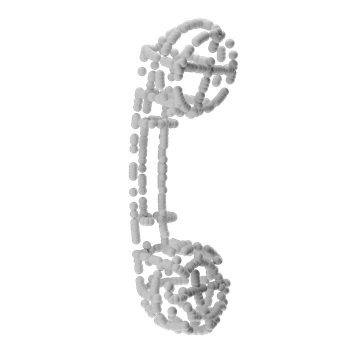}
    \includegraphics[width=\eightsize\textwidth]{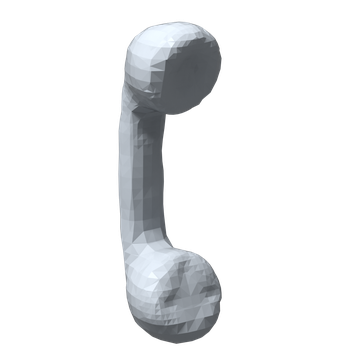}
    \includegraphics[width=\eightsize\textwidth]{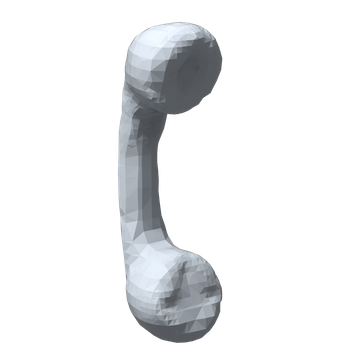}
    \includegraphics[width=\eightsize\textwidth]{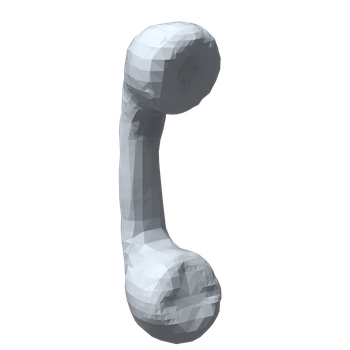}
    \includegraphics[width=\eightsize\textwidth]{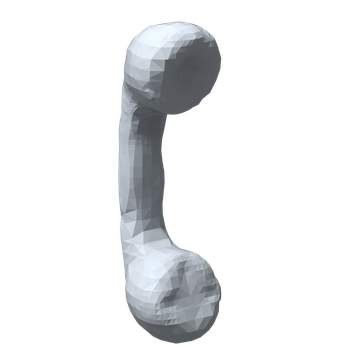}    \includegraphics[width=\eightsize\textwidth]{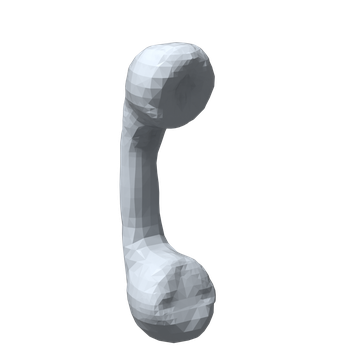}
    \includegraphics[width=\eightsize\textwidth]{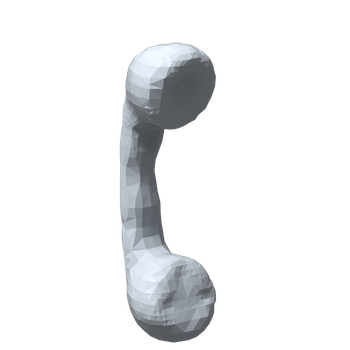}
    \includegraphics[width=\eightsize\textwidth]{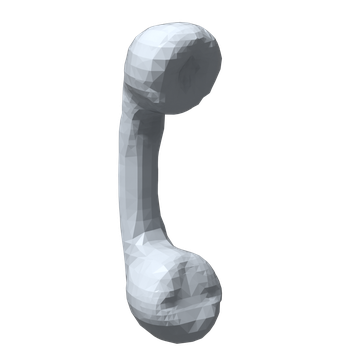} 
    \includegraphics[width=\eightsize\textwidth]{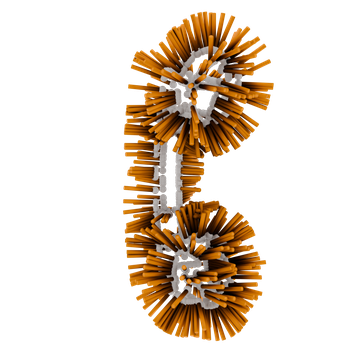}\\
    \makebox[\eightsize\textwidth]{\footnotesize 500 points }
    \makebox[\eightsize\textwidth]{\footnotesize 00:03:22}
    \makebox[\eightsize\textwidth]{\footnotesize 00:00:07}
    \makebox[\eightsize\textwidth]{\footnotesize 00:00:23}
    \makebox[\eightsize\textwidth]{\footnotesize 00:00:18}
     \makebox[\eightsize\textwidth]{\footnotesize 00:00:54}
    \makebox[\eightsize\textwidth]{\footnotesize 00:00:01}
    \makebox[\eightsize\textwidth]{\footnotesize 00:00:01}
    \makebox[\eightsize\textwidth]{\footnotesize }\\    \includegraphics[width=\eightsizetrebol\textwidth]{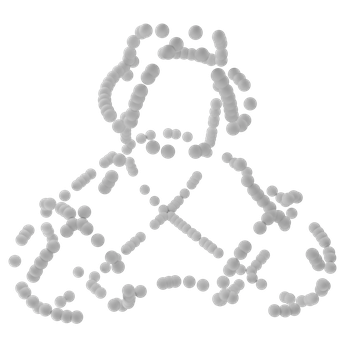}
    \includegraphics[width=\eightsize\textwidth]{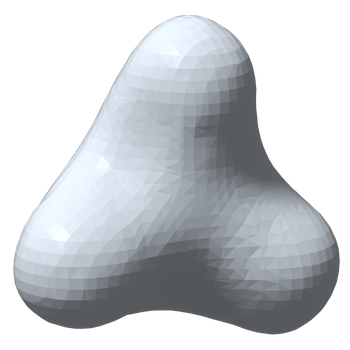}
    \includegraphics[width=\eightsize\textwidth]{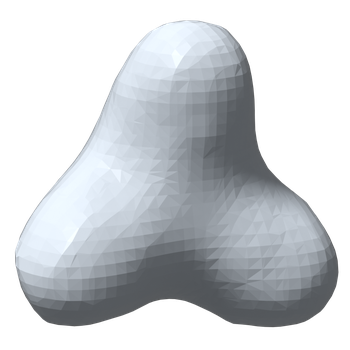}
    \includegraphics[width=\eightsize\textwidth]{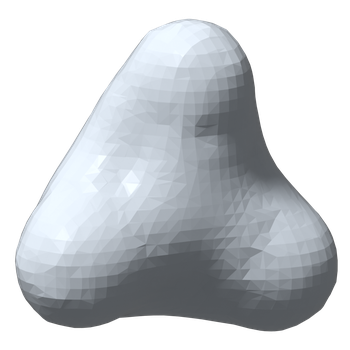}
    \includegraphics[width=\eightsize\textwidth]{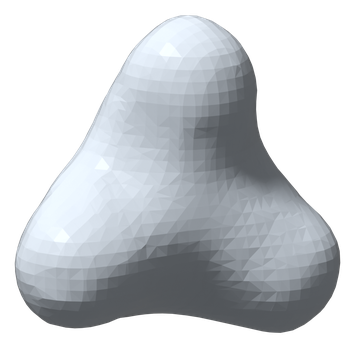}    \includegraphics[width=\eightsize\textwidth]{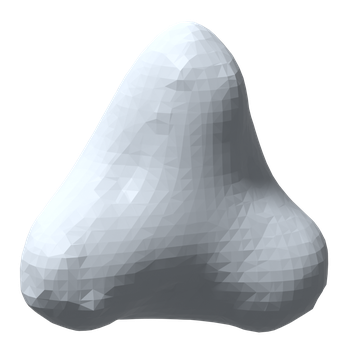}    \includegraphics[width=\eightsize\textwidth]{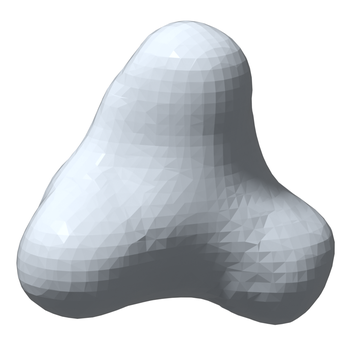}    \includegraphics[width=\eightsize\textwidth]{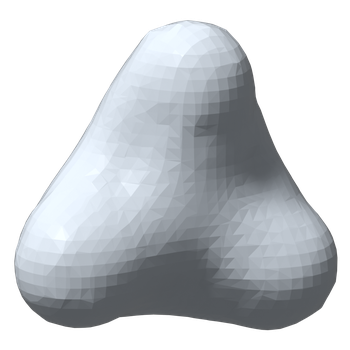}   \includegraphics[width=\eightsize\textwidth]{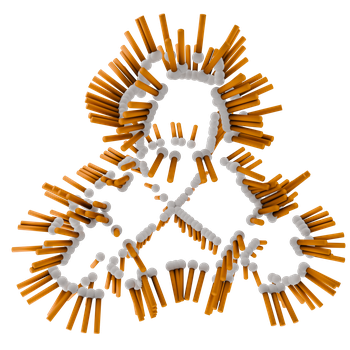}\\
    \makebox[\eightsize\textwidth]{\footnotesize 250 points }
    \makebox[\eightsize\textwidth]{\footnotesize 00:02:31}
    \makebox[\eightsize\textwidth]{\footnotesize 00:00:12}
    \makebox[\eightsize\textwidth]{\footnotesize 00:00:11}
    \makebox[\eightsize\textwidth]{\footnotesize 00:00:10}
    \makebox[\eightsize\textwidth]{\footnotesize 00:01:01}
    \makebox[\eightsize\textwidth]{\footnotesize 00:00:01} 
    \makebox[\eightsize\textwidth]{\footnotesize 00:00:02}
    \makebox[\eightsize\textwidth]{\footnotesize Normals }\\
    \makebox[\eightsize\textwidth]{\footnotesize Input}
    \makebox[\eightsize\textwidth]{\footnotesize VIPSS}
    \makebox[\eightsize\textwidth]{\footnotesize iPSR}
    \makebox[\eightsize\textwidth]{\footnotesize GCNO}
    \makebox[\eightsize\textwidth]{\footnotesize BIM}
    \makebox[\eightsize\textwidth]{\footnotesize SNO}
    \makebox[\eightsize\textwidth]{\footnotesize WNNC}
    \makebox[\eightsize\textwidth]{\footnotesize DWG+sPSR}    \makebox[\eightsizenormals\textwidth]{\footnotesize oriented by DWG}\\
    \label{fig:wireframe}
    \caption{\textcolor{black}{Results on wireframe data~\cite{VIPSS}, characterized by sparse yet structured point distributions, demonstrate the efficacy of VIPSS, iPSR, DWG, and other recent GWN-based methods on such models. Our method, however, outpaces these in terms of speed. Note that classic approaches, such as \cite{kobbelt} and \cite{mullen2010signing}, face difficulties with this type of input.}}
\end{figure*}

\begin{figure*}[t] \centering
   \newcommand{\sevensizea}{0.1195} \includegraphics[width=\sevensizea\textwidth]{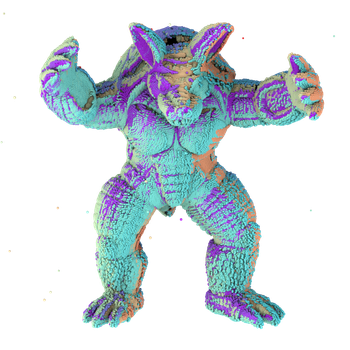}
    \includegraphics[width=\sevensizea\textwidth]{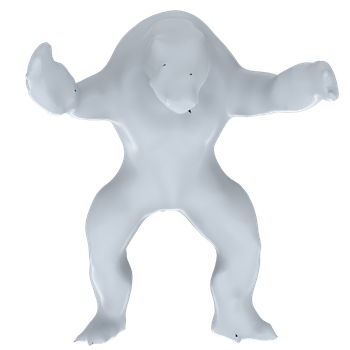}
    \includegraphics[width=\sevensizea\textwidth]{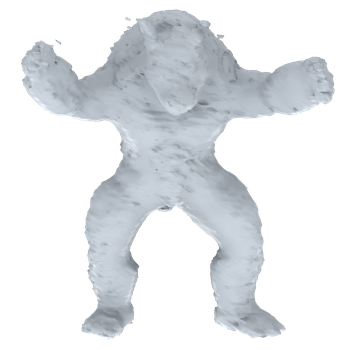}
    \includegraphics[width=\sevensizea\textwidth]{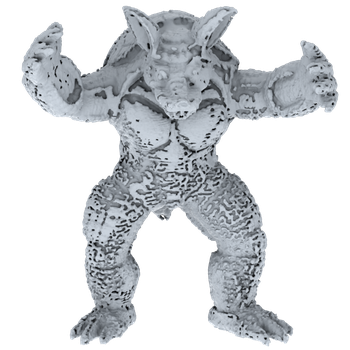}
    \includegraphics[width=\sevensizea\textwidth]{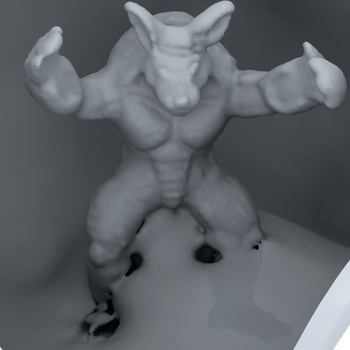}
    \includegraphics[width=\sevensizea\textwidth]{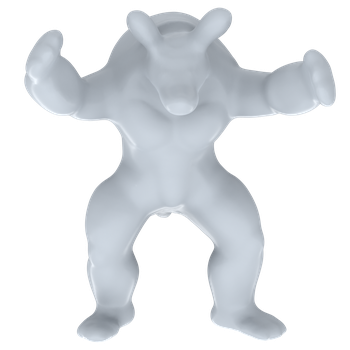}
    \includegraphics[width=\sevensizea\textwidth]{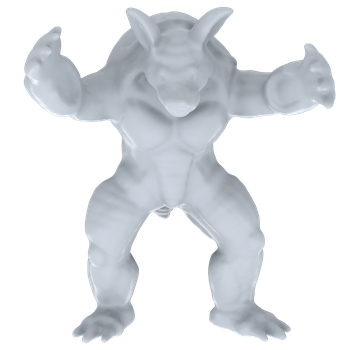}
    \\
    \makebox[\sevensizea\textwidth]{\footnotesize 639K points, 24 scans}
    \makebox[\sevensizea\textwidth]{\footnotesize 00:29:01}
    \makebox[\sevensizea\textwidth]{\footnotesize 00:01:59}
    \makebox[\sevensizea\textwidth]{\footnotesize 00:06:24}
    \makebox[\sevensizea\textwidth]{\footnotesize 03:21:58}
    \makebox[\sevensizea\textwidth]{\footnotesize 00:00:25}
    \makebox[\sevensizea\textwidth]{\footnotesize 00:00:13} \\
     
    \includegraphics[width=\sevensizea\textwidth]{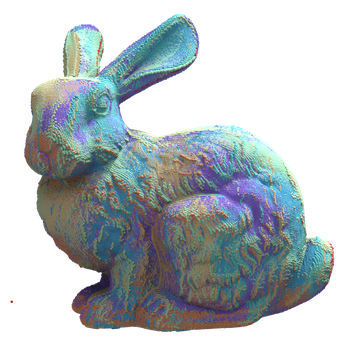}
    \includegraphics[width=\sevensizea\textwidth]{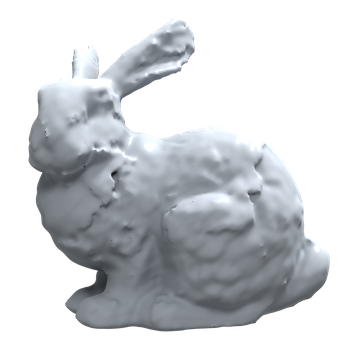}
    \includegraphics[width=\sevensizea\textwidth]{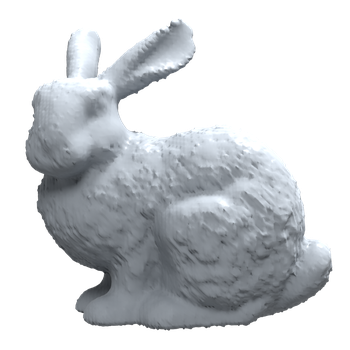}
    \includegraphics[width=\sevensizea\textwidth]{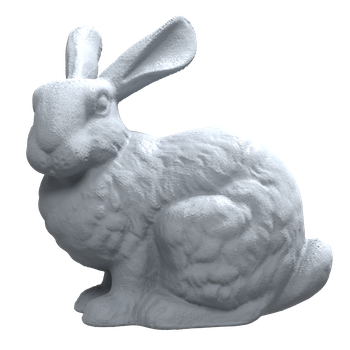}
    \includegraphics[width=\sevensizea\textwidth]{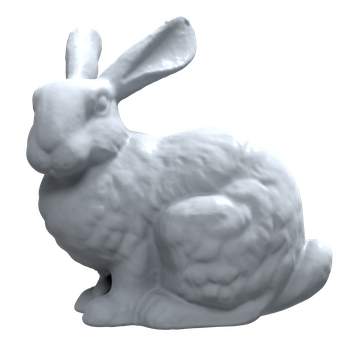}
    \includegraphics[width=\sevensizea\textwidth]{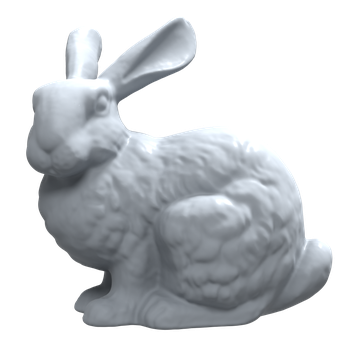}
    \includegraphics[width=\sevensizea\textwidth]{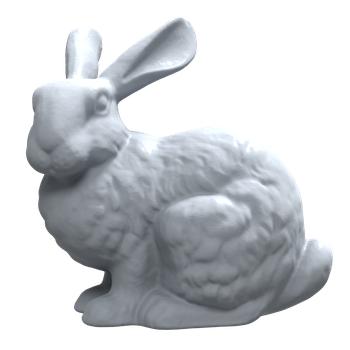}
    \\
      \makebox[\sevensizea\textwidth]{\footnotesize 362K points, 10 scans}
    \makebox[\sevensizea\textwidth]{\footnotesize 00:49:23}
    \makebox[\sevensizea\textwidth]{\footnotesize 00:09:07}
    \makebox[\sevensizea\textwidth]{\footnotesize 00:01:55}
    \makebox[\sevensizea\textwidth]{\footnotesize 00:51:17}
    \makebox[\sevensizea\textwidth]{\footnotesize 00:00:24}
    \makebox[\sevensizea\textwidth]{\footnotesize 00:00:11} \\
     
    \includegraphics[width=\sevensizea\textwidth]{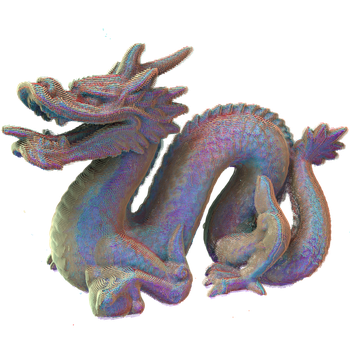}
    \includegraphics[width=\sevensizea\textwidth]{figures/NA.png}
    \includegraphics[width=\sevensizea\textwidth]{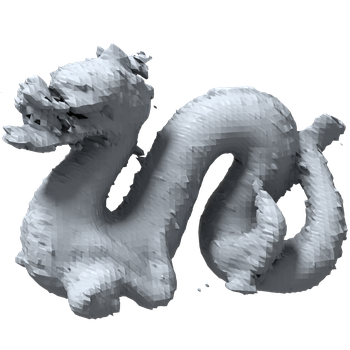}
    \includegraphics[width=\sevensizea\textwidth]{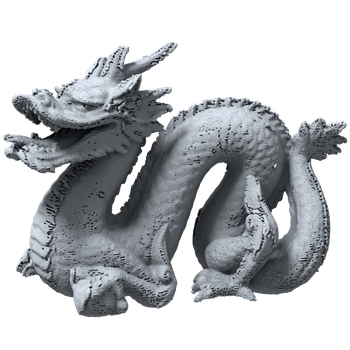}
    \includegraphics[width=\sevensizea\textwidth]{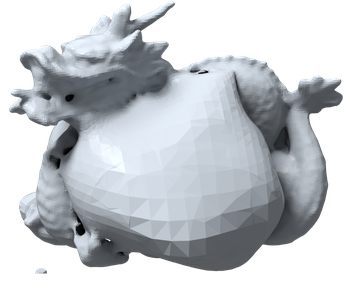}
    \includegraphics[width=\sevensizea\textwidth]{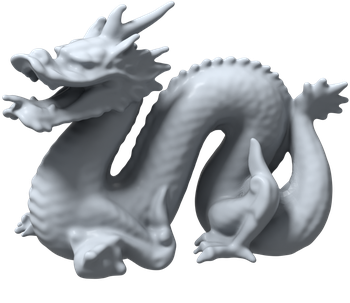}
    \includegraphics[width=\sevensizea\textwidth]{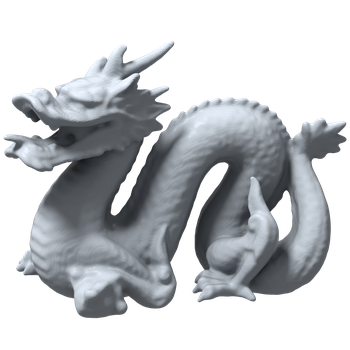}
    \\
    \makebox[\sevensizea\textwidth]{\footnotesize 1002K points, 31 scans}
    \makebox[\sevensizea\textwidth]{\footnotesize }
    \makebox[\sevensizea\textwidth]{\footnotesize 00:08:31}
    \makebox[\sevensizea\textwidth]{\footnotesize 00:16:07}
    \makebox[\sevensizea\textwidth]{\footnotesize 05:37:31}
    \makebox[\sevensizea\textwidth]{\footnotesize 00:00:45}
    \makebox[\sevensizea\textwidth]{\footnotesize 00:00:29} \\
     
     \includegraphics[width=\sevensizea\textwidth]{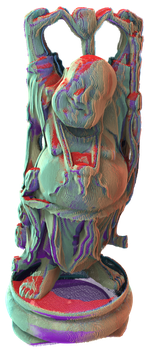}
    \includegraphics[width=\sevensizea\textwidth]{figures/NA.png}
    \includegraphics[width=\sevensizea\textwidth]{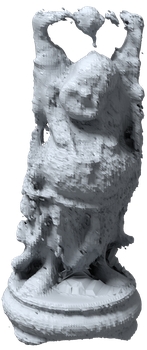}
    \includegraphics[width=\sevensizea\textwidth]{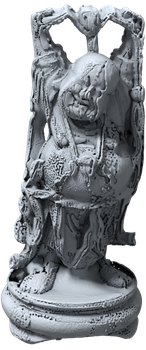}
    \includegraphics[width=\sevensizea\textwidth]{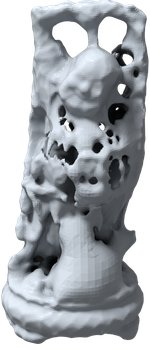}
     \includegraphics[width=\sevensizea\textwidth]{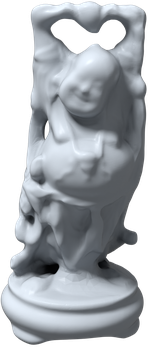}
    \includegraphics[width=\sevensizea\textwidth]{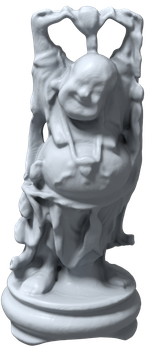}
    \\
     \makebox[\sevensizea\textwidth]{\footnotesize 1274K points, 18 scans}
    \makebox[\sevensizea\textwidth]{\footnotesize }
    \makebox[\sevensizea\textwidth]{\footnotesize 00:11:25}
    \makebox[\sevensizea\textwidth]{\footnotesize 00:21:49}
    \makebox[\sevensizea\textwidth]{\footnotesize 05:10:07}
    \makebox[\sevensizea\textwidth]{\footnotesize 00:00:52}
    \makebox[\sevensizea\textwidth]{\footnotesize 00:00:32} \\
        
    \makebox[\sevensizea\textwidth]{\footnotesize Input}
    \makebox[\sevensizea\textwidth]{\footnotesize Hornung and Kobbelt}
    \makebox[\sevensizea\textwidth]{\footnotesize Mullen et al.}
    \makebox[\sevensizea\textwidth]{\footnotesize iPSR}
    \makebox[\sevensizea\textwidth]{\footnotesize SNO}
    \makebox[\sevensizea\textwidth]{\footnotesize WNNC}
    \makebox[\sevensizea\textwidth]{\footnotesize DWG}
    \\
        \vspace{-0.1in}
    \caption{\textcolor{black}{Results on real scans from the Stanford 3D Scanning Repository. Each model consists of multiple scans rendered in distinct colors to emphasize differences. To simulate typical registration misalignments, small translations and rotations were introduced between scans. Baseline methods often struggle with these misalignments, resulting in topological and geometrical errors such as disconnected components, undesired high genus, incorrectly filling holes, double-layered sections, and over-smoothed  geometries. Moreover, SNO is sensitive to outliers, often producing outputs that unnecessarily encapsulate these outliers. WNNC, which focuses on point orientation, requires an additional step for surface reconstruction. In contrast, our method directly generates the surface. Our method directly generates surfaces with higher quality and does so in less time.  }    } 
    \label{fig:scan}
\end{figure*}

\paragraph{Baselines and Metrics} We conducted a quantitative comparison of our approach with leading 3D reconstruction methods, including \textcolor{black}{ two classic approaches ~\cite{kobbelt} and~\cite{mullen2010signing},
and six contemporary methods:}
iPSR~\cite{hou2022iterative}, PGR~\cite{lin2022surface}, GCNO~\cite{xu2023globally},  BIM~\cite{BIM}, \textcolor{black}{SNO~\cite{Huang2024Stochastic}} and \textcolor{black}{WNNC~\cite{Lin2024fast}}. 
\textcolor{black}{For \citet{mullen2010signing}'s method, due to the absence of  an official implementation, we utilized a third-party, open-source version\footnote{\textcolor{black}{\url{https://github.com/ajyang99/geometry-processing-signing-the-unsigned}}}. For the remaining baseline methods, we employed the official implementations provided by the authors.} \textcolor{black}{\textbf{We have carefully tuned the parameters for all methods to ensure the reliability of the reconstruction results for each technique.} }
\textcolor{black}{We evaluated the two parallel methods, DWG and WNNC, on an NVIDIA RTX 4090 GPU and the other methods on a high-end PC equipped with an Intel i9 CPU and 120 GB of memory. } We assessed performance metrics such as Chamfer distance for surface quality, running time, and peak memory consumption on representative models. \textcolor{black}{These models include models with thin structures (Figure~\ref{fig:thinstructure}), outliers (Figure~\ref{fig:outliers}), wireframe data (Figure~\ref{fig:wireframe}), and real-world inputs exhibiting overlapping and misaligned scans (Figure~\ref{fig:scan}). To generate noisy models, we introduced Gaussian noise to each point, with the intensity of the noise proportional to the \textbf{diagonal} of the bounding box of the input point cloud.} 

\begin{figure*}
\newcommand{\finalpagesize}{0.11350}
\newcommand{\finalpagesizethai}{0.12350}
\newcommand{\finalpagesizethaitwo}{0.11250}
\newcommand{\thaiheight}{0.225}
\newcommand{\alienheight}{0.185}
\newcommand{\lucynaheight}{0.125}
\newcommand{\lucyheight}{0.1955}
\newcommand{\lucynaheightna}{0.1505}
\newcommand{\statuteheight}{0.130}
\newcommand{\finalpagesizethaicloseup}{0.11350}
\newcommand{\finalpagesizelucy}{0.110}
\centering
\includegraphics[height=\thaiheight\textwidth]{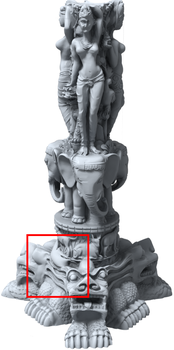}
\includegraphics[height=\thaiheight\textwidth]{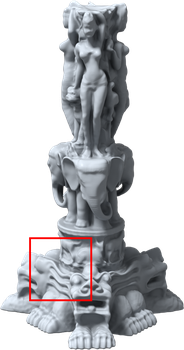}
\includegraphics[height=\thaiheight\textwidth]{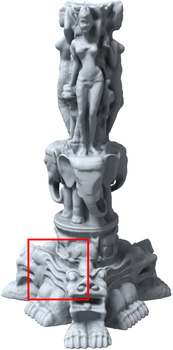}
\includegraphics[height=\thaiheight\textwidth]{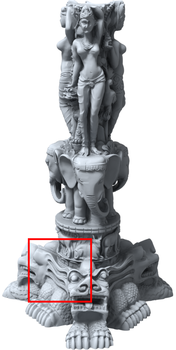}
\includegraphics[height=\thaiheight\textwidth]{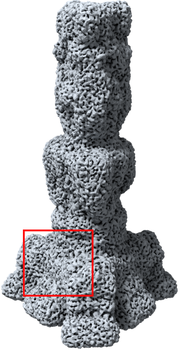}
\includegraphics[height=\thaiheight\textwidth]{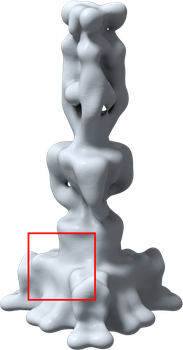}
\includegraphics[height=\thaiheight\textwidth]{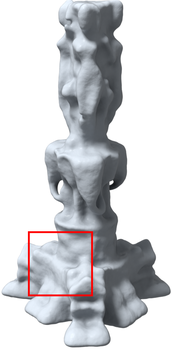}
\includegraphics[height=\thaiheight\textwidth]{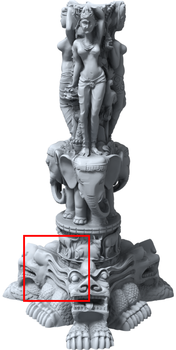}\\
\includegraphics[width=\finalpagesizethaicloseup\textwidth]{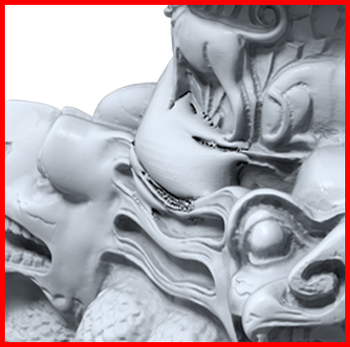}
\includegraphics[width=\finalpagesizethaicloseup\textwidth]{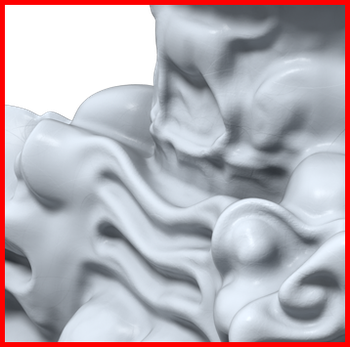}
\includegraphics[width=\finalpagesizethaicloseup\textwidth]{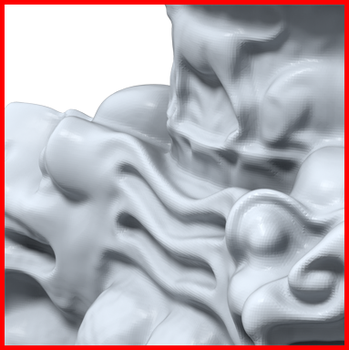}
\includegraphics[width=\finalpagesizethaicloseup\textwidth]{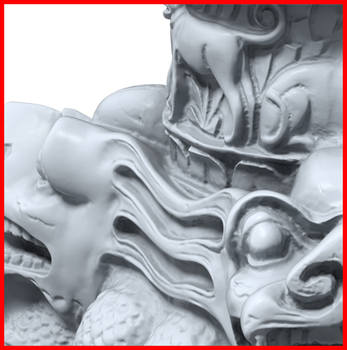}
\includegraphics[width=\finalpagesizethaicloseup\textwidth]{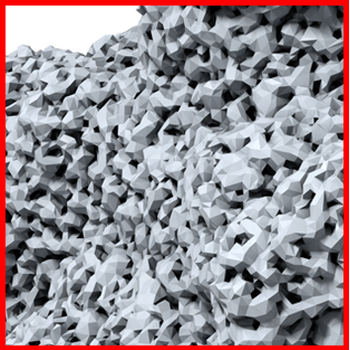}
\includegraphics[width=\finalpagesizethaicloseup\textwidth]{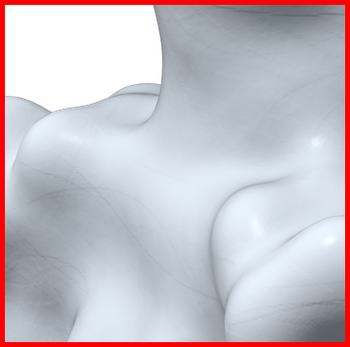}
\includegraphics[width=\finalpagesizethaicloseup\textwidth]{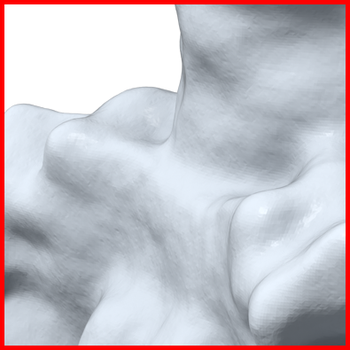}
\includegraphics[width=\finalpagesizethaicloseup\textwidth]{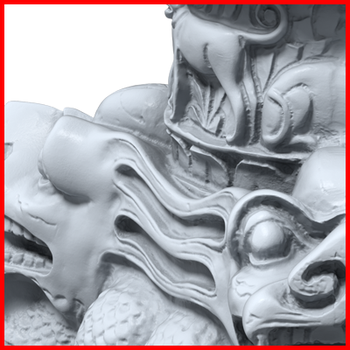}\\
\makebox[\finalpagesize\textwidth]{\scriptsize   11:48:38, 0.00120}
\makebox[\finalpagesize\textwidth]{\scriptsize   00:28:10, 0.00171}
\makebox[\finalpagesize\textwidth]{\scriptsize   00:05:53, 0.00164}
\makebox[\finalpagesize\textwidth]{\scriptsize   00:19:17, 0.00122}
\makebox[\finalpagesize\textwidth]{\scriptsize   04:01:01, 0.01121}
\makebox[\finalpagesize\textwidth]{\scriptsize   00:28:35, 0.00862}
\makebox[\finalpagesize\textwidth]{\scriptsize   00:03:51, 0.00579}
\makebox[\finalpagesize\textwidth]{\scriptsize   }\\
\makebox[\finalpagesize\textwidth]{\scriptsize   49332 MB}
\makebox[\finalpagesize\textwidth]{\scriptsize   75431 \& 10644 MB}
\makebox[\finalpagesize\textwidth]{\scriptsize   42151 MB}
\makebox[\finalpagesize\textwidth]{\scriptsize   42151 \& 17677 MB}
\makebox[\finalpagesize\textwidth]{\scriptsize   15933 MB}
\makebox[\finalpagesize\textwidth]{\scriptsize   30463 \& 10119 MB}
\makebox[\finalpagesize\textwidth]{\scriptsize   21048 MB}
\makebox[\finalpagesize\textwidth]{\scriptsize   }\\
\includegraphics[height=\alienheight\textwidth]{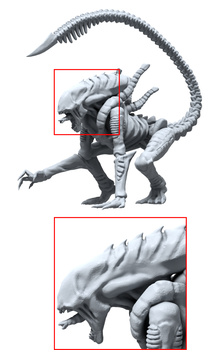}
\includegraphics[height=\alienheight\textwidth]{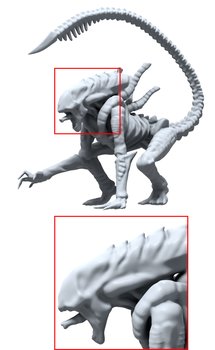}
\includegraphics[height=\alienheight\textwidth]{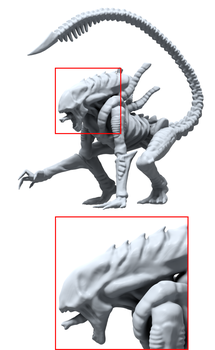}
\includegraphics[height=\alienheight\textwidth]{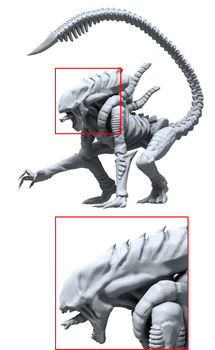}
\includegraphics[height=\alienheight\textwidth]{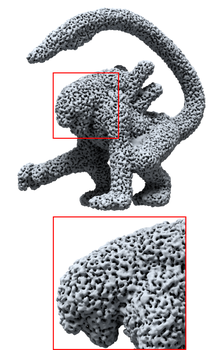}
\includegraphics[height=\alienheight\textwidth]{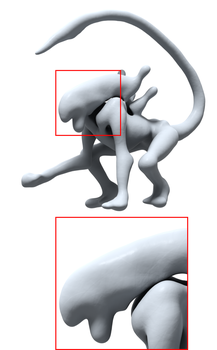}
\includegraphics[height=\alienheight\textwidth]{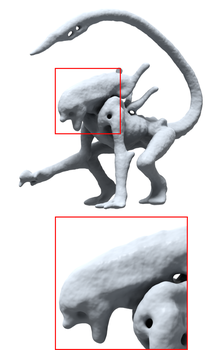}
\includegraphics[height=\alienheight\textwidth]{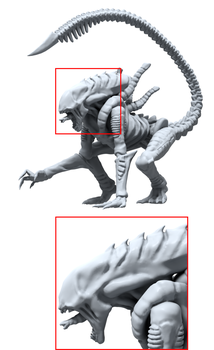}\\
\makebox[\finalpagesize\textwidth]{\scriptsize   06:41:41, 0.00115}
\makebox[\finalpagesize\textwidth]{\scriptsize   00:24:05, 0.00155}
\makebox[\finalpagesize\textwidth]{\scriptsize   00:04:53, 0.00145}
\makebox[\finalpagesize\textwidth]{\scriptsize   00:15:06, 0.00117}
\makebox[\finalpagesize\textwidth]{\scriptsize   03:00:11, 0.01247}
\makebox[\finalpagesize\textwidth]{\scriptsize   00:22:16, 0.00920}
\makebox[\finalpagesize\textwidth]{\scriptsize   00:04:23, 0.00566}
\makebox[\finalpagesize\textwidth]{\scriptsize   }\\
\makebox[\finalpagesize\textwidth]{\scriptsize   43047 MB}
\makebox[\finalpagesize\textwidth]{\scriptsize   68345 \& 8496 MB}
\makebox[\finalpagesize\textwidth]{\scriptsize   34325 MB}
\makebox[\finalpagesize\textwidth]{\scriptsize   34325 \& 16413 MB}
\makebox[\finalpagesize\textwidth]{\scriptsize   16722 MB}
\makebox[\finalpagesize\textwidth]{\scriptsize   27815 \& 8087 MB}
\makebox[\finalpagesize\textwidth]{\scriptsize   19408 MB}
\makebox[\finalpagesize\textwidth]{\scriptsize   }\\
\includegraphics[height=\lucyheight\textwidth]{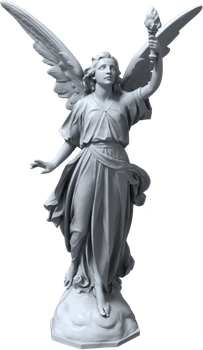}
\includegraphics[height=\lucyheight\textwidth]{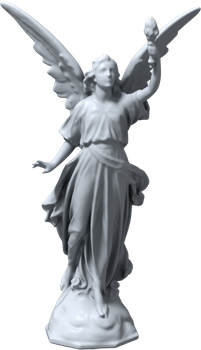}
\includegraphics[height=\lucyheight\textwidth]{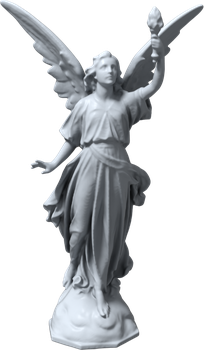}
\includegraphics[height=\lucyheight\textwidth]{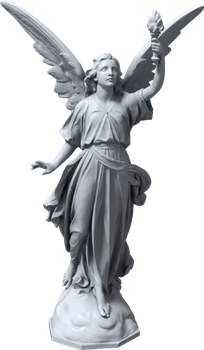}
\includegraphics[height=\lucyheight\textwidth]{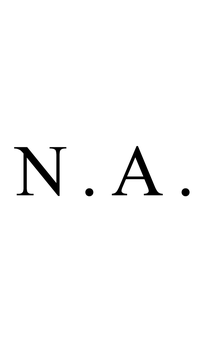}
\includegraphics[height=\lucyheight\textwidth]{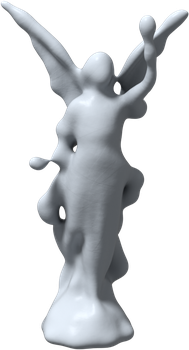}
\includegraphics[height=\lucyheight\textwidth]{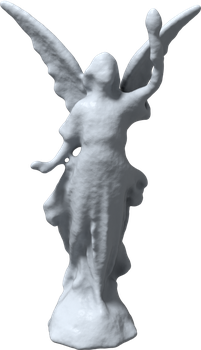}
\includegraphics[height=\lucyheight\textwidth]{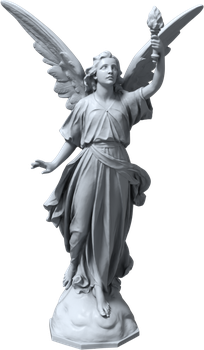}\\
\makebox[\finalpagesize\textwidth]{\scriptsize   04:24:47, 0.00100}
\makebox[\finalpagesize\textwidth]{\scriptsize   00:22:04, 0.00138}
\makebox[\finalpagesize\textwidth]{\scriptsize   00:04:13, 0.00172}
\makebox[\finalpagesize\textwidth]{\scriptsize   00:11:49, 0.00100}
\makebox[\finalpagesize\textwidth]{\scriptsize  }
\makebox[\finalpagesize\textwidth]{\scriptsize   00:20:36, 0.00828}
\makebox[\finalpagesize\textwidth]{\scriptsize   00:02:47, 0.00504}
\makebox[\finalpagesize\textwidth]{\scriptsize  }\\
\makebox[\finalpagesize\textwidth]{\scriptsize   33105 MB}
\makebox[\finalpagesize\textwidth]{\scriptsize   56263 \& 8522 MB}
\makebox[\finalpagesize\textwidth]{\scriptsize   30977 MB}
\makebox[\finalpagesize\textwidth]{\scriptsize   30977 \& 16432 MB}
\makebox[\finalpagesize\textwidth]{\scriptsize   }
\makebox[\finalpagesize\textwidth]{\scriptsize  22188 \& 8116 MB}
\makebox[\finalpagesize\textwidth]{\scriptsize  15057 MB}
\makebox[\finalpagesize\textwidth]{\scriptsize  }\\
\includegraphics[height=\statuteheight\textwidth]{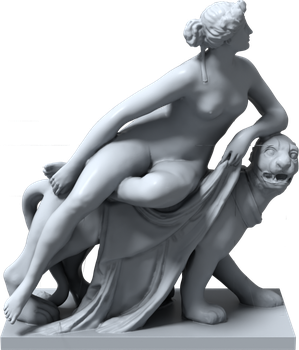}
\includegraphics[height=\statuteheight\textwidth]{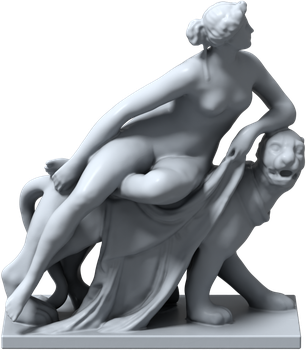}
\includegraphics[height=\statuteheight\textwidth]{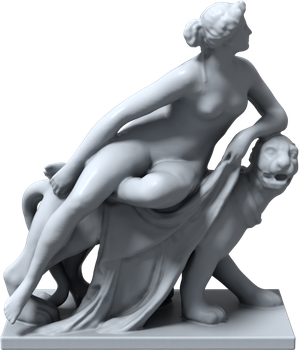}
\includegraphics[height=\statuteheight\textwidth]{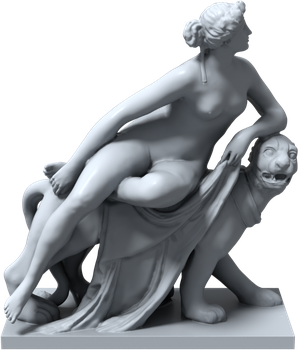}
\includegraphics[height=\statuteheight\textwidth]{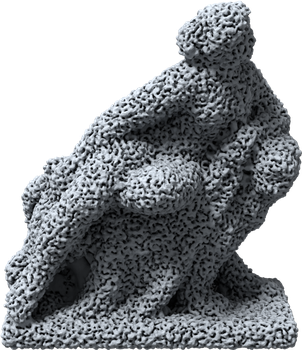}
\includegraphics[height=\statuteheight\textwidth]{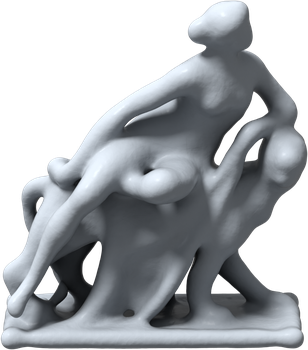}
\includegraphics[height=\statuteheight\textwidth]{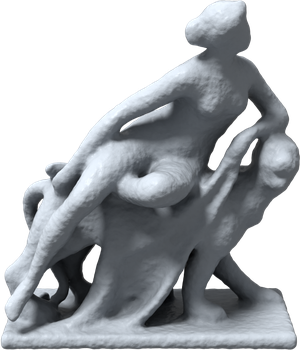}
\includegraphics[height=\statuteheight\textwidth]{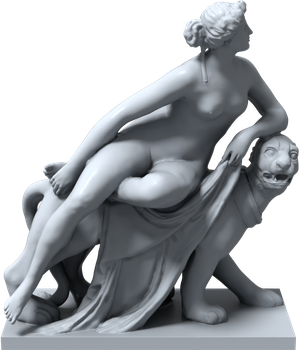}\\
\makebox[\finalpagesize\textwidth]{\scriptsize   05:28:04, 0.00163}
\makebox[\finalpagesize\textwidth]{\scriptsize   00:16:24, 0.00174}
\makebox[\finalpagesize\textwidth]{\scriptsize   00:03:44, 0.00174}
\makebox[\finalpagesize\textwidth]{\scriptsize   00:10:00, 0.00163}
\makebox[\finalpagesize\textwidth]{\scriptsize   12:10:48, 0.01261}
\makebox[\finalpagesize\textwidth]{\scriptsize   00:16:02, 0.00603}
\makebox[\finalpagesize\textwidth]{\scriptsize   00:04:29, 0.00472}
\makebox[\finalpagesize\textwidth]{\scriptsize   }\\
\makebox[\finalpagesize\textwidth]{\scriptsize   27324 MB}
\makebox[\finalpagesize\textwidth]{\scriptsize   63163 \& 6621 MB}
\makebox[\finalpagesize\textwidth]{\scriptsize   22134 MB}
\makebox[\finalpagesize\textwidth]{\scriptsize   22134 \& 13478 MB}
\makebox[\finalpagesize\textwidth]{\scriptsize   24012 MB}
\makebox[\finalpagesize\textwidth]{\scriptsize   37161 \& 6202 MB}
\makebox[\finalpagesize\textwidth]{\scriptsize   17405 MB}
\makebox[\finalpagesize\textwidth]{\scriptsize   }\\
\makebox[\finalpagesize\textwidth]{\scriptsize iPSR (0\% noise)}
 \makebox[\finalpagesize\textwidth]{\scriptsize \textcolor{black}{WNNC (0\% noise)}}
 \makebox[\finalpagesize\textwidth]{\scriptsize DWG (0\% noise)}
 \makebox[\finalpagesize\textwidth]{\scriptsize DWG+sPSR (0\% noise)}
 \makebox[\finalpagesize\textwidth]{\scriptsize iPSR (0.75\% noise)}
 \makebox[\finalpagesize\textwidth]{\scriptsize \textcolor{black}{WNNC (0.75\% noise)}}
 \makebox[\finalpagesize\textwidth]{\scriptsize DWG (0.75\% noise)}
 \makebox[\finalpagesize\textwidth]{\scriptsize GT }\\ 
 \vspace{-0.1in}
\caption{Comparison with iPSR and WNNC on large-scale models, under noise-free conditions and with 0.75\% noise (relative to the diagonal of the bounding box). The values displayed below each model indicate the runtime, peak memory consumption, and Chamfer distance. WNNC operates in two stages for point orientation and surface reconstruction and reports the peak memory usage separately for both GPUs and CPUs.}
\label{fig:largescalemodels}
\end{figure*}

\paragraph{Runtime Performance and Accuracy} \textcolor{black}{Table~\ref{tab:statistics} reports the runtime performance and accuracy on five small-scale models (up to 100K points) and five large-scale models (10 to 20 million points). Each model underwent testing under noise-free conditions and with a 0.75\% noise level. For small-scale models, DWG completes reconstructions in just one or two seconds, significantly outperforming all baselines in terms of speed. In scenarios involving large-scale models, where only iPSR and WNNC managed to produce results within a reasonable timeframe. Both iPSR and WNNC depend on sparse linear system solver: iPSR utilizes a finite-element method~\cite{kazhdan2013screened} for solving Poisson's equation, whereas WNNC employs a GPU-accelerated steepest decent solver, which significantly outpaces the solver used by iPSR. In contrast, DWG employs a straightforward approach that alternates between computing the screened GWN field and updating point normals using the field's gradient. This methodology avoids reliance on any complex numerical solver, which significantly enhances DWG's speed and stability compared to both iPSR and WNNC. Specifically, with the same octree depth, DWG is typically 30-120 times quicker than iPSR and 4-10 times quicker than WNNC on test models containing 10 to 20 million points. This speed advantage will become more pronounced on larger models and on more advanced GPUs with greater memory capacity. Furthermore, DWG consistently shows comparable or higher accuracy in reconstruction results across both small and large-scale models compared to all baseline methods. Visual results can be found in 
Figures~\ref{fig:largescalemodels} and~\ref{fig:thinstructure}. } 




\newcommand{\largefigsize}{0.12}
\newcommand{\smallboxsize}{0.85}

\newcommand{\thricesmallfigsize}{0.405}
\newcommand{\rheightsize}{0.4}
\newcommand{\vspacesize}{-2}
\newcommand{\hspacesize}{-0.5}

\begin{figure*}[htbp]
\centering
\newcommand{\smallfigsize}{0.0475}
\setlength\tabcolsep{0.25pt}
\begin{tabular}{cc|cccccc|cccccc|cccccc}
\multicolumn{2}{c|}{} & 
\multicolumn{6}{c|}{iPSR $(d,\lambda)$ } & 
\multicolumn{6}{c|}{DWG $(d,\lambda)$} &
\multicolumn{6}{c}{WNNC $(ws_{\min},ws_{\max})$}\\
\includegraphics[width=\smallfigsize\textwidth]{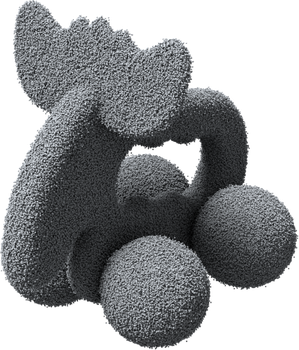} & 
\includegraphics[width=\smallfigsize\textwidth]{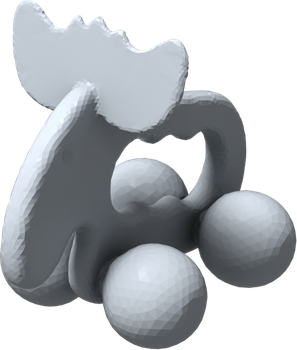} &
\includegraphics[width=\smallfigsize\textwidth]{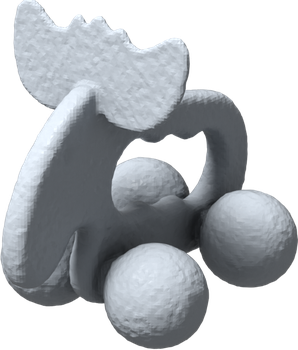} &
\includegraphics[width=\smallfigsize\textwidth]{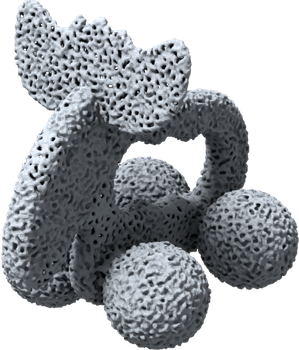} &
\includegraphics[width=\smallfigsize\textwidth]{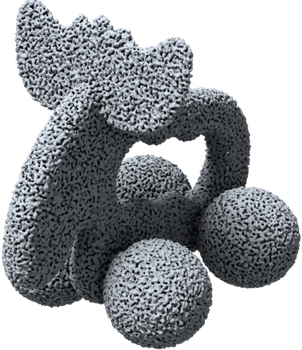} &
\includegraphics[width=\smallfigsize\textwidth]{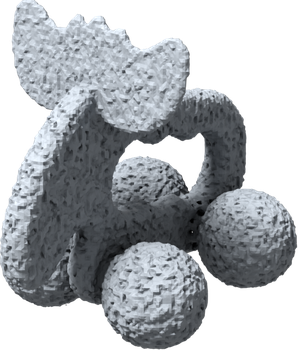} & 
\includegraphics[width=\smallfigsize\textwidth]{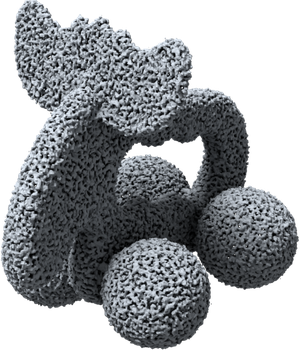} &
\includegraphics[width=\smallfigsize\textwidth]{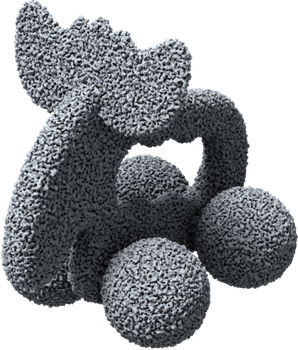} & 
\includegraphics[width=\smallfigsize\textwidth] {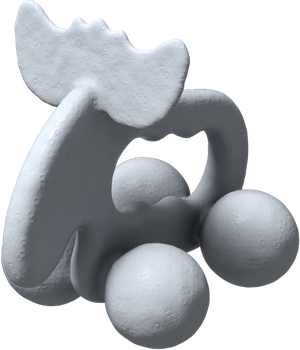} & 
\includegraphics[width=\smallfigsize\textwidth]{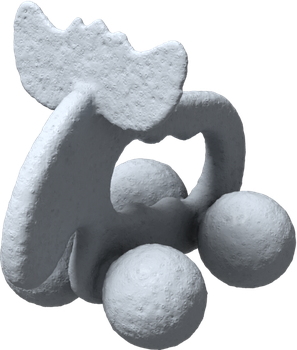} & 
\includegraphics[width=\smallfigsize\textwidth]{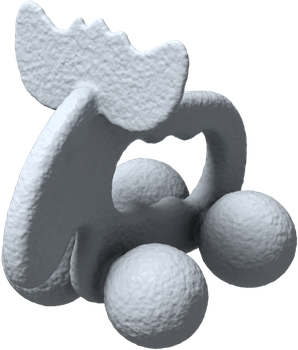} &
\includegraphics[width=\smallfigsize\textwidth]{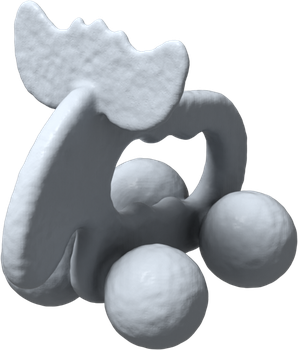} &
\includegraphics[width=\smallfigsize\textwidth]{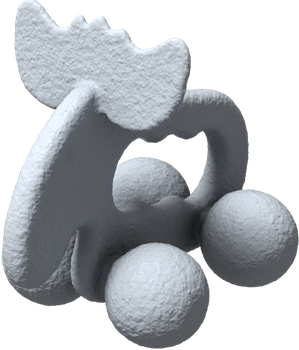} &
\includegraphics[width=\smallfigsize\textwidth]{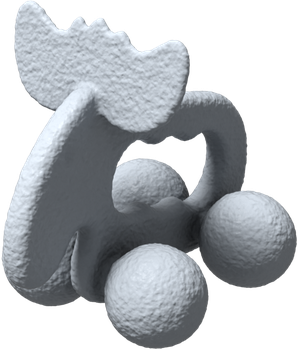} &
\includegraphics[width=\smallfigsize\textwidth]{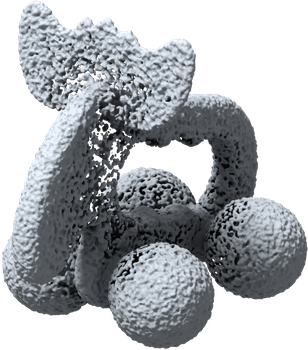}& 
\includegraphics[width=\smallfigsize\textwidth]{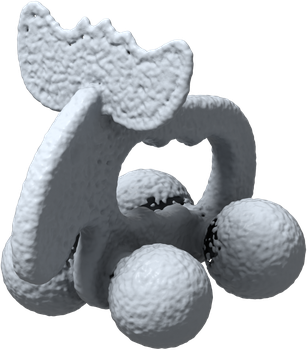}& \includegraphics[width=\smallfigsize\textwidth]{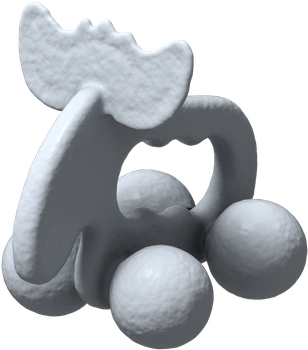}& \includegraphics[width=\smallfigsize\textwidth]{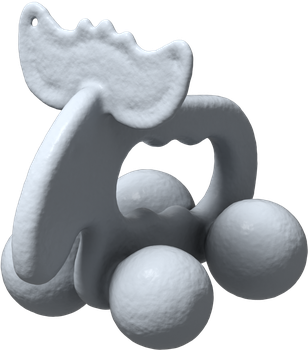}& 
\includegraphics[width=\smallfigsize\textwidth]{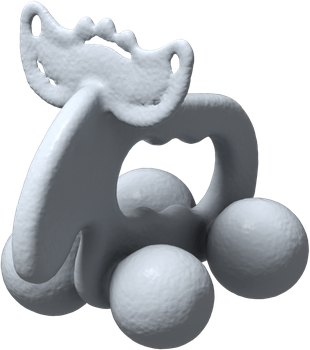}& 
\includegraphics[width=\smallfigsize\textwidth]{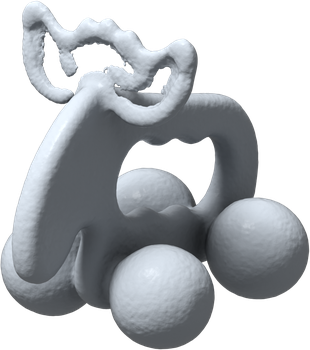}\\
\includegraphics[width=\smallfigsize\textwidth]{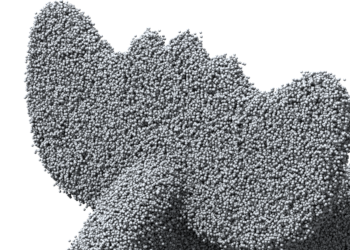} & 
\includegraphics[width=\smallfigsize\textwidth]{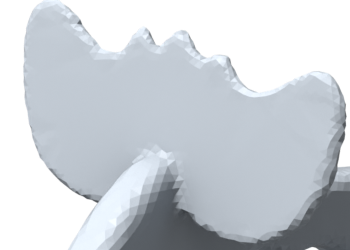} &
\includegraphics[width=\smallfigsize\textwidth]{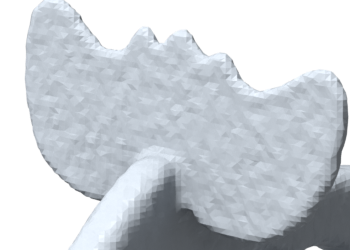} &
\includegraphics[width=\smallfigsize\textwidth]{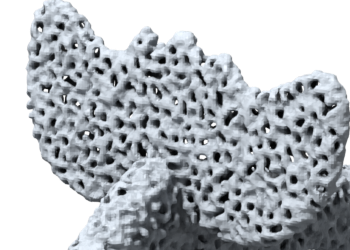} &
\includegraphics[width=\smallfigsize\textwidth]{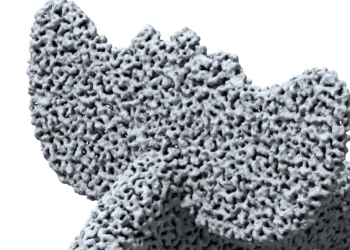} &
\includegraphics[width=\smallfigsize\textwidth]{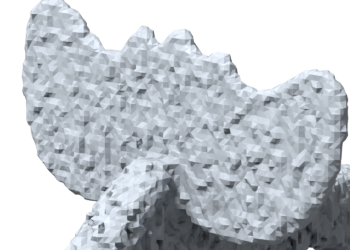} & 
\includegraphics[width=\smallfigsize\textwidth]{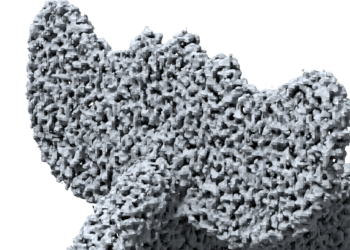} &
\includegraphics[width=\smallfigsize\textwidth]{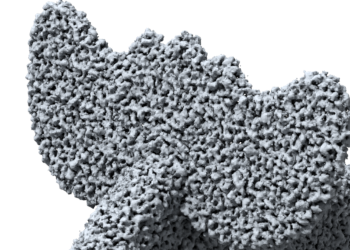} & 
\includegraphics[width=\smallfigsize\textwidth] {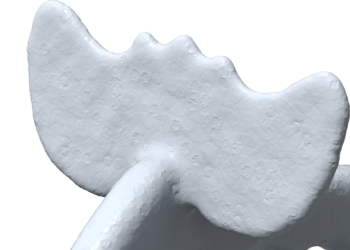} & 
\includegraphics[width=\smallfigsize\textwidth]{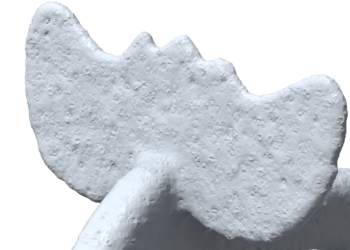} & 
\includegraphics[width=\smallfigsize\textwidth]{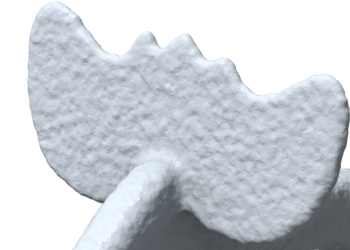} &
\includegraphics[width=\smallfigsize\textwidth]{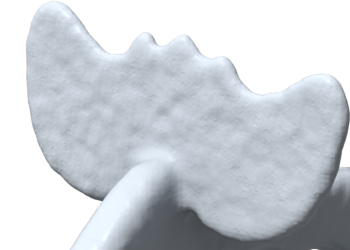} &
\includegraphics[width=\smallfigsize\textwidth]{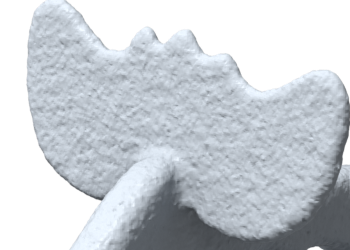} &
\includegraphics[width=\smallfigsize\textwidth]{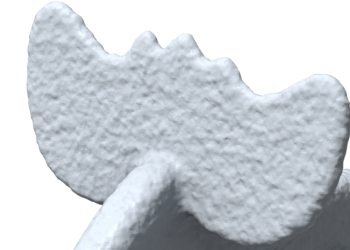} &
\includegraphics[width=\smallfigsize\textwidth]{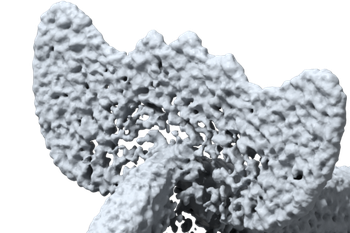}& 
\includegraphics[width=\smallfigsize\textwidth]{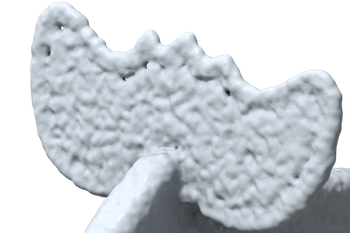}& \includegraphics[width=\smallfigsize\textwidth]{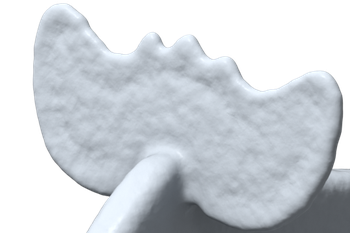}& \includegraphics[width=\smallfigsize\textwidth]{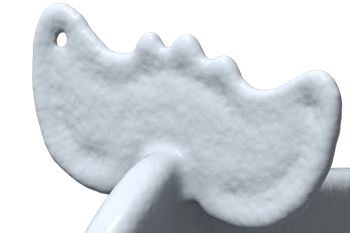}& 
\includegraphics[width=\smallfigsize\textwidth]{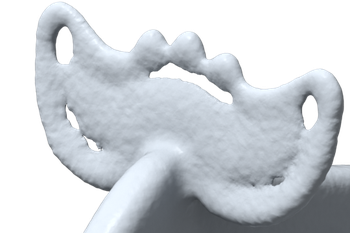}& 
\includegraphics[width=\smallfigsize\textwidth]{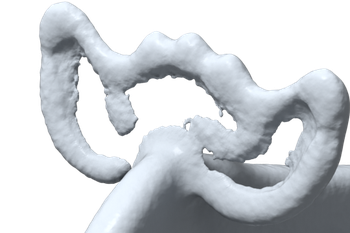}\\
{\scriptsize Input} & 
{\scriptsize GT} & 
{\scriptsize $(7, 10$)} & 
{\scriptsize $(8, 10$)} & 
{\scriptsize $(9, 10$)} & 
{\scriptsize $(7, 100)$} & 
{\scriptsize $(8, 100)$} & 
{\scriptsize $(9, 100)$} & 
{\scriptsize $(7, 10$)} & 
{\scriptsize $(8, 10$)} & 
{\scriptsize $(9, 10$)} & 
{\scriptsize $(7, 100)$} & 
{\scriptsize $(8, 100)$} & 
{\scriptsize $(9, 100)$} & {\scriptsize $\#1$} & {\scriptsize $\#2$} & {\scriptsize $\#3$} & {\scriptsize $\#4$} & {\scriptsize $\#5$} & {\scriptsize $\#6$}\\
\includegraphics[width=\smallfigsize\textwidth]{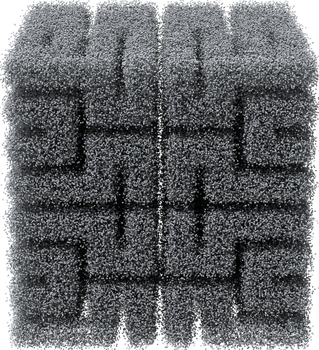} & 
\includegraphics[width=\smallfigsize\textwidth]{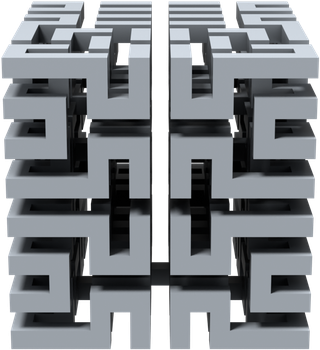} &
\includegraphics[width=\smallfigsize\textwidth]{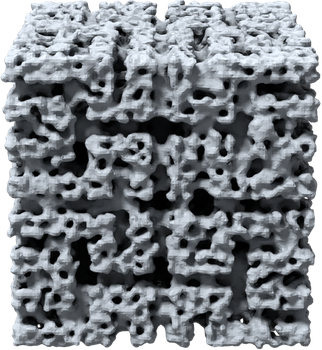} &
\includegraphics[width=\smallfigsize\textwidth]{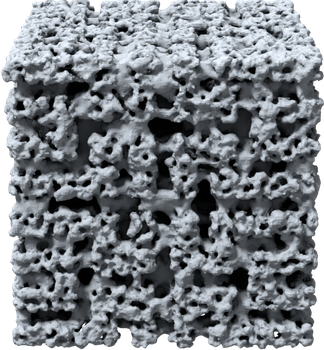} &
\includegraphics[width=\smallfigsize\textwidth]{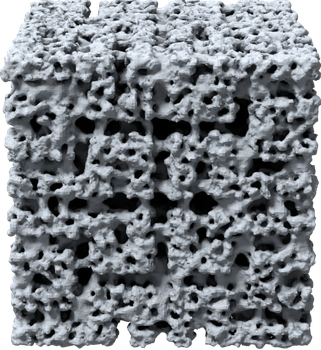} &
\includegraphics[width=\smallfigsize\textwidth]{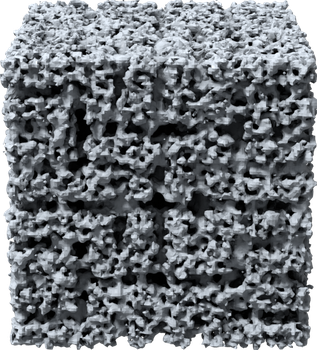} &
\includegraphics[width=\smallfigsize\textwidth]{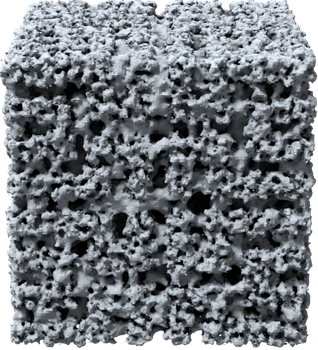} &
\includegraphics[width=\smallfigsize\textwidth]{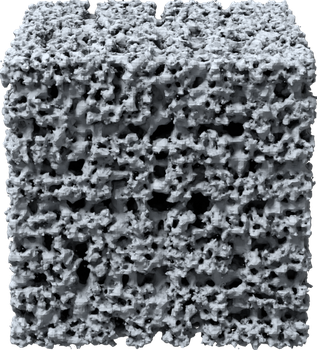} &
\includegraphics[width=\smallfigsize\textwidth]{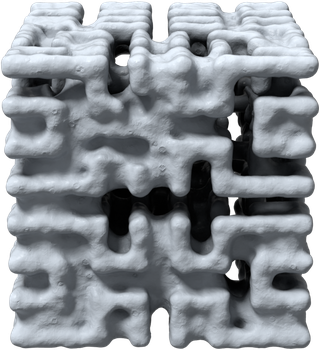} &
\includegraphics[width=\smallfigsize\textwidth]{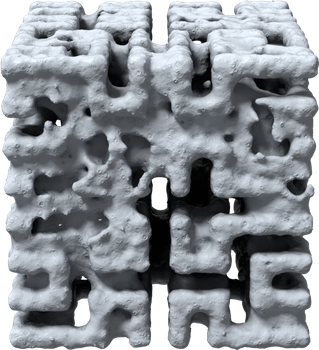} &
\includegraphics[width=\smallfigsize\textwidth]{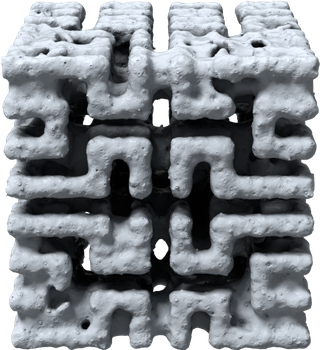} &
\includegraphics[width=\smallfigsize\textwidth]{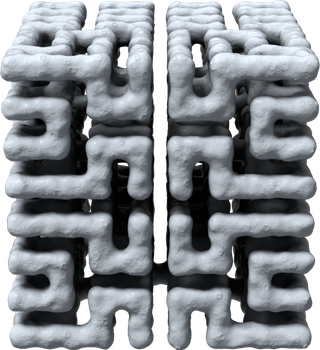} &
\includegraphics[width=\smallfigsize\textwidth]{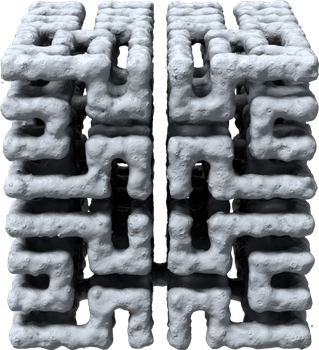} &
\includegraphics[width=\smallfigsize\textwidth]{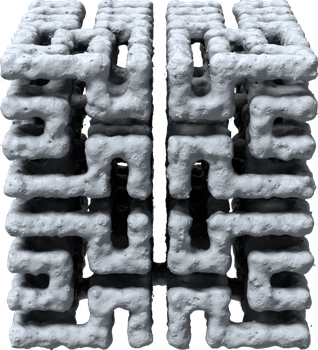} & \includegraphics[width=\smallfigsize\textwidth]{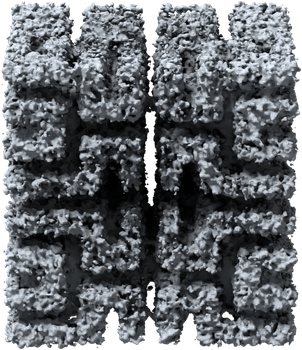}& \includegraphics[width=\smallfigsize\textwidth]{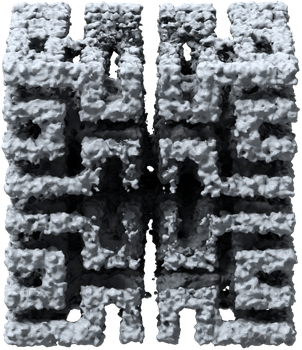}& \includegraphics[width=\smallfigsize\textwidth]{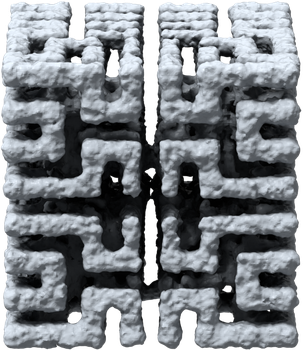}& \includegraphics[width=\smallfigsize\textwidth]{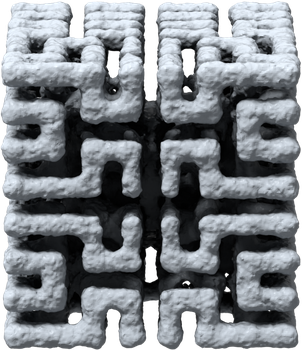}& \includegraphics[width=\smallfigsize\textwidth]{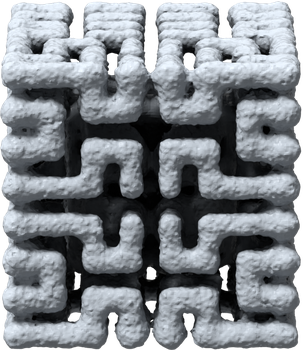}&
\includegraphics[width=\smallfigsize\textwidth]{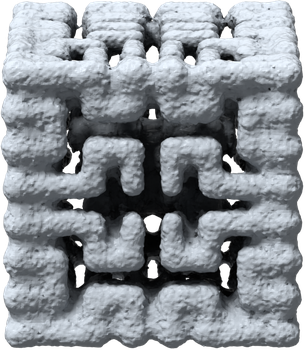}\\
\includegraphics[width=\smallfigsize\textwidth]{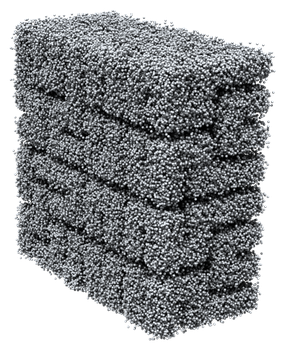} & 
\includegraphics[width=\smallfigsize\textwidth]{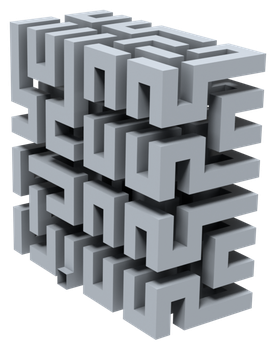} &
\includegraphics[width=\smallfigsize\textwidth]{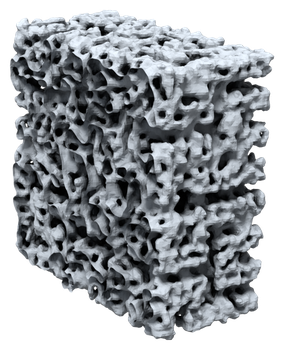} &
\includegraphics[width=\smallfigsize\textwidth]{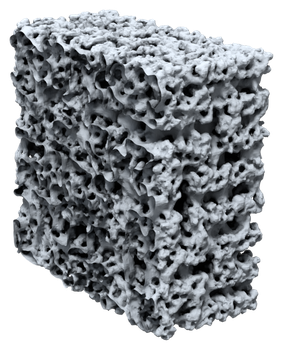} &
\includegraphics[width=\smallfigsize\textwidth]{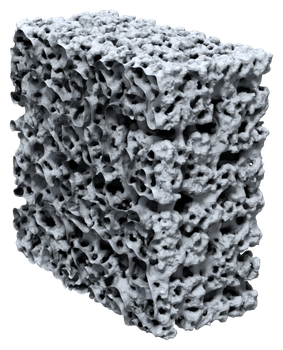} &
\includegraphics[width=\smallfigsize\textwidth]{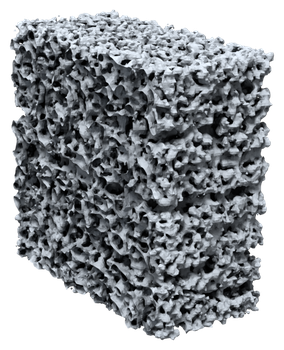} &
\includegraphics[width=\smallfigsize\textwidth]{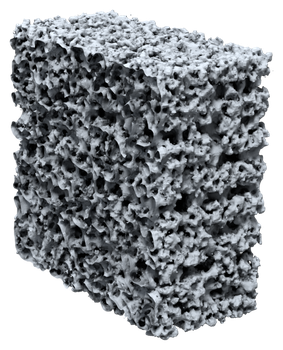} &
\includegraphics[width=\smallfigsize\textwidth]{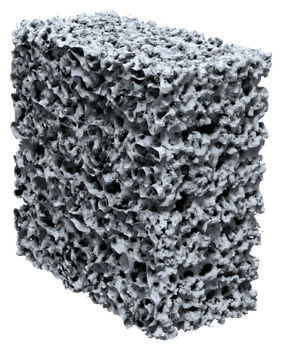} &
\includegraphics[width=\smallfigsize\textwidth]{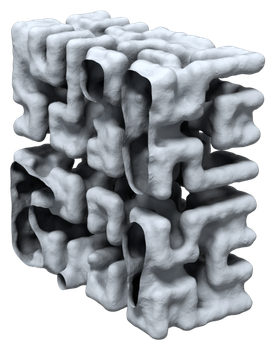} &
\includegraphics[width=\smallfigsize\textwidth]{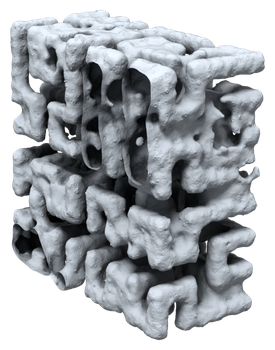} &
\includegraphics[width=\smallfigsize\textwidth]{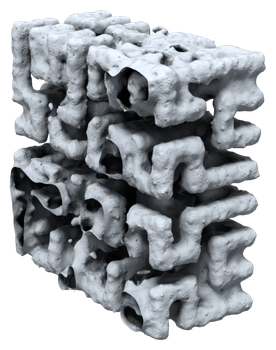} &
\includegraphics[width=\smallfigsize\textwidth]{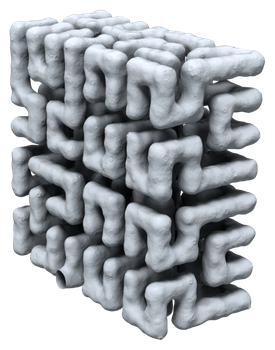} &
\includegraphics[width=\smallfigsize\textwidth]{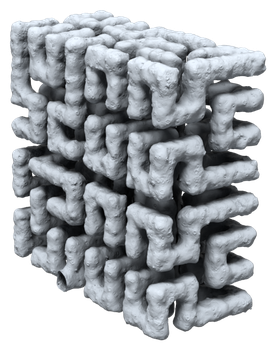} &
\includegraphics[width=\smallfigsize\textwidth]{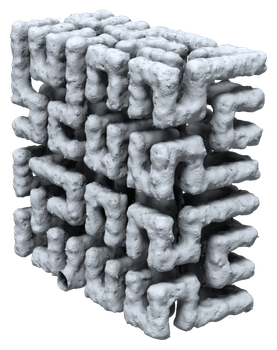} & \includegraphics[width=\smallfigsize\textwidth]{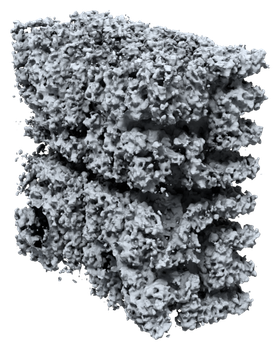}&
\includegraphics[width=\smallfigsize\textwidth]{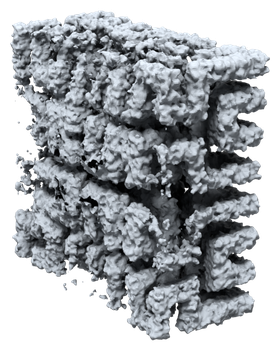}& \includegraphics[width=\smallfigsize\textwidth]{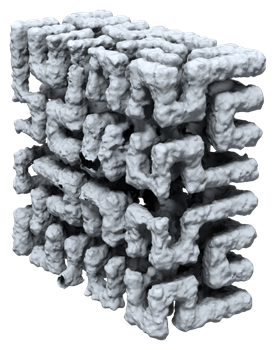}& \includegraphics[width=\smallfigsize\textwidth]{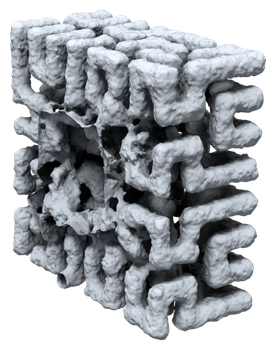}& \includegraphics[width=\smallfigsize\textwidth]{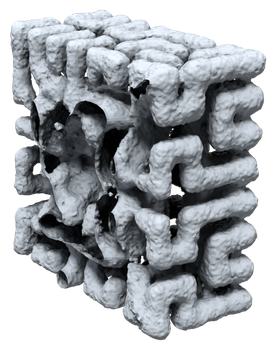}&
\includegraphics[width=\smallfigsize\textwidth]{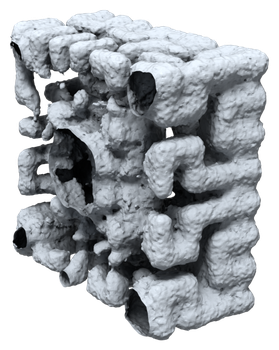}\\
{\scriptsize Input} & 
{\scriptsize GT} & 
{\scriptsize $(7, 10)$} & 
{\scriptsize $(8, 10$} & 
{\scriptsize $(9, 10$} & 
{\scriptsize $(7, 400)$} & 
{\scriptsize $(8, 400)$} & 
{\scriptsize $(9, 400)$} & 
{\scriptsize $(7, 10)$} & 
{\scriptsize $(8, 10)$} & 
{\scriptsize $(9, 10)$} & 
{\scriptsize $(7, 400)$} & 
{\scriptsize $(8, 400)$} & 
{\scriptsize $(9, 400)$} & {\scriptsize $\#1$} & {\scriptsize $\#2$} & {\scriptsize $\#3$} & {\scriptsize $\#4$} & {\scriptsize $\#5$} & {\scriptsize $\#6$}\\
\end{tabular}
\caption{\textcolor{black}{Robustness comparison of DWG, iPSR and WNNC on the Elk ($n=1400$K) and Hilbert Cube ($n=730$K) models, both tested at a noise level of 0.75\%. DWG and iPSR employ varying octree depths $d$ and screening coefficients $\lambda$ to combat noise, with shallower octrees and higher screening coefficients generally improving robustness. WNNC addresses noise by setting a distance threshold $ws$ that modifies the Poisson kernels. For any query point $\mathbf{q}$ and input point $\mathbf{p}_i$, if their distance is less than or equal to $ws$, it is adjusted to $ws$. This threshold $ws$ is progressively decreased from $ws_{\max}$ to $ws_{\min}$ during the iterative procedure. WNNC provides six configurations (\#1 to \#6) for $ws_{\min}$ and $ws_{\max}$ to handle different noise levels, ranging from low to high: $(0.002,0.016)$, $(0.01, 0.04)$, $(0.02, 0.08)$, $(0.03, 0.12)$, $(0.04,0.16)$ and $(0.05,0.2)$. iPSR struggles with high levels of noise, often producing outputs with an unnecessarily high genus across various parameter settings. Meanwhile, while WNNC successfully reconstructs the Elk model under parameter settings for moderate noise levels, it fails to handle the Hilbert Cube model under any configuration. The complex thin structures of this model, which are revealed in the cut views, pose significant challenges to both iPSR and WNNC. In contrast, DWG consistently outperforms both iPSR and WNNC, demonstrating superior robustness to noise and an enhanced ability to handle models with thin structures, coupled with more consistent parameter settings. }}
\label{fig:robustness}
\end{figure*}

\begin{figure*}[t] \centering
    \newcommand{\ninesizel}{0.095}\includegraphics[width=\ninesizel\linewidth]{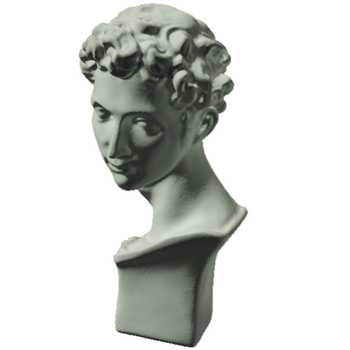} 
    \includegraphics[width=\ninesizel\linewidth]{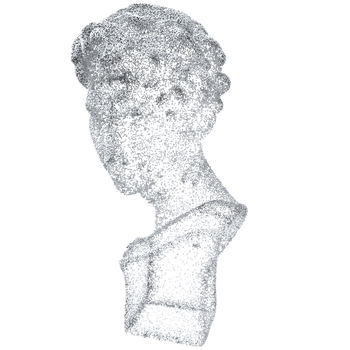}
    \includegraphics[width=\ninesizel\linewidth]{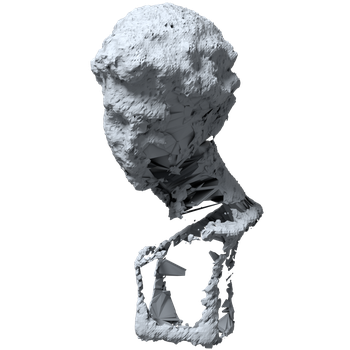} 
    \includegraphics[width=\ninesizel\linewidth]{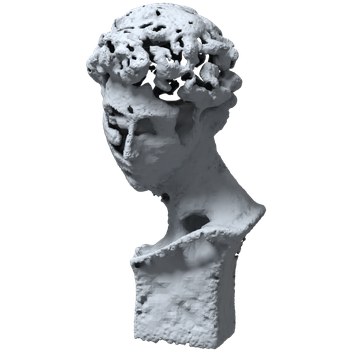} 
    \includegraphics[width=\ninesizel\linewidth]{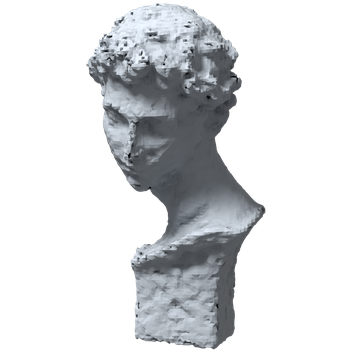} 
    \includegraphics[width=\ninesizel\linewidth]{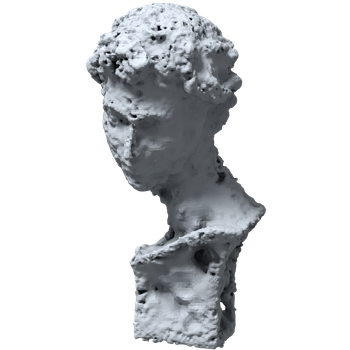}
    \includegraphics[width=\ninesizel\linewidth]{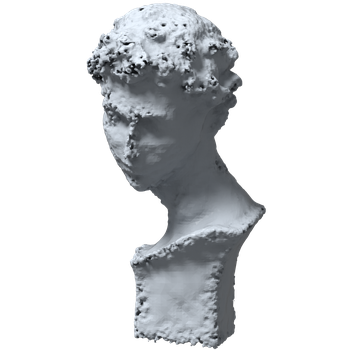} 
    \includegraphics[width=\ninesizel\linewidth]{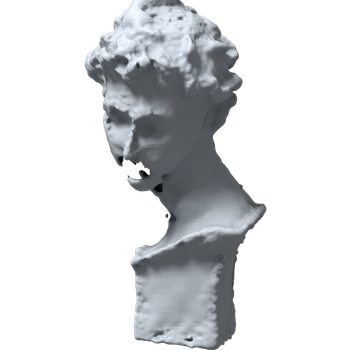} 
    \includegraphics[width=\ninesizel\linewidth]{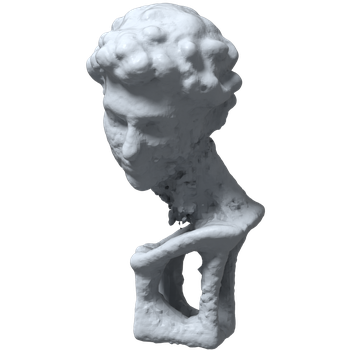} 
    \includegraphics[width=\ninesizel\linewidth]{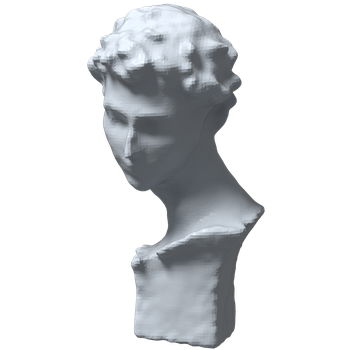} 
    \\
    \includegraphics[width=\ninesizel\linewidth]{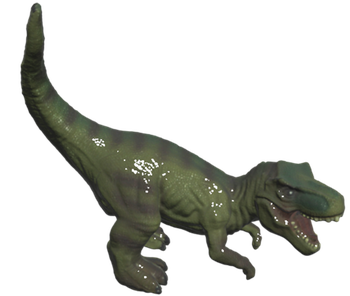} 
    \includegraphics[width=\ninesizel\linewidth]{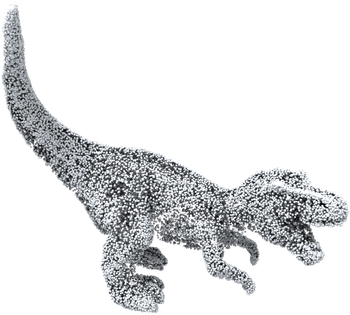}
    \includegraphics[width=\ninesizel\linewidth]{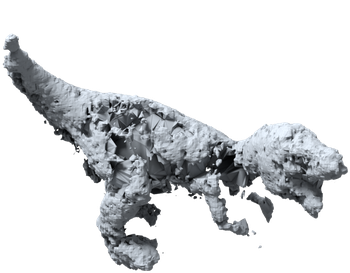} 
    \includegraphics[width=\ninesizel\linewidth]{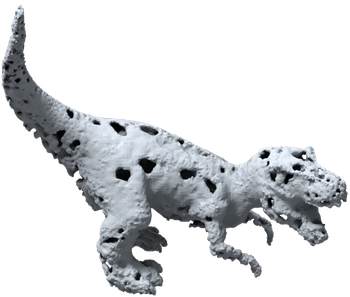} 
    \includegraphics[width=\ninesizel\linewidth]{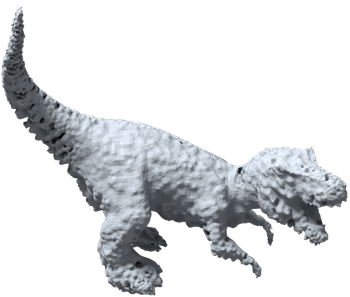} 
    \includegraphics[width=\ninesizel\linewidth]{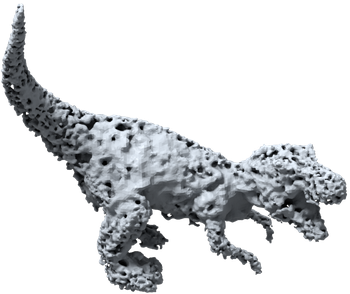}
    \includegraphics[width=\ninesizel\linewidth]{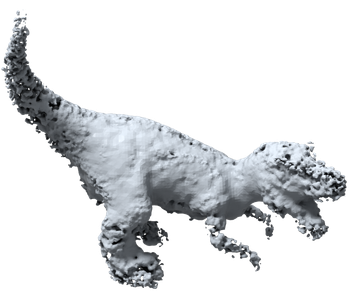} 
    \includegraphics[width=\ninesizel\linewidth]{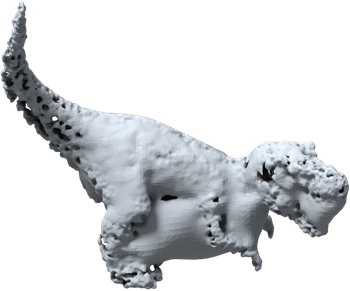} 
    \includegraphics[width=\ninesizel\linewidth]{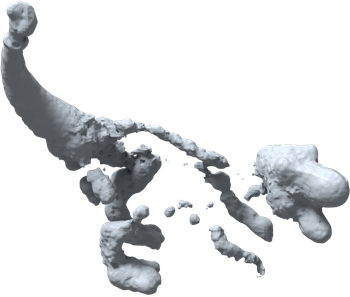} 
    \includegraphics[width=\ninesizel\linewidth]{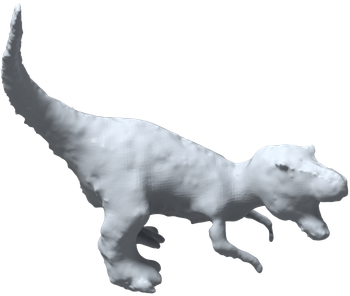} 
    \\
    \includegraphics[width=\ninesizel\linewidth]{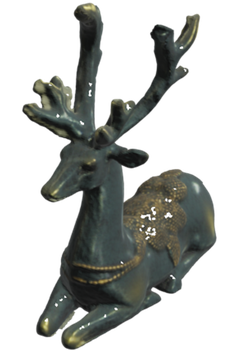} 
    \includegraphics[width=\ninesizel\linewidth]{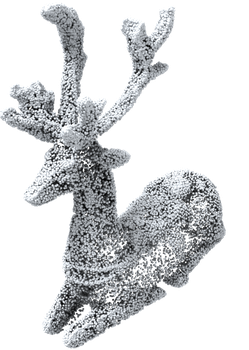}
    \includegraphics[width=\ninesizel\linewidth]{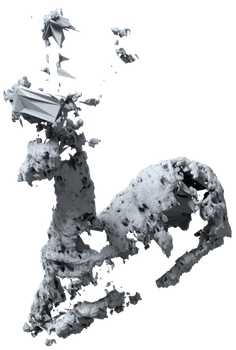} 
    \includegraphics[width=\ninesizel\linewidth]{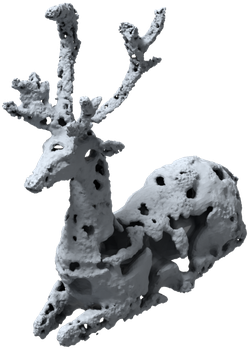} 
    \includegraphics[width=\ninesizel\linewidth]{figures/NA.png} 
    \includegraphics[width=\ninesizel\linewidth]{figures/NA.png}
    \includegraphics[width=\ninesizel\linewidth]{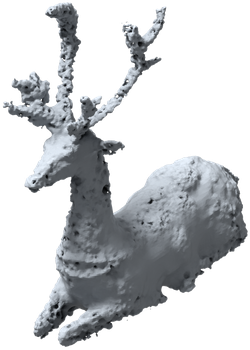} 
    \includegraphics[width=\ninesizel\linewidth]{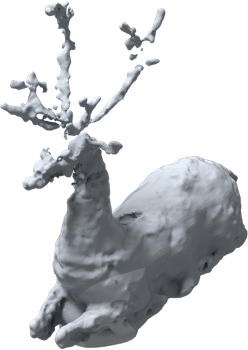} 
    \includegraphics[width=\ninesizel\linewidth]{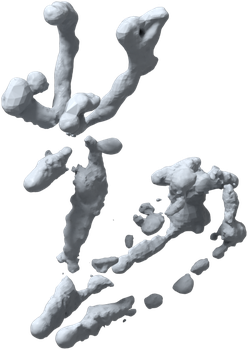} 
    \includegraphics[width=\ninesizel\linewidth]{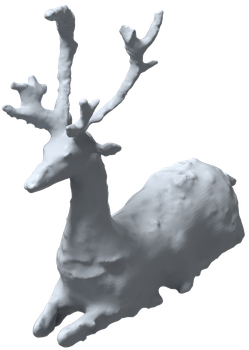} 
    \\
    \makebox[\ninesizel\linewidth]{ \scriptsize Reference image}
    \makebox[\ninesizel\linewidth]{\scriptsize 3D Gaussians}
    \makebox[\ninesizel\linewidth]{\scriptsize Mullen et al.}
    \makebox[\ninesizel\linewidth]{\scriptsize iPSR}
    \makebox[\ninesizel\linewidth]{\scriptsize PGR}
    \makebox[\ninesizel\linewidth]{\scriptsize GCNO}
    \makebox[\ninesizel\linewidth]{\scriptsize BIM}
    \makebox[\ninesizel\linewidth]{\scriptsize SNO}
    \makebox[\ninesizel\linewidth]{\scriptsize WNNC}
    \makebox[\ninesizel\linewidth]{\scriptsize Ours}
    \label{fig:3DGS}
    \caption{\textcolor{black}{Applying DWG and other baseline methods to point clouds generated through 3D Gaussian Splatting from multi-view images. These point clouds typically exhibit noise, outliers and non-uniform sampling. DWG consistently shows superior robustness to these issues compared to the baseline methods, offering more reliable and accurate reconstructions. }} 
\end{figure*}

\paragraph{Thin Structures} Models featuring high-genus and thin structures are notoriously difficult for 3D reconstruction, especially when the orientation information necessary to accurately separate spatially close faces is lacking. However, DWG handles these complexities more effectively than existing methods. This superior capability is primarily attributed to the inclusion of the screening coefficient $\lambda$, which acts as a decay factor. It ensures that the $\overline{w}^{(t)}_\lambda$ level set of the GWN field closely adheres to the underlying point cloud throughout the diffusion process. This close adherence is particularly beneficial for accurately separating thin structures, thereby enhancing the fidelity of the reconstructed surfaces (see Figure~\ref{fig:robustness}). 

\paragraph{Real Scans} \textcolor{black}{Reconstructing real scanned point cloud data presents challenges due to noise, registration errors, and overlaps between scans. We evaluated real-scanned models from the Stanford 3D Scanning Repository, utilizing registered range images transformed into point clouds. }\textcolor{black}{Experimental results show that registration misalignment often yields large errors in~\citep{kobbelt}'s method. \cite{mullen2010signing}'s method, while capable of reconstructing the overall shape of objects, fails to capture finer, thinner features such as the toes of the Armadillo, the feet and horns of the Dragon, and the hands of the Happy Buddha. Misalignments and overlaps between scans also cause iPSR to produce double-layered surfaces in the reconstructed object. The other baseline methods, due to their constraints on the input model size, cannot be applied to these real scans. DWG demonstrates improved accuracy in surface reconstruction from the real scan data. See Figure~\ref{fig:scan} for the visual results.}

\textcolor{black}{ \paragraph{Combining with 3DGS} We tested DWG and other baseline methods on point clouds generated by the widely-used 3DGS technique~\cite{kerbl3Dgaussians, Huang2DGS2024} for multi-view 3D reconstruction. These point clouds often display non-uniform distributions, where areas rich in texture and geometric details are densely sampled, whereas flat regions with less texture are sparsely sampled. By employing non-uniform coefficients $a_i$, DWG demonstrates enhanced robustness and superior reconstruction accuracy compared to baseline methods.}

\begin{figure*}[!htbp]
    \centering
\includegraphics[width=0.95in]{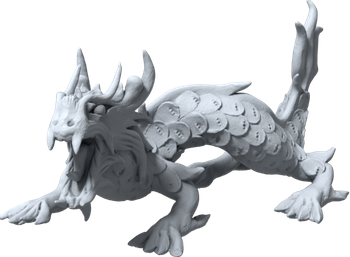}
    \includegraphics[width=1.350in]{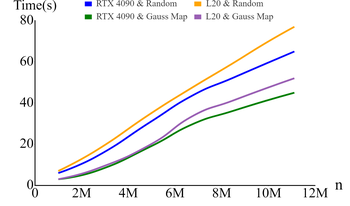}
    \includegraphics[width=1.250in]{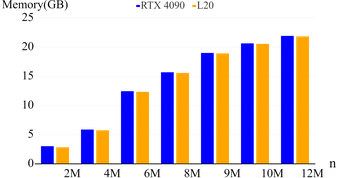} 
     \includegraphics[width=0.35in]{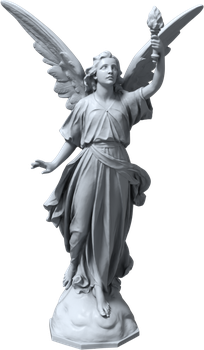}
    \includegraphics[width=1.350in]{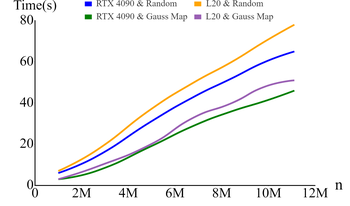}
    \includegraphics[width=1.250in]{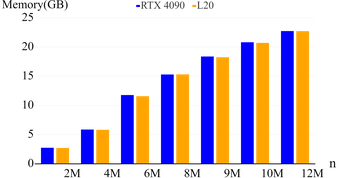} 
    \\
     \makebox[0.95in]{}
     \makebox[1.350in]{\scriptsize Time}
     \makebox[1.250in]{\scriptsize Memory}
     \makebox[0.35in]{}
     \makebox[1.350in]{\scriptsize Time}
     \makebox[1.250in]{\scriptsize Memory}
    \vspace{-0.1in}
    \caption{Parallelization and scalability assessment. The horizontal axis represents the number of input points ($n$), while the vertical axes show the peak memory consumption (in gigabytes) and the running time (in seconds) for the parallel diffusion stage. \textcolor{black}{Despite the theoretical time complexity of $O(In\log n)$, the practical runtime exhibits empirical linear complexity. This efficiency arises from DWG's straightforward, numerical-package-free methodology and the extensive parallelization enabled by a large number of GPU cores. }\textcolor{black}{Both random and Gauss map initializations are tested, with the Gauss map typically reducing runtime by 30\%-60\% compared to random initialization.}}   \label{fig:parallelization}
\end{figure*}

\paragraph{Parallelization and Scalability} DWG exhibits linear space complexity of $O(n)$ and a time complexity of $O(n\log n)$ per iteration. Computational results consistently show that DWG converges within 10-50 iterations from a random initialization. These space and time complexities make DWG superior to other methods, enabling it to handle large-scale models efficiently. To demonstrate this, we evaluated DWG's runtime performance on an NVIDIA RTX 4090 GPU (with over 16K CUDA cores and 24 GB of memory) and an NVIDIA L20 GPU (with over 11K CUDA cores and 48 GB of memory). In Figure~\ref{fig:parallelization}, we plot the memory consumption and running time with respect to model size $n$. The space complexity plots exhibit roughly linear trends, confirming that the number of non-empty leaf nodes in the octree is $l=O(n)$. We observe that DWG's actual growth in time complexity is  approximately linear for $n \leq 10^7$, benefiting substantially from the extensive parallelization capabilities of modern hardware.

\paragraph{Memory Consumption}
\textcolor{black}{VIPSS~\cite{VIPSS}, PGR~\cite{lin2022surface} and AGR~\cite{Ma2024APGR} rely on solving dense linear systems, resulting in $O(n^2)$ space complexity. \citep{Kai_linear}'s method employs $O(n)$ sparse linear systems, which also lead to $O(n^2)$ space complexity. Both GCNO~\cite{xu2023globally} and BIM~\cite{BIM} utilize a Voronoi diagram to partition the space around the input point cloud $\mathcal{P}$, resulting in $O(m+n)$ space complexity, where $m (\gg n)$ is the number of Voronoi vertices. The octree-based algorithms, including \citet{mullen2010signing}'s method, SNO~\cite{Huang2024Stochastic},  iPSR~\cite{hou2022iterative}, WNNC~\cite{Lin2024fast} and DWG, use octrees to discretize the space. As the number of octree nodes is empirically $O(n\log n)$, these methods have $O(n\log n)$ space complexity. Computational results show that DWG exhibits lower memory consumption compared to both WNNC and iPSR, especially on large-scale models.}

\section{Conclusion \& Future Works}
\label{sec:conclusion}

This paper tackles the important task of efficiently reconstructing watertight surfaces from large-scale, unoriented point clouds. Our method, DWG, sets a new benchmark in runtime performance while also achieving superior accuracy and robustness compared to existing methods. This  demonstrates its potential for practical applications involving complex, real-world models. 

\begin{figure*}[htbp]
    \centering
     \newcommand{\sevensizea}{0.125}
     \includegraphics[width=\sevensizea\textwidth]{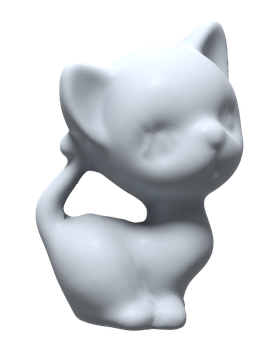}
    \includegraphics[width=\sevensizea\textwidth]{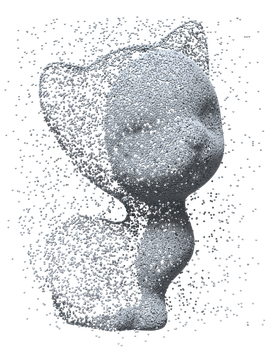}
    \includegraphics[width=\sevensizea\textwidth]{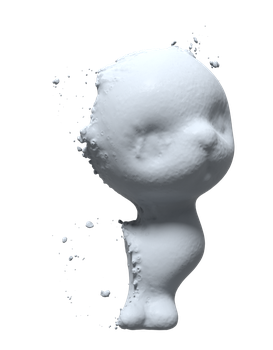}
    \includegraphics[width=\sevensizea\textwidth]{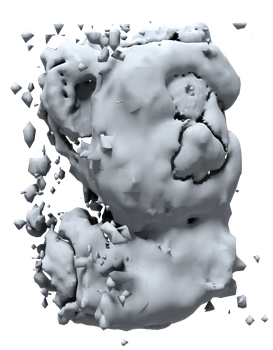}
    \includegraphics[width=\sevensizea\textwidth]{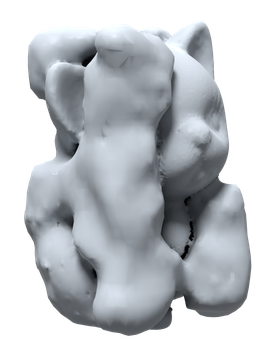}
    \includegraphics[width=\sevensizea\textwidth]{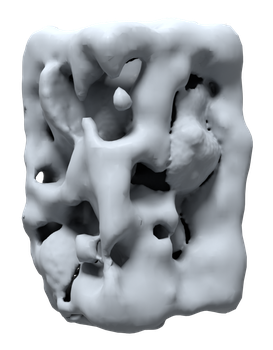}
     \includegraphics[width=\sevensizea\textwidth]{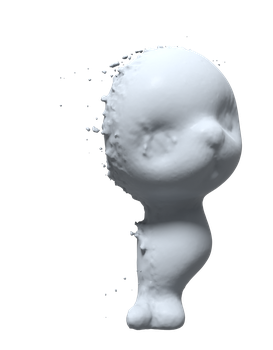}
    \\
    \makebox[\sevensizea\textwidth]{GT}
    \makebox[\sevensizea\textwidth]{Input}
    \makebox[\sevensizea\textwidth]{DWG }
     \makebox[\sevensizea\textwidth]{PGR}
    \makebox[\sevensizea\textwidth]{iPSR}
    \makebox[\sevensizea\textwidth]{SNO}
    \makebox[\sevensizea\textwidth]{WNNC}\\
    \vspace{-0.1in}
    \caption{\textcolor{black}{A failed case resulting from a highly uneven point distribution and outliers. We non-uniformly sampled 50K points from the Kitten model, concentrating the majority of points on one side and added 10\% outliers. This scenario, characterized by an uneven distribution of points combined with the presence of outliers, poses significant challenges to all existing methods, leading to incomplete or inaccurate reconstructions.}}
    \label{fig:failure}
\end{figure*}

There are several areas where DWG's effectiveness could be further enhanced: \textbf{1) Improving the Initialization Stage:} While the parallel diffusion process is fully executed on GPUs, the majority of the initialization process, including octree construction and computing the geodesic Voronoi area for each input point, is performed on CPUs. Our reported running times cover both the initialization and the parallel diffusion stages. Particularly for large-scale models, the initialization stage constitutes a substantial portion of the total processing time. For example, in the case of the Thai statue model, which contains 20 million points, the initialization phase at an octree depth of $d=13$ takes 276 seconds, which exceeds the 205 seconds required for the diffusion process. This considerable overhead highlights the need to fully transition the initialization stage to GPUs to enhance overall performance efficiency significantly. \textcolor{black}{\textbf{2) Handling of Highly Uneven Point Distributions:} DWG struggles with point clouds that exhibit highly uneven point distributions, where the GWN field cannot be robustly computed. Figure~\ref{fig:failure} illustrates such failed cases. These scenarios pose significant challenges to all existing 3D reconstruction methods, highlighting areas for further methodological improvements.}  \textbf{3) Enhancing Geometric Details:} DWG does not effectively reconstruct fine geometric details--a domain where the PSR family excels~\cite{kazhdan2013screened}. Although using the normals predicted by DWG in conjunction with sPSR has proven effective in addressing this issue, it remains an intriguing question whether the diffusion framework can inherently produce surfaces with rich geometric details. \textcolor{black}{\textbf{4) Proving Convergence:} Experimental results show that DWG, despite occasionally producing unsatisfactory results in cases of highly uneven point distributions, consistently  converges to a steady state across various initializations, including random, Gauss map and PCA. However, a rigorous theoretical proof of the convergence is still lacking, presenting an opportunity for future research to establish formal guarantees for the convergence behavior of DWG. }


\bibliography{reference}

\newcommand{\eachsize}{0.08}
\newcommand{\eachsizethree}{0.09}
\newcommand{\eachsizefour}{0.075}
\newcommand{\eachsizefive}{0.0705}
\newcommand{\nasize}{0.050}
\begin{figure*}[t] \centering
\setlength\tabcolsep{1pt}
\begin{footnotesize}
\begin{tabular}{ccccccccccc}
\includegraphics[width=\eachsize\textwidth]{figures/clear_models_hj/clear_pc/Bunny_1w_0.png}& 
\includegraphics[width=\eachsize\textwidth]{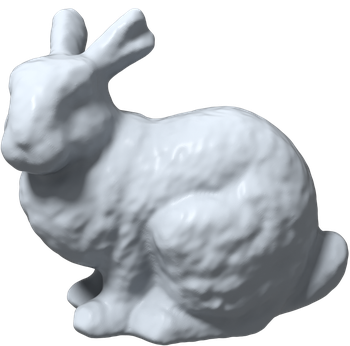}&
\includegraphics[width=\eachsize\textwidth]{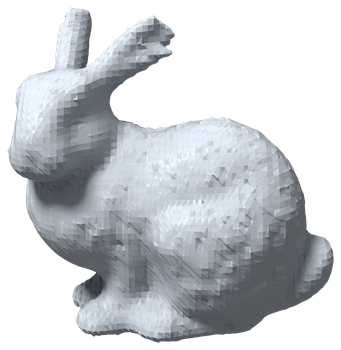}&
\includegraphics[width=\eachsize\textwidth]{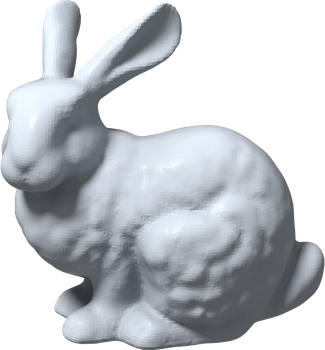}&
\includegraphics[width=\eachsize\textwidth]{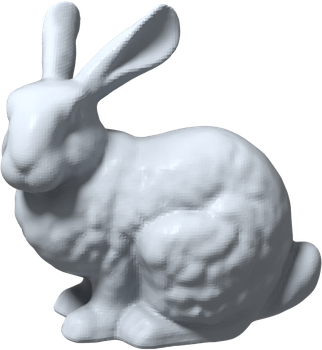}&
\includegraphics[width=\eachsize\textwidth]{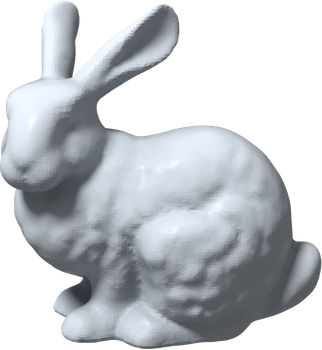}&
\includegraphics[width=\eachsize\textwidth]{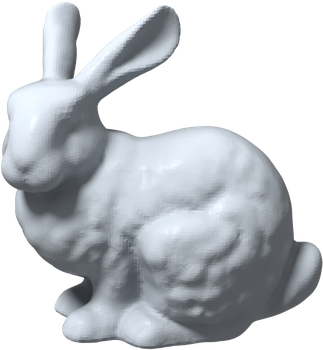}&
\includegraphics[width=\eachsize\textwidth]{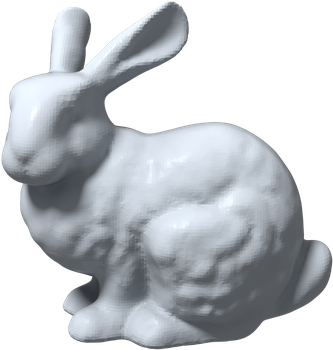}&
    \includegraphics[width=\eachsize\textwidth]{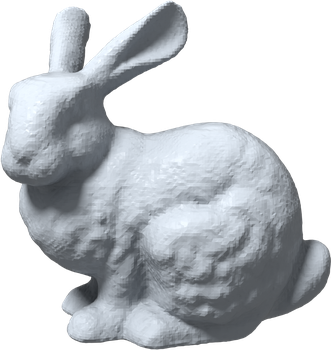}&
\includegraphics[width=\eachsize\textwidth]{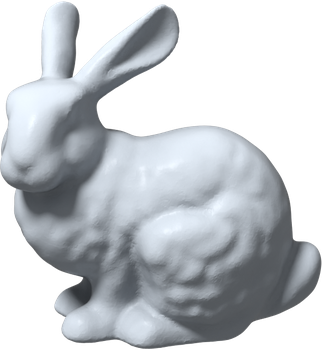}&
\includegraphics[width=\eachsize\textwidth]{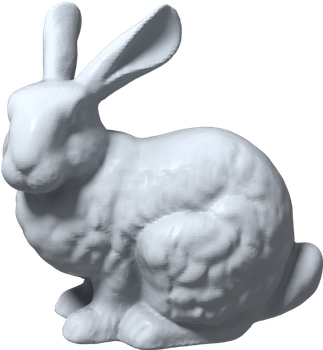}\\
 \includegraphics[width=\eachsize\textwidth]{figures/0.75_noise/0.75_noise_pc/Bunny_1w_0.75.png} &
 \includegraphics[width=\eachsize\textwidth]{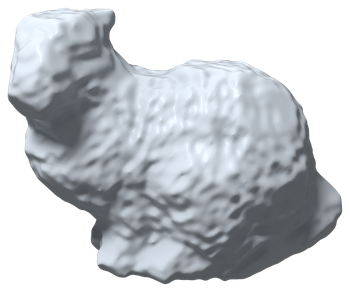}&
 \includegraphics[width=\eachsize\textwidth]{figures/0.75_noise/1/Bunny_1w_0_signed.png}&
    \includegraphics[width=\eachsize\textwidth]{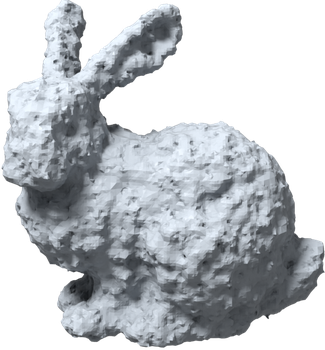} &
    \includegraphics[width=\eachsize\textwidth]{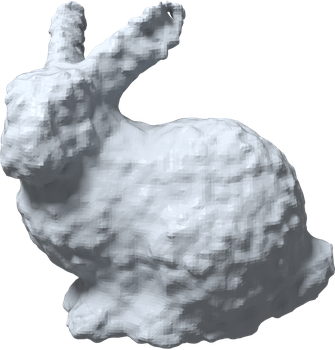} &
    \includegraphics[width=\eachsize\textwidth]{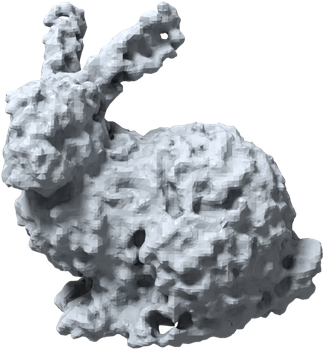} &
    \includegraphics[width=\eachsize\textwidth]{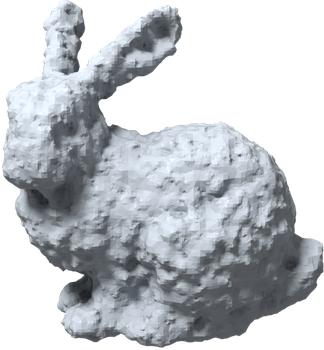} &
    \includegraphics[width=\eachsize\textwidth]{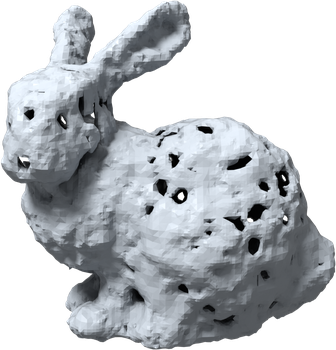}&
    \includegraphics[width=\eachsize\textwidth]{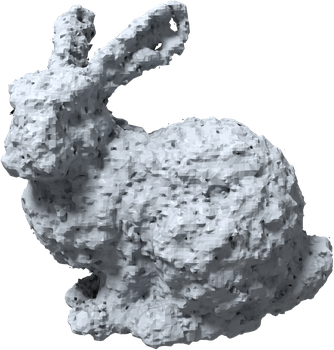}&
    \includegraphics[width=\eachsize\textwidth]{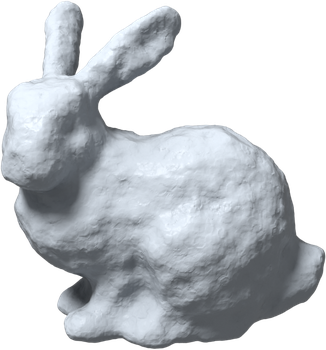} &
    \includegraphics[width=\eachsize\textwidth]{figures/0.75_noise/1/bunny.png}\\
\includegraphics[width=\eachsize\textwidth]{figures/clear_models_hj/clear_pc/A2_3w_0.png}&
\includegraphics[width=\eachsize\textwidth]{figures/NA.png}&
\includegraphics[width=\eachsize\textwidth]{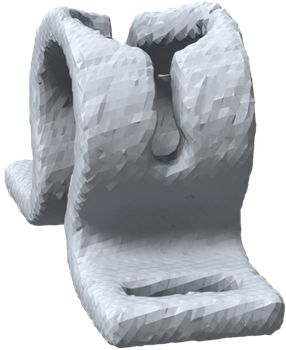}&
\includegraphics[width=\eachsize\textwidth]{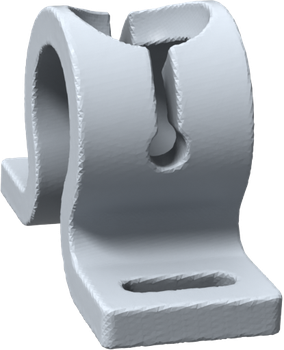}&
\includegraphics[width=\eachsize\textwidth]{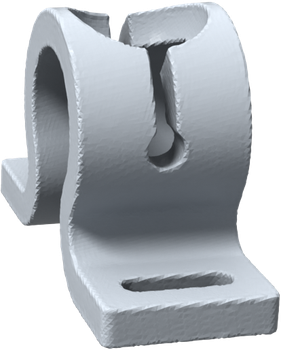}&
\includegraphics[width=\eachsizethree\textwidth]{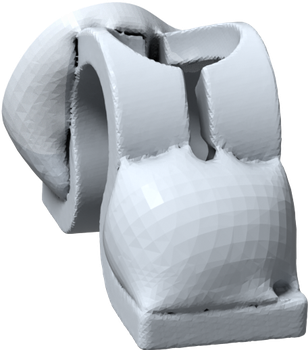}&
\includegraphics[width=\eachsize\textwidth]{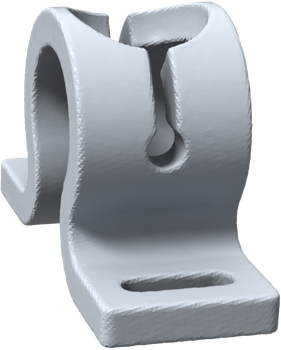}&
\includegraphics[width=\eachsize\textwidth]{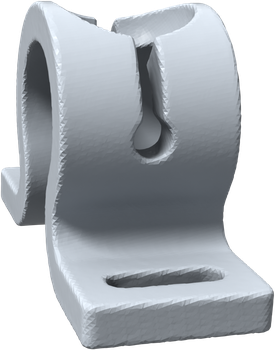}&
    \includegraphics[width=\eachsize\textwidth]{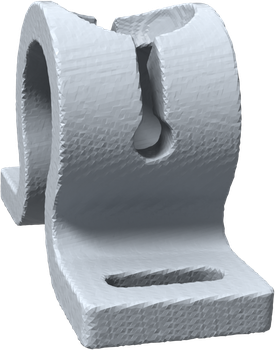}&
\includegraphics[width=\eachsize\textwidth]{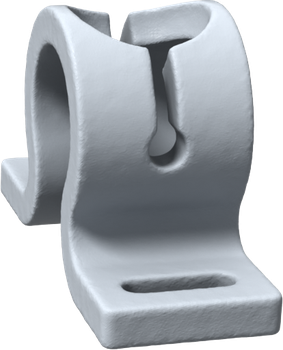}&
\includegraphics[width=\eachsize\textwidth]{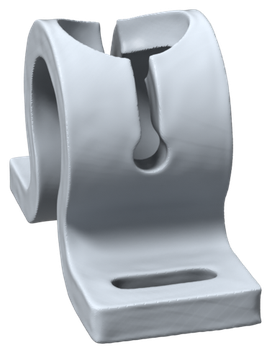}\\ 
    \includegraphics[width=\eachsize\textwidth]{figures/0.75_noise/0.75_noise_pc/A2_3w_0.75.png} &
    \includegraphics[width=\eachsize\textwidth]{figures/NA.png}&
    \includegraphics[width=\eachsize\textwidth]{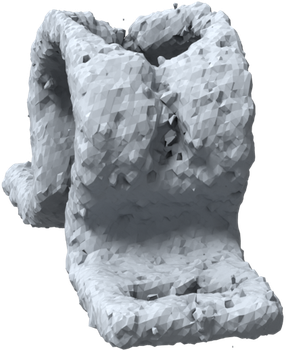}&
    \includegraphics[width=\eachsize\textwidth]{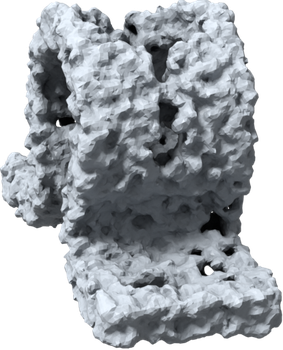} &
    \includegraphics[width=\eachsize\textwidth]{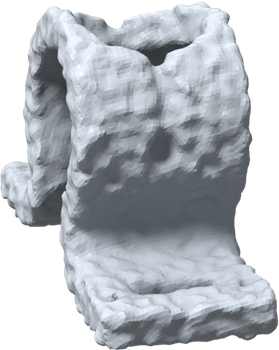} &
    \includegraphics[width=\eachsize\textwidth]{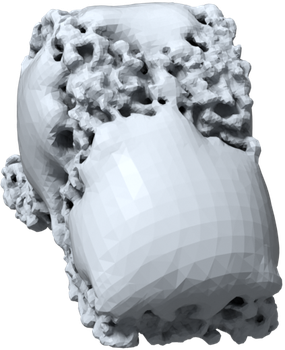} &
    \includegraphics[width=\eachsize\textwidth]{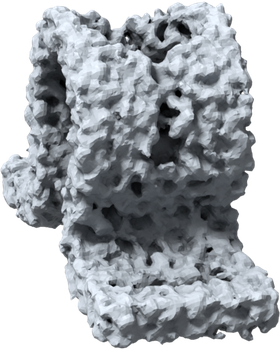} &
    \includegraphics[width=\eachsize\textwidth]{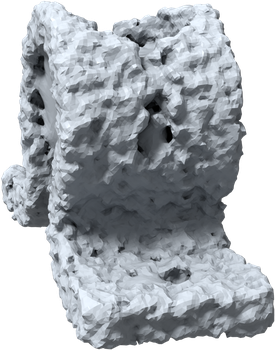}&
    \includegraphics[width=\eachsize\textwidth]{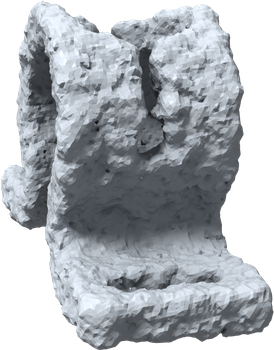}&
    \includegraphics[width=\eachsize\textwidth]{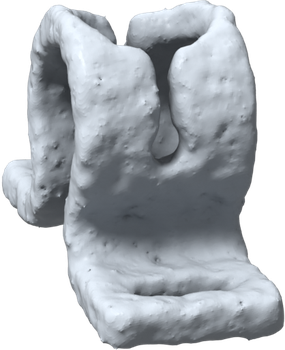} &
    \includegraphics[width=\eachsize\textwidth]{figures/0.75_noise/2/MechanicalPart-aspect.png}\\
    \includegraphics[width=\eachsizefour\textwidth]{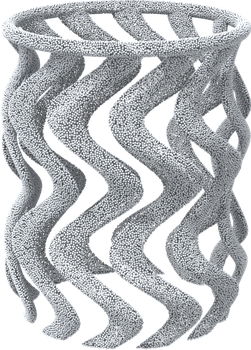}&
    \includegraphics[width=\eachsize\textwidth]{figures/NA.png}&
    \includegraphics[width=\eachsize\textwidth]{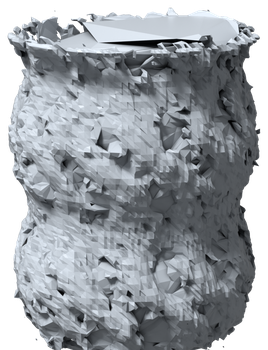}&
    \includegraphics[width=\eachsizefour\textwidth]{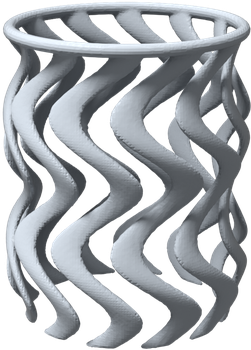}&
    \includegraphics[width=\eachsizefour\textwidth]{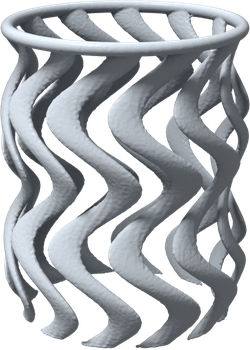}&
    \includegraphics[width=\eachsizefive\textwidth]{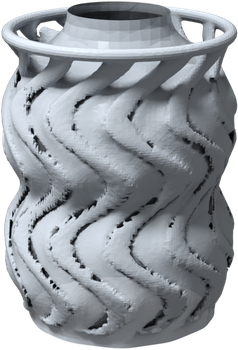}&
    \includegraphics[width=\eachsizefour\textwidth]{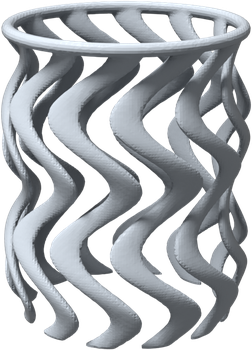}&
    \includegraphics[width=\eachsizefour\textwidth]{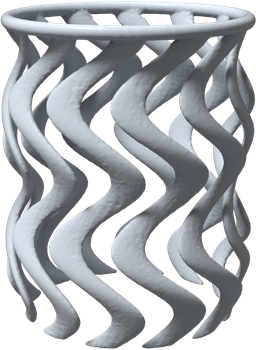}&
    \includegraphics[width=\eachsizefour\textwidth]{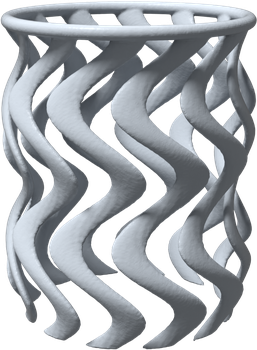}&
    \includegraphics[width=\eachsizefour\textwidth]{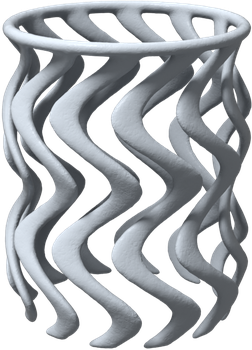}&
    \includegraphics[width=\eachsizefour\textwidth]{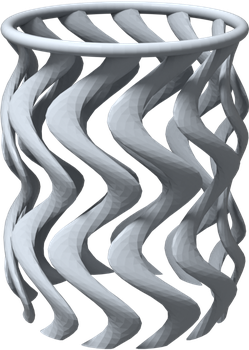}\\
    \includegraphics[width=\eachsize\textwidth]{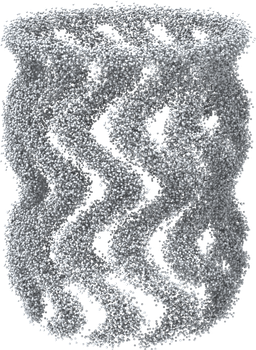}&
    \includegraphics[width=\eachsize\textwidth]{figures/NA.png}&
    \includegraphics[width=\eachsize\textwidth]{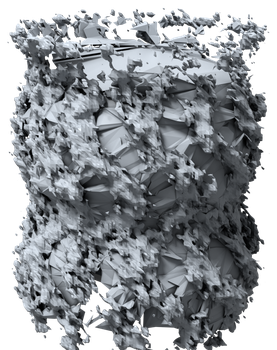}&
    \includegraphics[width=\eachsizefour\textwidth]{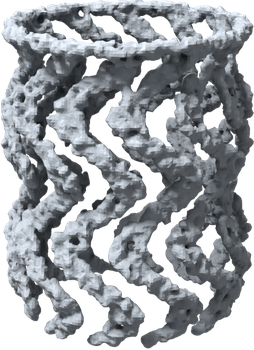} &
    \includegraphics[width=\eachsizefour\textwidth]{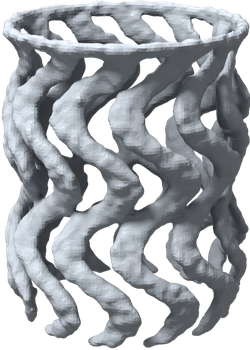} &
    \includegraphics[width=\eachsizefour\textwidth]{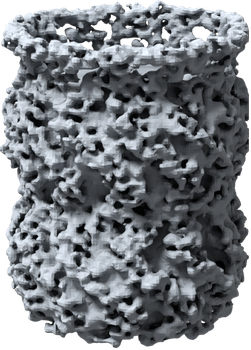} &
    \includegraphics[width=\eachsizefour\textwidth]{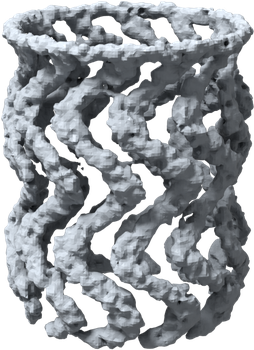} &
    \includegraphics[width=\eachsizefour\textwidth]{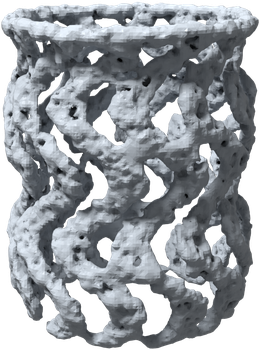}&
    \includegraphics[width=\eachsizefour\textwidth]{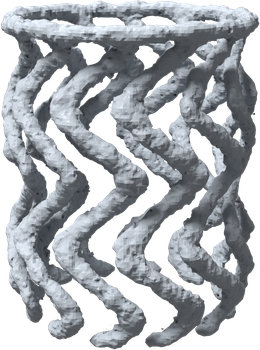}&
    \includegraphics[width=\eachsizefour\textwidth]{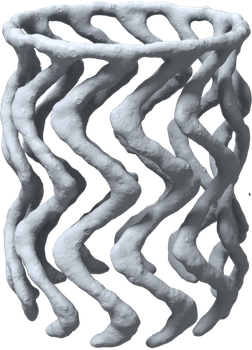} &
    \includegraphics[width=\eachsizefour\textwidth]{figures/0.75_noise/linkCupTop.png}\\
    \includegraphics[width=\eachsize\textwidth]{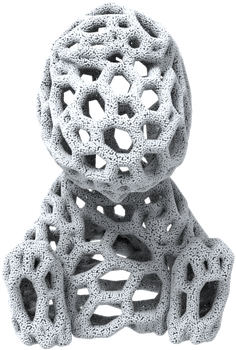}&
    \includegraphics[width=\eachsize\textwidth]{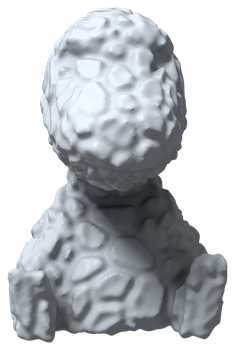}&
    \includegraphics[width=\eachsizefour\textwidth]{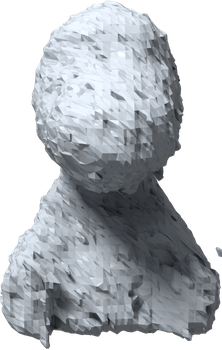}&
    \includegraphics[width=\eachsize\textwidth]{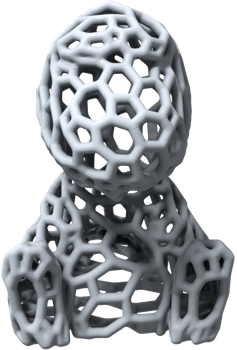}&
    \includegraphics[width=\eachsize\textwidth]{figures/NA.png}&
    \includegraphics[width=\eachsize\textwidth]{figures/NA.png}&
    \includegraphics[width=\eachsize\textwidth]{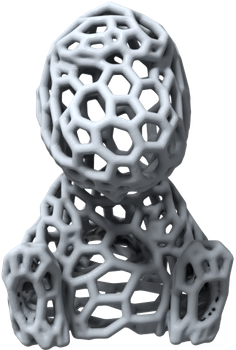}&
    \includegraphics[width=\eachsize\textwidth]{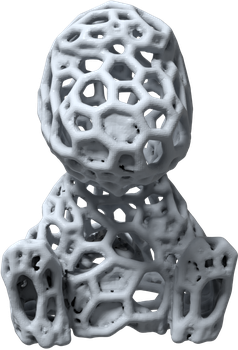}&
    \includegraphics[width=\eachsize\textwidth]{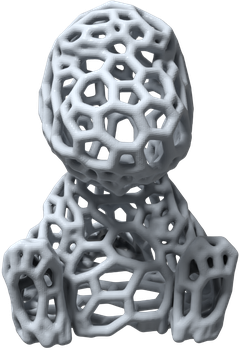}&
    \includegraphics[width=\eachsize\textwidth]{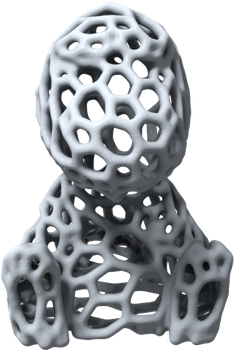}&
    \includegraphics[width=\eachsize\textwidth]{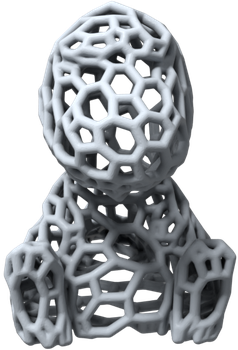}\\
    \includegraphics[width=\eachsize\textwidth]{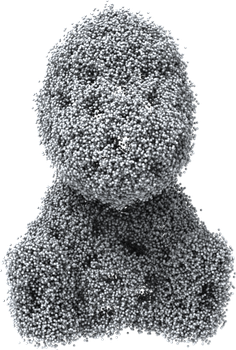} &
    \includegraphics[width=\eachsize\textwidth]{figures/NA.png}&
    \includegraphics[width=\eachsize\textwidth]{figures/NA.png}&
    \includegraphics[width=\eachsize\textwidth]{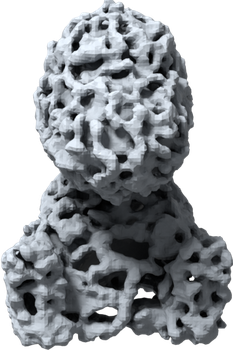} &
    \includegraphics[width=\eachsize\textwidth]{figures/NA.png} & 
    \includegraphics[width=\eachsize\textwidth]{figures/NA.png} &
    \includegraphics[width=\eachsize\textwidth]{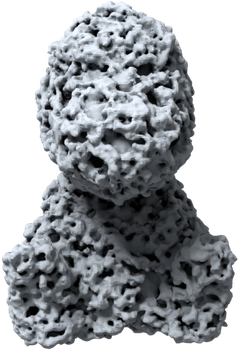} &
    \includegraphics[width=\eachsize\textwidth]{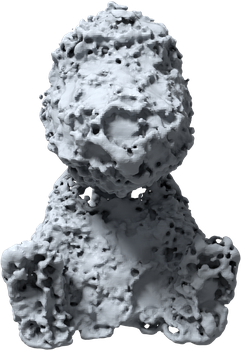}&
    \includegraphics[width=\eachsize\textwidth]{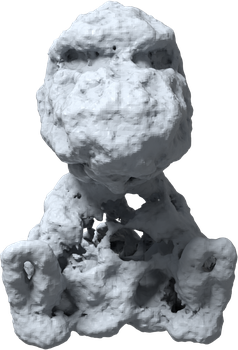}&
    \includegraphics[width=\eachsize\textwidth]{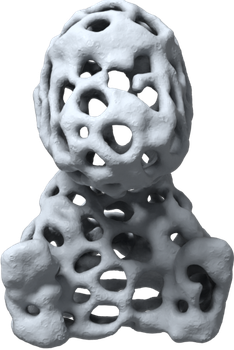} &
    \includegraphics[width=\eachsize\textwidth]{figures/0.75_noise/5/dino.png}\\
    Input & Hornung and Kobbelt & Mullen et al. & iPSR & PGR & GCNO & BIM & SNO & WNNC & DWG & GT\\
\end{tabular}
    \end{footnotesize}
    \vspace{-0.10in}
    \caption{Visual results on small-scale models tested under both noise-free conditions and with a noise level of 0.75\% relative to the diagonal of the bounding box. While all baseline methods perform well on the noise-free Bunny model, they struggle with high noise levels and thin structures. } 
    \label{fig:thinstructure}
\end{figure*}

\end{document}